\pdfoutput=1
\documentclass[11pt,a4paper]{article}
\usepackage{jheppub}
\usepackage{microtype}
\usepackage{dcolumn}
\usepackage{booktabs}

\usepackage{xspace}

\def\bfseries{\fontseries \bfdefault \selectfont \boldmath}

\newcommand{\Ba}[1]{\ensuremath{^{#1}\mathrm{Ba}}\xspace}
\newcommand{\Bi}[1]{\ensuremath{^{#1}\mathrm{Bi}}\xspace}
\newcommand{\Co}[1]{\ensuremath{^{#1}\mathrm{Co}}\xspace}
\newcommand{\Cs}[1]{\ensuremath{^{#1}\mathrm{Cs}}\xspace}
\newcommand{\K}[1]{\ensuremath{^{#1}\mathrm{K}}\xspace}
\newcommand{\Kr}[1]{\ensuremath{^{#1}\mathrm{Kr}}\xspace}

\newcommand{\Rb}[1]{\ensuremath{^{#1}\mathrm{Rb}}\xspace}
\newcommand{\Rn}[1]{\ensuremath{^{#1}\mathrm{Rn}}\xspace}
\newcommand{\Th}[1]{\ensuremath{^{#1}\mathrm{Th}}\xspace}
\newcommand{\Tl}[1]{\ensuremath{^{#1}\mathrm{Tl}}\xspace}

\newcommand{\Xe}[1]{\ensuremath{^{#1}\mathrm{Xe}}\xspace}

\newcommand{\Cu}[1]{\ensuremath{^{#1}\mathrm{Cu}}\xspace}
\newcommand{\In}[1]{\ensuremath{^{#1}\mathrm{In}}\xspace}

\newcommand{\Deu}{\ensuremath{^{2}\mathrm{H}}\xspace}

\newcommand{\bbnonu}{\ensuremath{0\nu\beta\beta}\xspace}
\newcommand{\bbtwonu}{\ensuremath{2\nu\beta\beta}\xspace}
\newcommand{\bb}{\ensuremath{\beta\beta}\xspace}
\newcommand{\Qbb}{\ensuremath{Q_{\beta\beta}}\xspace}
\newcommand{\Qb}{\ensuremath{Q_{\beta}}\xspace}
\newcommand{\mbb}{\ensuremath{\langle m_{\beta\beta} \rangle}}

\newcommand{\Ttwonu}{\ensuremath{T_{1/2}^{2\nu}}\xspace}
\newcommand{\Tnonu}{\ensuremath{T_{1/2}^{0\nu}}\xspace}

\begin{document}

\title{Demonstration of neutrinoless double beta decay searches in gaseous xenon with NEXT}

\collaboration{The NEXT Collaboration}
\author[19]{P.~Novella}
\author[19]{M.~Sorel,}
\author[19]{A.~Us\'on,}
\author[2]{C.~Adams,}
\author[18]{H.~Almaz\'an,}
\author[26]{V.~\'Alvarez,}
\author[23]{B.~Aparicio,}
\author[22]{A.I.~Aranburu,}
\author[7]{L.~Arazi,}
\author[20]{I.J.~Arnquist,}
% ...
\affiliation[2]{
Argonne National Laboratory, Argonne, IL 60439, USA}
% ...
\affiliation[7]{
Unit of Nuclear Engineering, Faculty of Engineering Sciences, Ben-Gurion University of the Negev, P.O.B. 653, Beer-Sheva, 8410501, Israel}
% ...
\affiliation[18]{
Department of Physics and Astronomy, Manchester University, Manchester. M13 9PL, United Kingdom}
\affiliation[19]{
Instituto de F\'isica Corpuscular (IFIC), CSIC \& Universitat de Val\`encia, Calle Catedr\'atico Jos\'e Beltr\'an, 2, Paterna, E-46980, Spain}
\affiliation[20]{
Pacific Northwest National Laboratory (PNNL), Richland, WA 99352, USA}
% ...
\affiliation[22]{
Department of Applied Chemistry, Universidad del Pais Vasco (UPV/EHU), Manuel de Lardizabal 3, San Sebasti\'an / Donostia, E-20018, Spain}
\affiliation[23]{
Department of Organic Chemistry I, University of the Basque Country (UPV/EHU), Centro de Innovaci\'on en Qu\'imica Avanzada (ORFEO-CINQA), San Sebasti\'an / Donostia, E-20018, Spain}
% ...
\affiliation[26]{
Instituto de Instrumentaci\'on para Imagen Molecular (I3M), Centro Mixto CSIC - Universitat Polit\`ecnica de Val\`encia, Camino de Vera s/n, Valencia, E-46022, Spain}
% ...

\emailAdd{pau.novella@ific.uv.es}

\abstract{The NEXT experiment aims at the sensitive search of the neutrinoless double beta decay in \Xe{136}, using high-pressure gas electroluminescent time projection chambers. The NEXT-White detector is the first radiopure demonstrator of this technology, operated in the Laboratorio Subterr\'aneo de Canfranc. Achieving an energy resolution of 1\% FWHM at 2.6 MeV and further background rejection by means of the topology of the reconstructed tracks, NEXT-White has been exploited beyond its original goals in order to perform a neutrinoless double beta decay search. The analysis considers the combination of 271.6 days of \Xe{136}-enriched data and 208.9 days of \Xe{136}-depleted data. A detailed background modeling and measurement has been developed, ensuring the time stability of the radiogenic and cosmogenic contributions across both data samples. Limits to the neutrinoless mode are obtained in two alternative analyses: a background-model-dependent approach and a novel direct background-subtraction technique, offering results with small dependence on the background model assumptions. With a fiducial mass of only 3.50$\pm$0.01 kg of \Xe{136}-enriched xenon, 90\% C.L. lower limits to the neutrinoless double beta decay are found in the \Tnonu$>5.5\times10^{23}-1.3\times10^{24}$ yr range, depending on the method. The presented techniques stand as a proof-of-concept for the searches to be implemented with larger NEXT detectors.}

\maketitle
       
\clearpage

\section{Introduction}
\label{sec:intro}

%---- bb0nu ----%

The oscillation experiments in the last decades have demonstrated that neutrinos are not massless particles, as described in the Standard Model, and that the lepton flavor is not conserved. Revealing the nature of neutrino masses is therefore one of the major goals in particle physics. Regardless of the underlying decay mechanism, the observation of the neutrinoless double beta decay (\bbnonu) has been identified as the most practical way to establish that neutrinos are Majorana particles, that is, fermions equivalent to their anti-particles. When the beta decay is highly suppressed or energetically forbidden, even-even nuclei can undergo double beta decay (\bb), a process in which two bound neutrons are simultaneously transformed into two protons plus two electrons. The mode of this decay emitting two antineutrinos (\bbtwonu) has been directly observed in ten nuclides with half-lives in the range of $\sim$10$^{19}$--10$^{21}$~yr \cite{Barabash:2020nck}. However, the neutrinoless mode, which would violate lepton number by two units, has not been observed yet and the best limits to the half-life of this process exceed 10$^{26}$~yr \cite{KamLAND-Zen:2022tow,GERDA:2020emj}.

%---- NEXT ----%

The NEXT collaboration aims at the competitive search for the \bbnonu decay in \Xe{136} using high-pressure gas electroluminescent time projection chambers (TPCs). The xenon TPC provides both primary scintillation light (S1) and ionization electrons when charged particles interact in the active volume. The ionization electrons are drifted by an electric field towards the anode of the TPC, where they enter a more intense field region and produce secondary scintillation light (S2) by means of electroluminescence (EL). The TPC is equipped with two dedicated readout planes, located behind the cathode and the anode. While the amplitude of the S1 and S2 signals is registered by the so-called energy plane (EP), respectively providing the start time of the event and its total energy, the topological information of the involved tracks is measured in the tracking plane (TP). The NEXT-White detector \cite{NEXT:2018rgj} implements the first radiopure demonstrator of the NEXT technology, operating underground in the Laboratorio Subterr\'aneo de Canfranc (LSC). As demonstrated with NEXT-White, this technology offers an excellent energy resolution of $\sim$1\% FWHM at 2.6 MeV \cite{Renner:2019pfe} (above the $Q_{\beta\beta}$ of \Xe{136}, $2457.8\pm 0.4$~MeV \cite{Redshaw:2007un}) and topological information providing an efficient background rejection \cite{NEXT:2019gtz,NEXT:2020try,NEXT:2020jmz}, ultimately resulting in low background conditions \cite{Novella:2019cne}. This performance is expected to be further improved as presented in \cite{NEXT:2020amj}. In addition, NEXT also offers promising \Ba{136} (\Xe{136} daughter) tagging capabilities, which would lead to future background-free detectors \cite{Jones:2016qiq,McDonald:2017izm,Thapa:2019zjk,Rivilla:2020cvm,acssensors.0c02104,NEXT:2022ita}.

%---- bb analysis approaches ----%

Although NEXT-White was conceived as a prototype of the NEXT-100 detector \cite{NEXT:2015wlq} and a future ton-scale device \cite{NEXT:2020amj}, its outstanding performance has allowed the measurement of the half-life of the \bbtwonu decay in \Xe{136} \cite{NEXT:2021dqj}. In the \bbtwonu analysis, two techniques were exploited for the first time in the field of double beta decay searches: 1) a Richardson-Lucy deconvolution delivering high-definition tracks, in turn boosting the background rejection, and 2) a direct subtraction of the remaining backgrounds, other than \Xe{137}, by combining the data collected with xenon enriched in \Xe{136} and xenon depleted in this isotope. The latter is of particular interest for future \bb experiments, as it provides results with small dependence on the background model. On the other hand, this paper discusses the first \bbnonu search with a NEXT detector, demonstrating the capabilities of this technology even with a limited fiducial mass. Adopting the same data samples and analysis strategies used in \cite{NEXT:2021dqj}, this analysis incorporates for the first time the contribution of the cosmogenic backgrounds in NEXT, which are particularly relevant in the region of interest for the \bbnonu signal. The common data processing and analysis techniques developed for \bb analyses with the NEXT technology are presented in detail in this publication.

%---- Paper sections ----%

This paper is organized as follows. Sec.~\ref{sec:new} presents a description of the NEXT-White detector, as well as the operation conditions during the data taking periods devoted to this analysis. A description of the detector and data simulation is also provided. The reconstruction, calibration, and selection procedures, applied to both real data and Monte Carlo samples, are discussed in Sec.~\ref{sec:process}. A measurement of the radiogenic-induced and cosmogenic-induced backgrounds is performed in Sec.~\ref{sec:bg}, together with a measurement of their time stability. Finally, Sec.~\ref{sec:bb} presents the \bbnonu analysis and the limits obtained.

\section{The NEXT-White detector}
\label{sec:new}

NEXT-White implements the first radiopure large-scale demonstrator of the NEXT technology, operating underground in Hall A of the LSC from 2016 to 2021. While a series of small prototypes proved the detection principles \cite{NEXT:2012lrw,NEXT:2015rel}, the NEXT-White detector holds enough xenon mass to demonstrate the performance of the NEXT technology at large scales. It also offers the possibility to perform a significant measurement of the \bbtwonu half-life, as shown in \cite{NEXT:2021dqj}. Although this mass is not enough to realize a competitive search for the \bbnonu process, the excellent performance of NEXT-White has been fully exploited in order to provide a proof-of-concept for \bbnonu searches in the near-future NEXT detectors \cite{NEXT:2015wlq,NEXT:2020amj}.

\subsection{Detector description}
\label{sec:detector}

The NEXT-White detector has been described in detail in \cite{NEXT:2018rgj}. Inside a stainless steel pressure vessel, the active volume of the TPC is a cylindrical region of 530.3~mm along the drift direction with a radius of 208~mm. A total mass of $\sim$4.3 kg is contained within this volume when the detector is operated at 10 bar. This drift region is established between a gate grid and a transparent cathode located at the opposite sides of the TPC. A second, more intense, field region exists for signal amplification purposes. The EL region is defined between the gate grid and the anode, consisting of a 3~mm-thick fused silica plate coated with indium tin oxide (ITO) and located at a distance of 6~mm behind the gate. The field cage is made of copper rings inserted into a high-density polyethylene cylindrical shell, covered with polytetrafluoroethylene (PTFE) panels to enhance the light collection in the EP. The amplitude of the S1 and S2 signals is measured with an array of 12 Hamamatsu R11410-10 3-inch photomultiplier tubes (PMTs) located 13~cm behind the transparent cathode. The PMTs are distributed in a circular layout, with 3 of them in the center. As they cannot withstand the pressure, the PMTs are isolated from the active volume and optically coupled to the xenon gas through sapphire windows. The topology of the events is registered with a 10~mm-pitch array of 1792 SensL series-C 1 mm$^2$ silicon photomultipliers (SiPMs), located 2~mm behind the anode plate. The SiPM array consists of 28 square boards (8 x 8 pixels each) following an approximate hexagonal coverage. All the inner surfaces of the TPC and the two readout planes are coated with a thin wavelength-shifting layer of tetraphenyl butadiene (TPB) in order to shift the vacuum ultraviolet (VUV) light to the visible spectrum. A shield of 60--120~mm thick ultrapure copper is located inside the pressure vessel, surrounding the field cage and supporting the two readout planes.

\begin{figure}
  \begin{center}
    \includegraphics[width=0.47\textwidth]{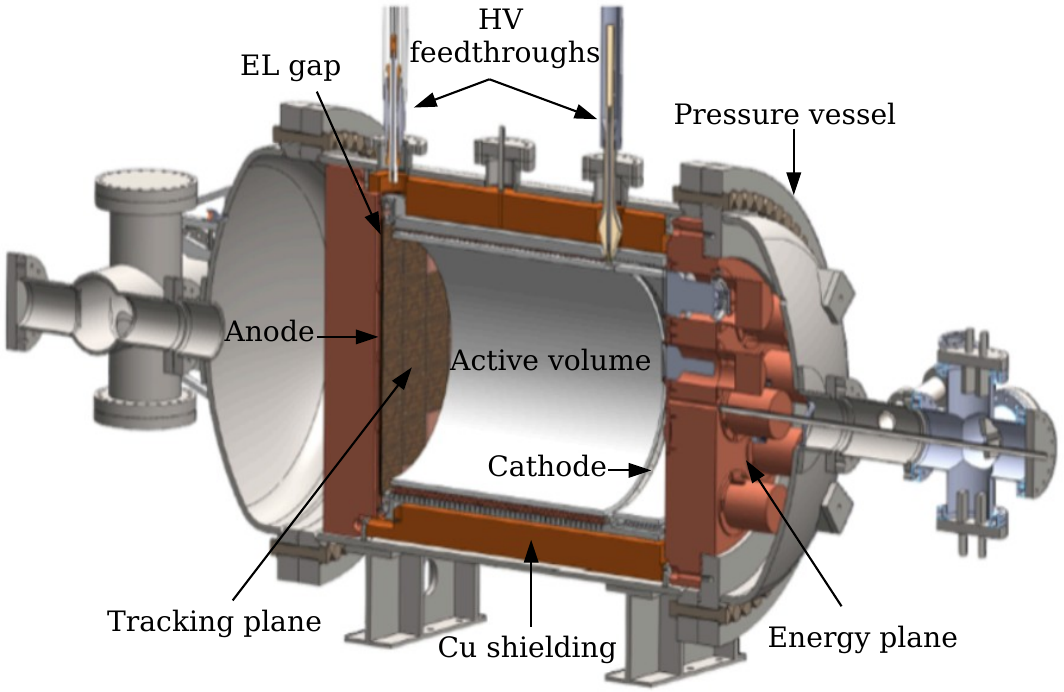}
    \includegraphics[width=0.47\textwidth]{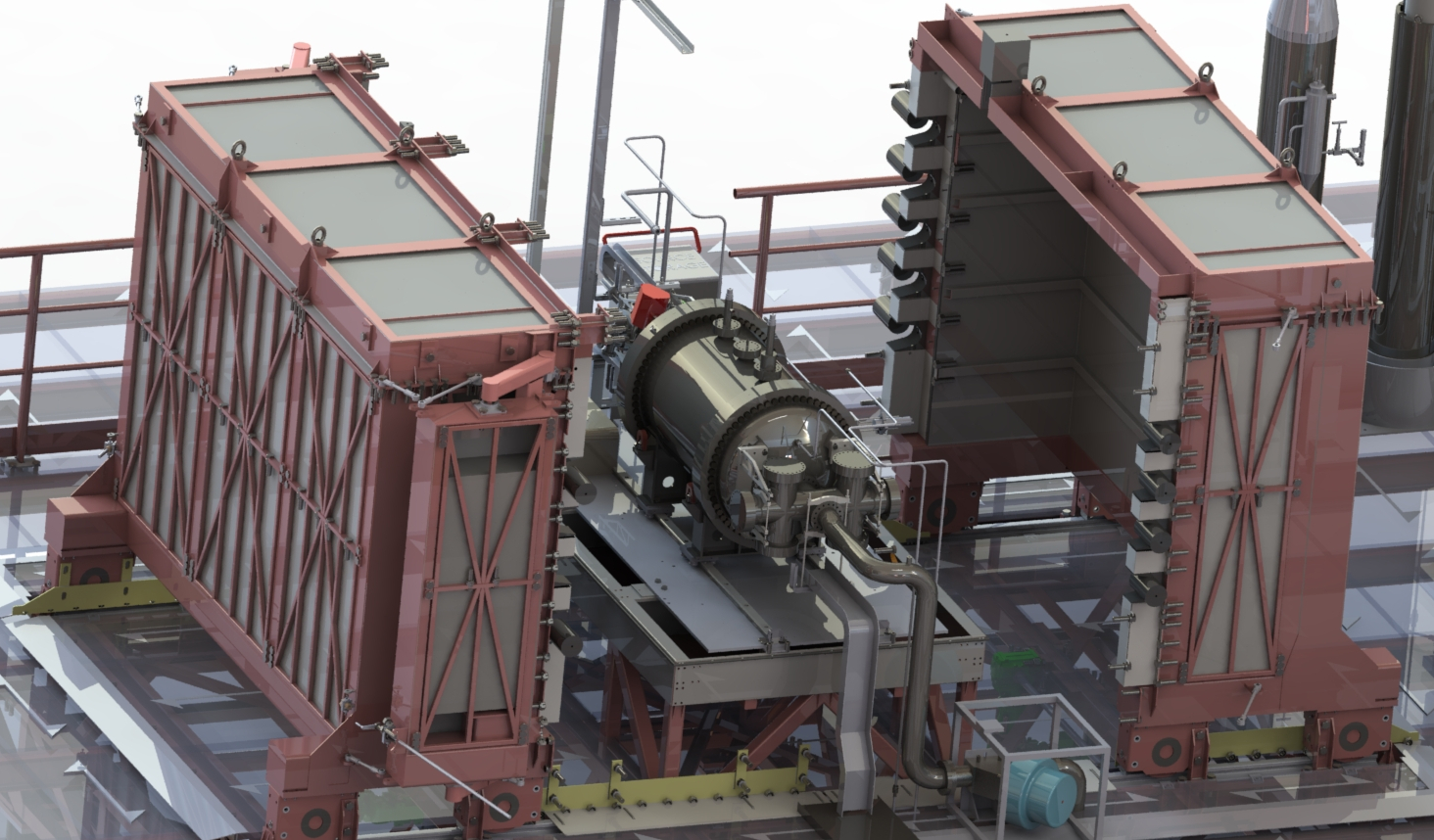} 
    \caption{Schematic view of NEXT-White (left) and render of the detector placed on the seismic platform and inside the movable outer lead castle in its open configuration (right). The inner lead castle is not shown for clarity.}
    \label{fig:new}
  \end{center}
\end{figure}

In order to provide further protection against external backgrounds, two lead-based structures surround the pressure vessel. A fixed inner lead castle (ILC) is placed on the same seismic platform as NEXT-White, while a movable outer lead castle (OLC) made of 20 cm thick bricks encloses the entire detector, the feedthroughs, and the platform. A schematic view of the NEXT-White detector, as well as the OLC, is shown in Fig.~\ref{fig:new}. A radon abatement system (RAS) by ATEKO A.S. flushes radon-free air into the air volume inside the OLC, reducing the \Rn{222} content by 4--5 orders of magnitude compared to LSC Hall A air \cite{Novella:2018ewv}. As demonstrated in \cite{Novella:2019cne}, the RAS has provided a virtually airborne-Rn-free environment for the operation of NEXT-White.

\subsection{Operation conditions}
\label{sec:data}

%--- operation conditions ---%

The data samples considered in the current analysis comprise different calibration and low-background data taking campaigns from 2019 to 2021. During this time, the same operation conditions have been kept, and the detector performance has remained stable. The gas pressure, drift field, and EL field have been set to $\sim$10.2~bar, 0.4~kV/cm, and 1.3~kV/(cm$\cdot$bar), respectively. Continuous detector calibration and monitoring have been carried out with a \Kr{83m} low-energy (41.5~keV) calibration source as shown in \cite{Martinez-Lema:2018ibw}. This is possible thanks to a dual-trigger implementation in the data acquisition system (DAQ), which allows for the collection of both low-energy (below $\sim$100 keV) and high-energy (above $\sim$400 keV) triggers within the same DAQ run, respectively registering \Kr{83m} and low-background or \Tl{208} calibration events. Although the rate of krypton events evolves in time (according to the activity of the parent \Rb{83} source), it is kept above 30 Hz. This relatively high rate induces a typical DAQ dead-time of 2--6\% which has been measured on a daily basis. According to the \Kr{83m} data, the electron drift velocity is found to be stable within 1\%, with a value around 0.92~mm/$\mu$s. The gas purity has been continuously improving due to the recirculation through a MonoTorr PS4-MT50-R SAES heated getter, operated at a flow rate of 100 slpm, with remaining impurities below 1.0 ppb. The electron drift lifetime has ranged from $\sim$5~ms at the beginning of the data taking to $\sim$13~ms at the end. As the maximum drift time in NEXT-White is $\sim$0.6~ms, this lifetime corresponds to a small electron attachment, which is corrected for within the time-dependent calibration procedure. According to the specific operation conditions, the light yield is $\sim$300 photo-electrons (p.e.) per keV. Typical time variations below 10\% are also corrected for, allowing to achieve a stable energy resolution around 4\% FWHM at 41.5~keV. The trigger efficiency for high energy events has been evaluated with dedicated data taking campaigns and an offline simulation applied to randomly selected waveforms. The trigger condition consists of a valid S2 signal (amplitude above 10$^5$ ADC, corresponding to $\sim$4000 p.e., and width above 2~$\mu$s) seen in time coincidence ($<1.3~\mu$s) by two of the central PMTs. This leads to a trigger threshold of about 200 keV. The trigger efficiency is found to reach a plateau of 97.6$\pm$0.2\% for events above $\sim$400 keV, with the inefficiency being due to the time coincidence requirement. For each recorded trigger, a DAQ event includes the 12 PMT waveforms sampled at 25~ns and the 1792 SiPM zero-suppressed waveforms sampled at 1~$\mu$s. The front-end electronics shapes, filters and amplifies the fast signals produced by the PMTs (less than 5 ns wide). The resulting bandwidth is about 3 MHz, stretching single-photoelectron signals to a width of about 150 ns. All sensor waveforms extend for a 1.6~ms buffer size, with 0.8~ms pre-trigger information.

%--- Isotopic compositions---%

\begin{figure}[ht]
  \begin{center}
    \includegraphics[width=0.49\textwidth]{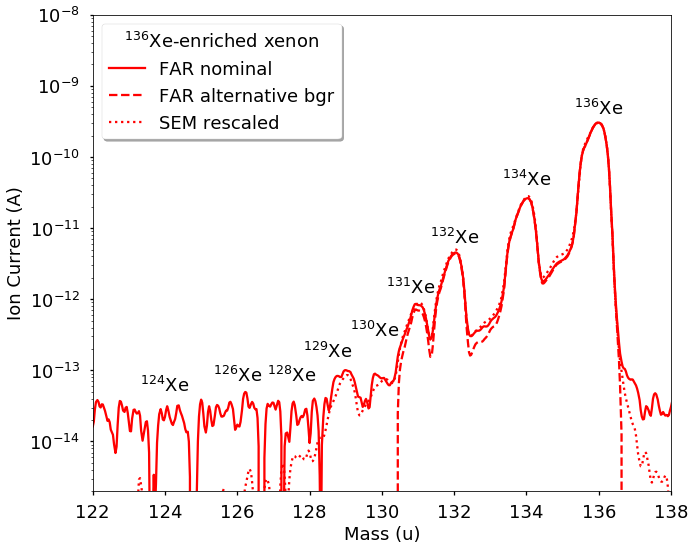} 
     \includegraphics[width=0.49\textwidth]{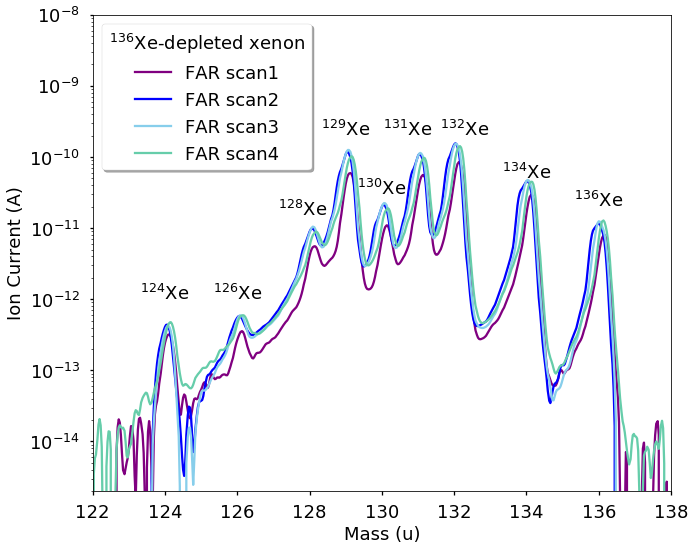}
     \caption{Isotopic composition of xenon gas. Left panel: RGA scans for \Xe{136}-enriched gas. Right panel: RGA scans for \Xe{136}-depleted gas. See text for details.} 
    \label{fig:isocomp}
  \end{center}
\end{figure}

The data used for the \bb analysis comprises low-background samples taken with xenon enriched in \Xe{136} and xenon depleted in this isotope. The isotopic compositions of the \Xe{136}-enriched and the \Xe{136}-depleted gas have been measured with a residual gas analyzer (RGA) from Pfeiffer Vacuum, focusing on the atomic mass range corresponding to the nine stable (or very long-lived) isotopes of xenon (\Xe{124}, \Xe{126}, \Xe{128}, \Xe{129}, \Xe{130}, \Xe{131}, \Xe{132}, \Xe{134} and \Xe{136}). The RGA measurements for the two xenon samples are shown in Fig.~\ref{fig:isocomp}. For the measurements of the \Xe{136}-enriched gas, the RGA has been operated in both Faraday plate (FAR) and Secondary Electron Multiplier (SEM) modes. In FAR mode, the RGA electrode captures directly the charge of xenon ions. For relatively large concentrations (hence, ion currents) as in the \Xe{136} case, this is the most robust measurement, and is used as default. In SEM mode, the RGA coating emits many electrons per incoming ion. This mode therefore yields better signal-to-noise, but the absolute calibration is more complicated. A background subtraction has been performed according to two successive measurements of the RGA vacuum. The \Xe{136} isotopic fraction is found to be $90.9\pm 0.4\%$, where the systematic uncertainty is derived from the differences in the FAR and SEM measurements and different integration windows. For the measurements of the \Xe{136}-depleted gas, only four scans in FAR mode have been made. In this case, we define the isotopic fraction central values from the average over the four scans, and the isotopic fraction errors from the RMS between the four scans. For \Xe{136}, a fraction of $2.6\pm0.2\%$ is obtained.

The first low-background (OLC closed, and no high-energy calibration sources deployed) data taking period with \Xe{136}-enriched gas (hereafter Run-V) has been conducted from February 2019 to June 2020. During this run, the detector performance has remained stable despite the two gas recoveries that took place in order to perform minor maintenance interventions. A total exposure of 271.6 days has been reached. A second low-background period with \Xe{136}-depleted gas (hereafter Run-VI) was carried out from October 2020 to June 2021, for a total run time of 208.9 days. According to the \Kr{83m} rate variations, the integrated DAQ live-time during Run-V and Run-VI is 97.04$\pm$0.01\% and 97.86$\pm$0.01\%, respectively. Although the performance of NEXT-White has not been significantly impacted by the change of gas, the operation pressure has been slightly different between Run-V and Run-VI due to the uncertainty in the amount of gas used in the refilling process. The evolution of the gas density over time has been monitored across Run-V and Run-VI. Beyond minor time variations ($<$1\%) within each period, the cryogenic recovery of the gas and the refilling of the detector induce typical variations of few percent in the gas density. In particular, the integrated electron number density in the gas during Run-VI is 1.9$\pm$0.2\% larger than in Run-V. This induces a relative increase in the observed background event rates for $E>1$~MeV of 2.4$\pm$0.6\% (as derived from MC studies) due to the reduction in the gamma-ray attenuation length and the larger probability of multi-Compton interactions. The average gas temperature in the active volume of NEXT-White is known within a 0.5 K uncertainty, inferred from the temperature spread among the sensors mounted at various locations in the detector and Hall A of the LSC. This translates into a 0.2\% uncertainty in the total number of Xe atoms.

\subsection{Simulation}
\label{sec:sim}

A full simulation of NEXT-White has been implemented in GEANT4. We use the G4EmStandardPhysics\_option4 physics list, with the fluorescence and Auger emission activated. In the range of energies of the NEXT experiment (below 10 MeV), the following models are adopted for the photon and $e^{-}/e^{+}$ interactions \cite{G4Phys}: BetheHeitler5D model for the gamma-induced pair production, Monarsh University model (G4LowEPComptonModel) for the Compton scattering, Livermore models for photo-electric effect and Rayleigh scattering,  Goudsmit-Sounderson model for multiple Coulomb scattering, eBremSB model for bremsstrahlung, Penelope model for ionization, and eplus2gg model for positron annihilation. Both production cuts and step size limiters are used and set to 1 mm. While larger step sizes lead to observable differences between data and MC, 1 mm is found to be a good compromise between simulation time and performance. It is also well above the spatial resolution of the NEXT-White detector, according to the diffusion and the SiPM pitch size.

\begin{figure}
  \begin{center}
    \includegraphics[width=0.80\textwidth]{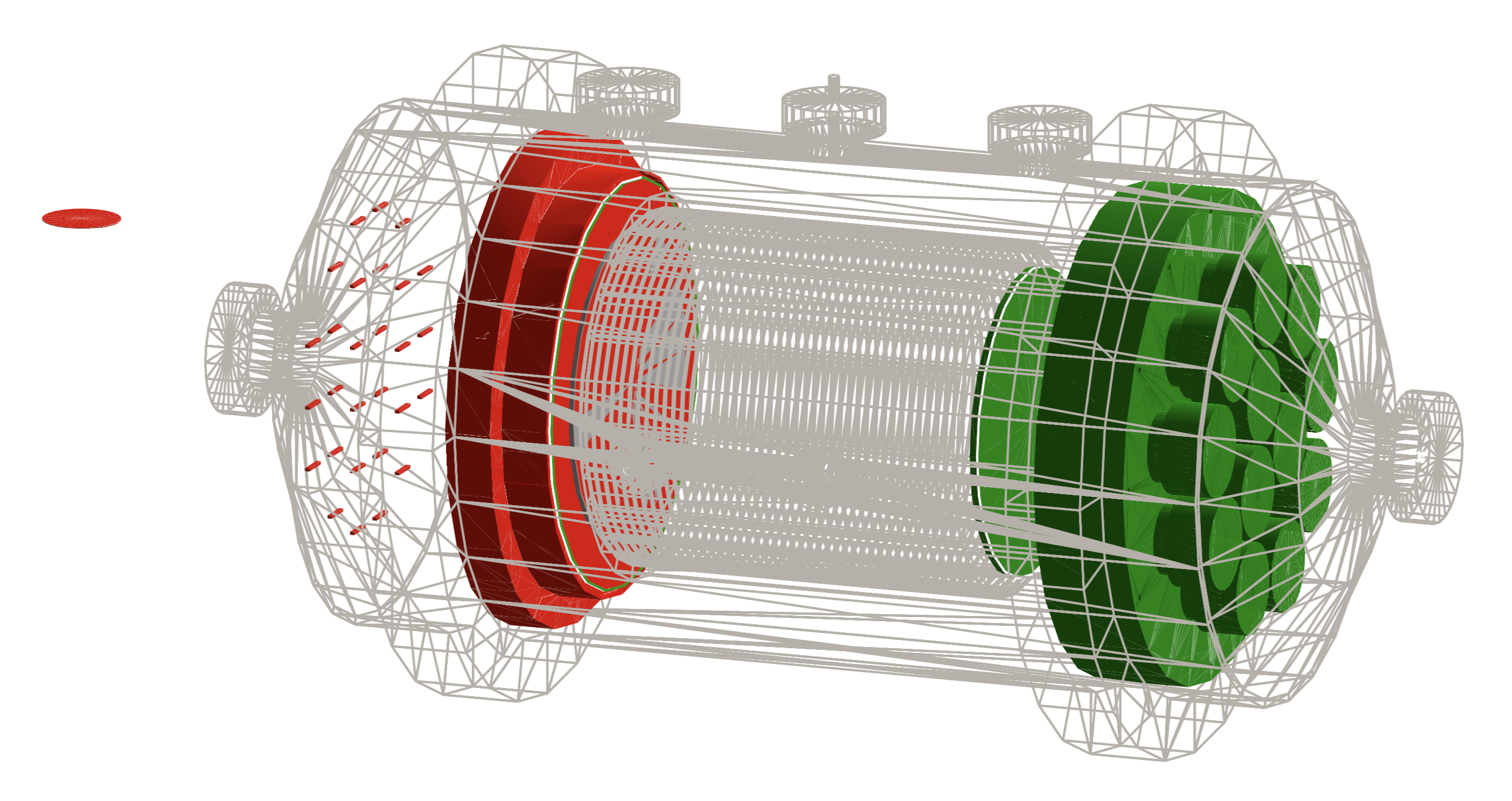} 
    \caption{GEANT4 description of the NEXT-White geometry. The red, green, and gray volumes in the figure correspond to the ANODE, CATHODE, and OTHER effective volumes, respectively. The disk placed outside the pressure vessel and assigned to the ANODE volume corresponds to SiPM electronics components. For clarity, the geometry of the shielding structures is not shown.}
    \label{fig:nextwhite_fitvolumes}
  \end{center}
\end{figure}

A total of 23 GEANT4 volumes have been described, representing the main components of the detector as well as the shielding structures. For analysis purposes, the 23 volumes are grouped into three effective spatial categories: ``ANODE'', ``CATHODE'' and ``OTHER''. The ANODE and CATHODE categories include all GEANT4 volumes placed in, or near, the two detector end-caps. The OTHER category includes inner volumes in the detector barrel region, the pressure vessel, and external materials such as the ones in the shielding structure. A visual representation of these effective volumes is shown in Fig.~\ref{fig:nextwhite_fitvolumes}. A number of event generators have also been implemented so that \bb, radiogenic background, muon and calibration events can be simulated. Starting from these generators, the simulation of the particle interactions and propagation results in energy deposits in the active volume, which are stored for further processing. All simulations are performed at 10.1~bar pressure and 300~K temperature, with a GEANT4 maximum step size of 1~mm.

Starting from the GEANT4 events, a full simulation of the detector response is performed. First, a simulation of the electron drift and the light production, propagation, and detection is conducted, producing ideal PMT and SiPM waveforms. An electron drift velocity of 1 mm/$\mu$s is adopted. For the electron drift along the Z axis, a transverse and longitudinal diffusion of 1.072~mm/$\sqrt{{\rm cm}}$ and 0.267~mm/$\sqrt{{\rm cm}}$ is assumed, respectively, based on \cite{NEXT:2018kzp}. The electron drift lifetime is simulated according to a lifetime map in (X,Y) coordinates of a reference calibration run taken at the beginning of Run-V. For the light production, a nominal EL gain of 410~photons per ionization electron is assumed \cite{Monteiro:2007vz}. Second, the electronics effects (including signal shaping, gain fluctuations, and noise effects) are added to the ideal sensor waveforms so that they are comparable to the ones collected by the DAQ system of NEXT-White. With these simulated waveforms, Monte Carlo events are reconstructed and calibrated equivalently as if they were real data.

Four specific MC productions have been conducted for the current \bb analysis in NEXT-White. They account for the radiogenic and cosmogenic backgrounds, the \bb signals, and the calibration data. The radiogenic background model (see Sec.~\ref{sec:radio}), as well as the calibration MC (see Sec.~\ref{sec:reco}), relies on the simulation of isotope decays in the different detector volumes. The cosmogenic background model (see Sec.~\ref{sec:cosmo}) is built upon simulations of the muon flux at the LSC and neutron captures in the detector materials. Finally, the simulation of \bbtwonu and \bbnonu events in the active volume relies on the initial kinematics provided by the DECAY0 generator \cite{Ponkratenko:2000um}, considering the standard $0^+\to 0^+$ transition between the parent (\Xe{136}) and daughter (\Ba{136}) nuclei.

\begin{figure}
  \begin{center}
    \includegraphics[width=0.45\textwidth]{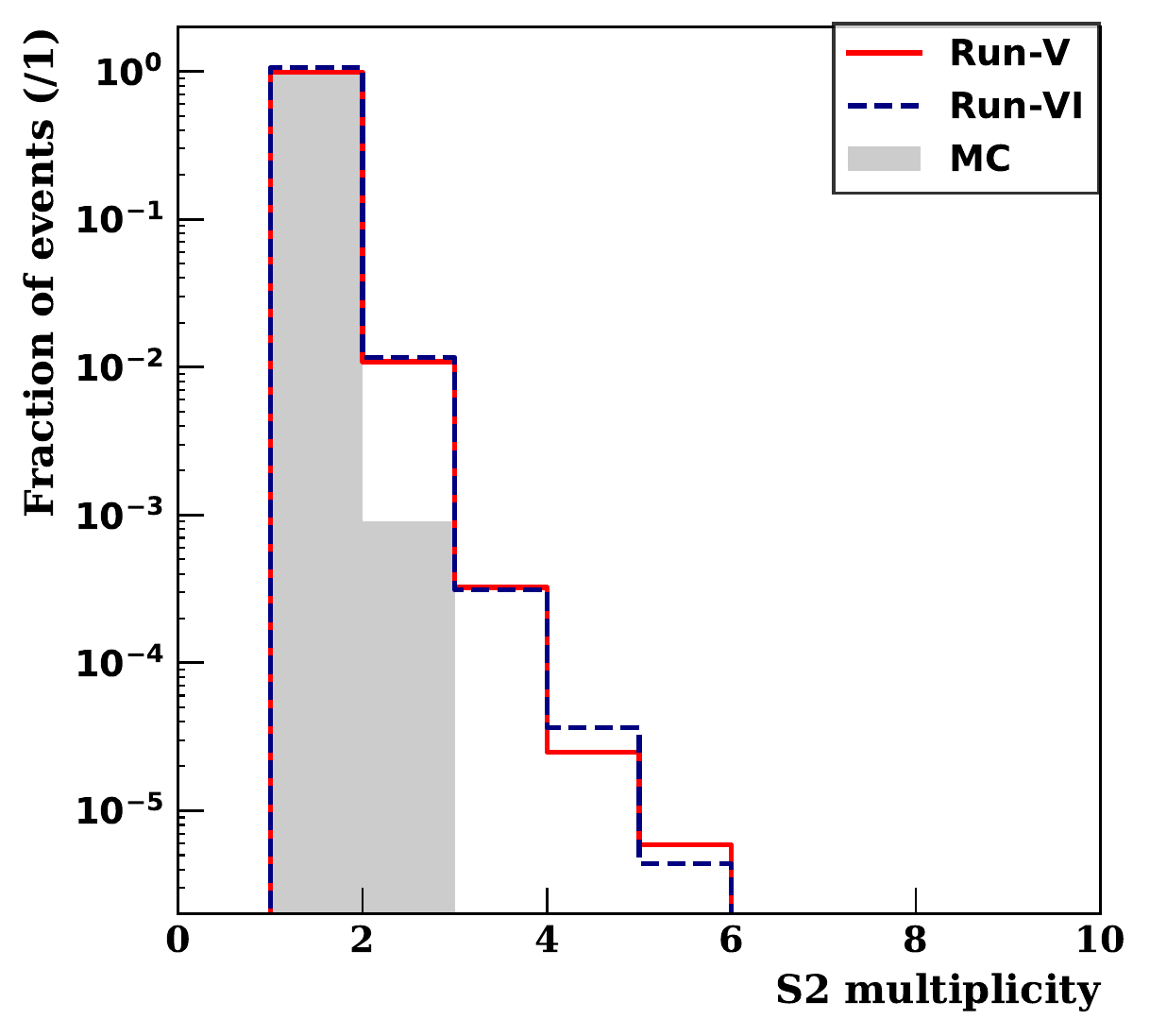}
    \includegraphics[width=0.45\textwidth]{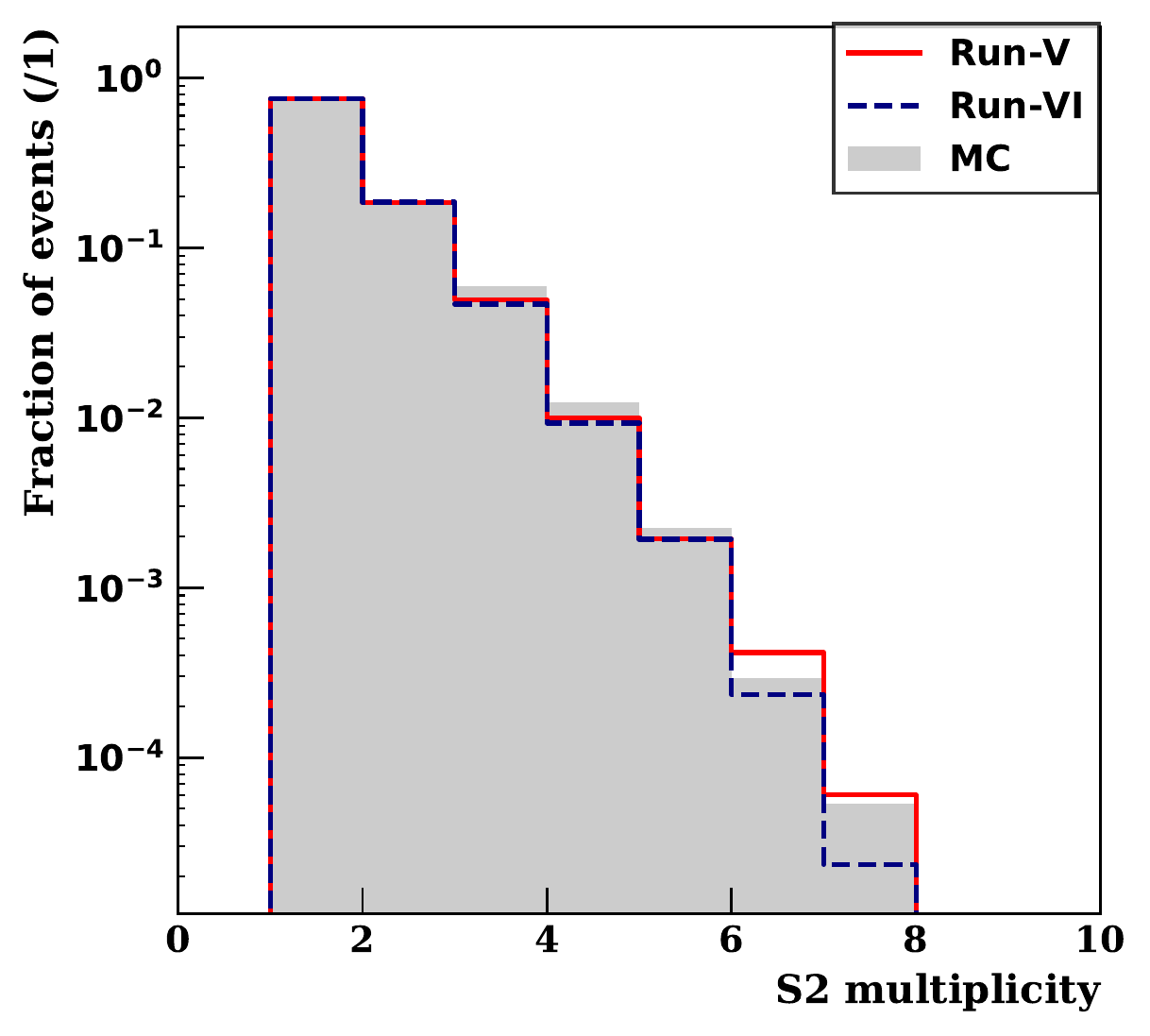} 
    \caption{Multiplicity of S2 signals in \Kr{83m} (left) and $^{208}$Tl (right) calibration data, as extracted from the PMT waveforms. Data results for Run-V and Run-VI (solid and dashed lines, respectively) are superimposed to the MC simulation (solid histogram).}
    \label{fig:ns2}
  \end{center}
\end{figure}

A low-level comparison between data and simulation has been performed in terms of the number of S2 signals per event in calibration samples. A comparison of the S2 multiplicity, as extracted from DAQ and simulated waveforms, is shown in Fig.~\ref{fig:ns2}. For low energy calibration events (\Kr{83m}), most of the events exhibit only one S2 signal, while a second one can also arise due to xenon X-rays. The negligible discrepancy between data and MC (events with larger multiplicities) is due to random coincidences. For high energy calibration data (\Tl{208}), the larger number of signals is due to multiple interactions, and it is well reproduced by the MC.

\section{Data processing}
\label{sec:process}

Both the real data and MC samples are processed following the same reconstruction, calibration, and selection procedures. The events are reconstructed and calibrated according to the techniques described in \cite{NEXT:2020try,Martinez-Lema:2018ibw}. A fully reconstructed and calibrated event accounts for a collection of three-dimensional (3D) hits with defined energy, grouped in a number of tracks according to connectivity criteria. In turn, each track has well-defined extremes with an associated energy. The processed events finally undergo different selection cuts. Specific selection criteria are defined for the measurement of the radiogenic and cosmogenic backgrounds, as well as for the \bb decay analysis. In order to avoid possible biases, the cuts are optimized by means of MC samples, while the selection efficiencies and discrepancies between data and MC are evaluated with calibration samples.  

\subsection{Event reconstruction and calibration}
\label{sec:reco}

In the first stage of the event reconstruction, the amplitudes of the PMT and SiPM signals are converted from ADC to p.e. according to the gains measured periodically by a set of LEDs installed in both the energy and the tracking planes. The 25-ns sampled PMT waveforms are then added into a global waveform. The times of the signals in the various channels are synchronized within 25 ns by the DAQ system. As the frontend electronics are only a few meters away from the sensors, no further timing calibration is required. A search for S1 (width below 125 ns) and S2 (width above 2 $\mu$s) pulses is performed within this waveform. Events with more than one S1 ($\sim$10\%), corresponding to coincident energy deposits or electronic noise, are rejected. The SiPM waveforms (1 MHz sampling) are then reduced to the corresponding time windows of the detected S2 signals. 3D hits are reconstructed from the PMT and SiPM S2 signals. The X and Y coordinates are obtained from the position of the fired SiPMs for each 2 $\mu$s slice (two samples) of the SiPM waveforms. The Z coordinate (drift direction) is derived from the time difference with respect to the S1 of the event. For each time slice, the S2 energy collected by the PMTs is divided among the reconstructed 3D hits, proportionally to the charge collected by the corresponding SiPM and with XY coordinates matching the SiPM position. The energy scale to convert from p.e. to keV is derived from the \Kr{83m} data. Corrections for electron drift lifetime, geometrical effects, and time variations are also applied, relying on \Kr{83m} data collected within a $\sim$24 h period.

In a second reconstruction stage, the blurring effect in the topological information induced by the electron diffusion and the EL light production is corrected by means of a Richardson-Lucy deconvolution, as shown in \cite{NEXT:2020try}. This iterative procedure relies on a point spread function obtained from the point-like events provided by the \Kr{83m} decays. A voxelization of the high-definition deconvolved hits is performed by grouping them into (5~mm)$^3$ volume elements. In turn, a breadth-first search algorithm \cite{Cormen2001_intro_algorithms} is applied in order to establish the connectivity criteria among them and build individual tracks with identified end points. The energies of these end-points are defined by the integration of energy of the hits contained within spheres of 18~mm radius. The energy of the track extremes ($E_\mathrm{b}$), also referred to as \emph{blobs}, allows us to identify the Bragg peaks corresponding to stopping electrons, thus offering a handle to distinguish between single-electron and double-electron tracks (see Sec.~\ref{sec:sel}). 

\begin{figure}
\includegraphics[width=\textwidth]{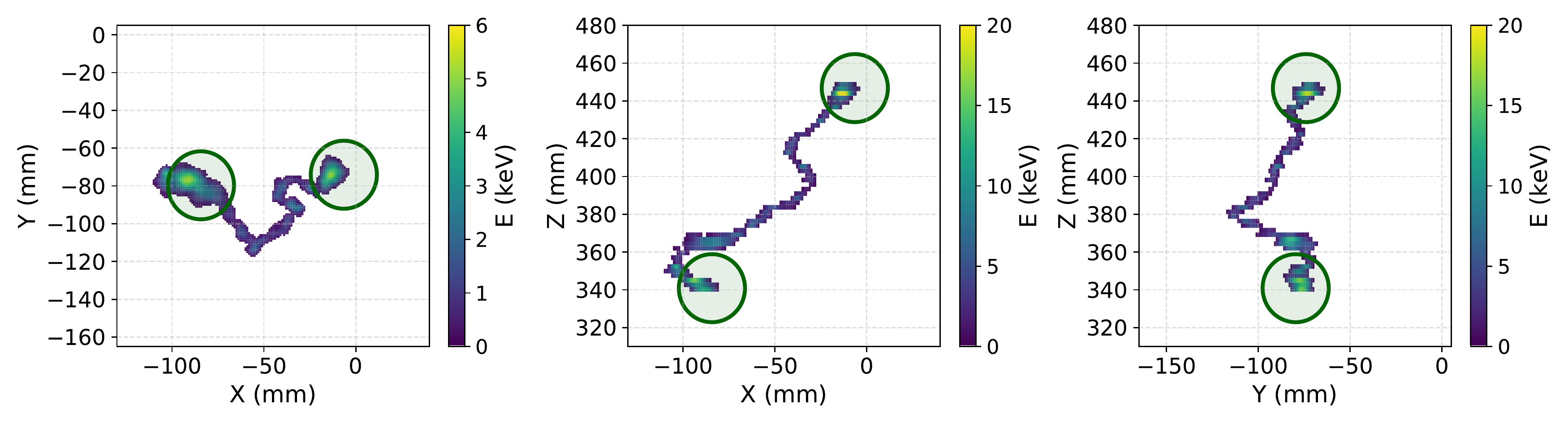}
\includegraphics[width=\textwidth]{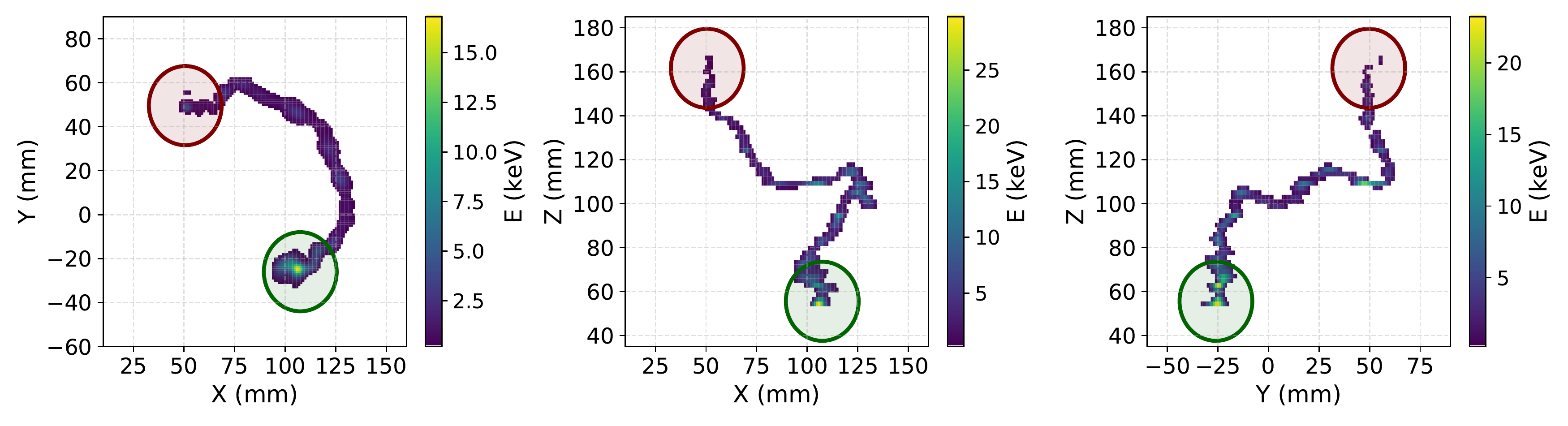}
\caption{\label{fig:tracks} Reconstructed 3D tracks of $\sim$2~MeV from two events collected during the low-background runs of NEXT-White. XY (left), XZ (middle), and YZ (right) projections are shown. According to the \bb selection defined in Sec.~\ref{sec:sel}, the top (bottom) track corresponds to a double-electron (single-electron) event. The energies of the track extremes correspond to the integration within the circles. For $E_\mathrm{b,min}\sim$377 keV (see Sec.~\ref{sec:sel} for details), two Bragg peaks are identified in the \bb candidate ($E_\mathrm{b}$ of 437 keV and 394 keV), while only one is present in the single-electron track ($E_\mathrm{b}$ of 602 keV and 71 keV).} 
\end{figure}

The total energy of tracks is obtained by summing the energy of all calibrated hits. In order to account for the non-linearities in the energy response, the energy scale for extended objects (as opposed to \Kr{83m} events) is derived from high energy calibration data. In particular, an empirical second-degree polynomial energy scale model is derived from the data collected deploying \Cs{137} and \Th{228} sources in dedicated ports on the NEXT-White pressure vessel. This model yields residuals below 0.3\% on the peak positions that appear in low-background data (\Co{60}, \K{40} and \Tl{208}), ranging from 1173 keV to  2615 keV. Overall, the reconstruction and calibration procedures allow obtaining an energy resolution of $\sim$1\% FWHM at 2615~keV \cite{Renner:2019pfe}, which is found to be stable across the different calibration campaigns. An example of two tracks of $\sim$2 MeV, collected during the low-background runs of NEXT-White, is presented in Fig.~\ref{fig:tracks}.

\subsection{Event selections and efficiencies}
\label{sec:sel}

Three consecutive selection stages are applied to the reconstructed events. First, a fiducial selection is performed in order to reject charged particles entering the active volume from the surfaces of the detector. Events are required to have only one S2 signal whose associated 3D hits are fully contained within the volume defined by 20$<$Z$<$510~mm and ${\mathrm R}=\sqrt{{\mathrm X}^2+{\mathrm Y}^2}<$195~mm. This volume accounts for a xenon fiducial mass of 3.50$\pm$0.01~kg, as derived from the average gas density. Hereafter, we refer to this selection as the cosmogenic selection, as the events fulfilling these conditions are used to model and measure the cosmogenic backgrounds in Sec.~\ref{sec:bg}. In a second stage, events are also required to have only one reconstructed track, as expected for \bb events. The events passing this selection are used to characterize and measure the radiogenic backgrounds in Sec.~\ref{sec:bg}. Therefore, hereafter, we refer to it as the radiogenic selection.

\begin{figure}
  \begin{center}
    \includegraphics[width=0.47\textwidth]{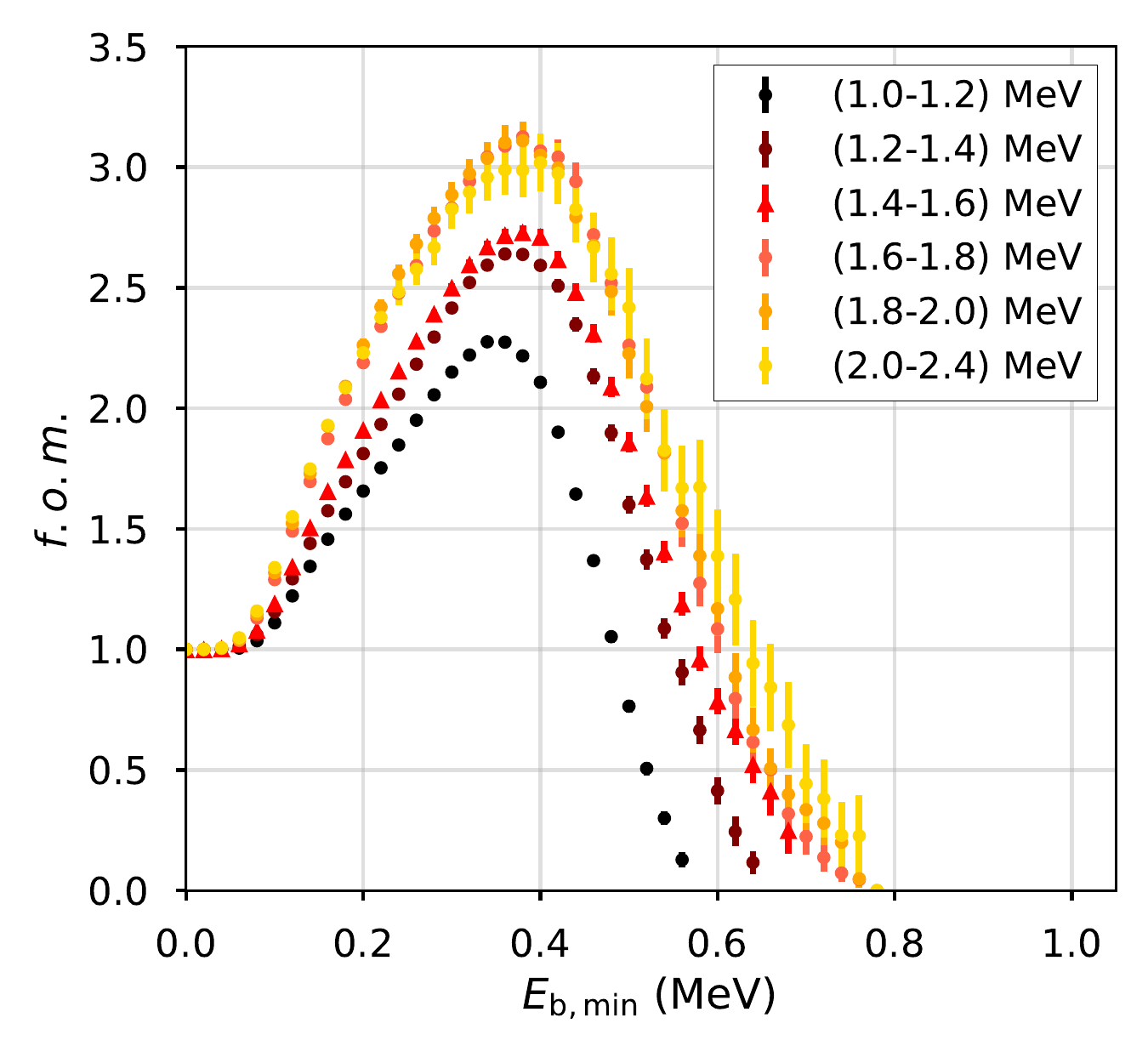}
    \includegraphics[width=0.47\textwidth]{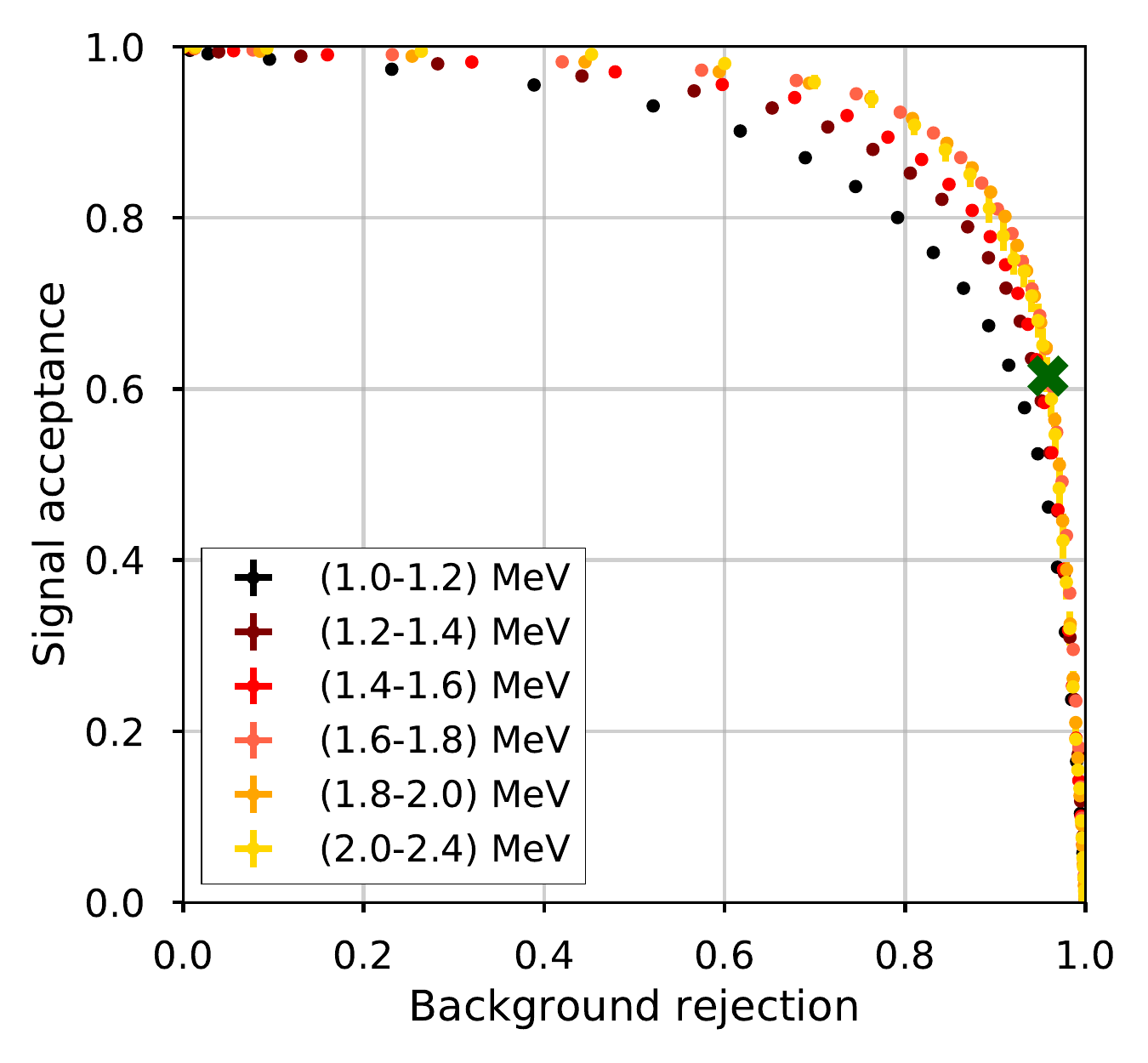}
    \caption{Blob energy cut efficiency. Left: figure of merit as a function of  $E_\mathrm{b,min}$. Right: background rejection versus signal acceptance. The green cross illustrates the values that optimize the f.o.m. in the region around the \Qbb value, which are specified in the text. In both panels, the results are presented for different energy ranges from 1 MeV to 2.4 MeV.}
    \label{fig:blobcut}
  \end{center}
\end{figure}

Finally, track-based topological cuts are added to the previous selection in order to identify double-electrons, as expected for \bb events. In this \bb selection, only tracks with no common hits in their end-point blobs are selected, assuring that the track extremes do not overlap. The less energetic blobs are then required to have an energy $E_\mathrm{b}$ above a given threshold,  $E_\mathrm{b,min}$, which depends on the total energy of the track. This condition ensures that the track ends in two Bragg peaks, corresponding to the stopping points of the two electrons. In turn, tracks not fulfilling this condition are identified as single-electron events. According to this \bb selection, the top (bottom) track displayed in Fig.~\ref{fig:tracks} corresponds to a double-electron (single-electron) candidate. The $E_\mathrm{b,min}$ threshold as a function of event energy is obtained with MC studies, by optimizing the figure of merit (f.o.m.) defined as the ratio of the \bb signal selection efficiency over the square root of the single-electron background acceptance. The left panel of Fig.~\ref{fig:blobcut} shows this f.o.m. as a function of $E_\mathrm{b,min}$, for different energy ranges from 1 MeV to 2.4 MeV. The maxima of the f.o.m. curves, ranging from $\sim$2.3 to $\sim$3.1, are consistent with the measurement of 2.94$\pm$0.28 for 1.6 MeV calibration events obtained in \cite{NEXT:2020try}. The background rejection versus the signal acceptance is presented for the same energy ranges in the right panel of Fig.~\ref{fig:blobcut}. In the energy region around the \Qbb value, a signal acceptance of $\sim$62\% is achieved for a background rejection of $\sim$96\%, as indicated in the plot by means of the green cross. As for energies below 1 MeV the topological discrimination worsens considerably, only tracks above this energy are considered in the current analysis.

%---- efficiencies

The efficiencies for the three selection stages are computed independently in Run-V and Run-VI by means of \Tl{208} calibration data, analyzing each specific cut. For the \bb selection, the efficiency is evaluated separately for double-electron and single-electron events. Along the lines of the analyses presented in \cite{NEXT:2019gtz,NEXT:2020try}, a double-electron-enriched data sample is obtained from the events with energy around the \Tl{208} double-escape peak at 1.6~MeV (produced by 2.6~MeV gamma-rays), as it is mostly populated by pair-production interactions. In turn, events outside this peak, dominated by Compton scattering interactions, provide a single-electron-enriched data sample. The overall efficiencies for the three selections, integrated from 1 MeV to 2.8 MeV, are presented in Tab.~\ref{tab:effs}. According to the \bb efficiencies for double-electron (signal-like) and single-electron (background-like) events, a figure of merit of $\sim$2.5 ($\sim$2.8) is achieved in Run-V (Run-VI). While the energy dependence of these efficiencies is well reproduced by simulated calibration samples, significant differences are observed in the integrated values. For the cosmogenic and radiogenic selections, these differences arise mostly from the requirement of having only one S1 signal (applied at reconstruction level, see Sec. 3.1). As backgrounds and time-dependent electronic noise are not simulated in the $^{208}$Tl MC, this cut has no impact in the simulation. On the contrary, $\sim$10\% of the calibration events are rejected in data due to this cut. For the $\beta\beta$ selection, the discrepancies arise either from the different gas conditions (small variations in pressure and temperature with respect to the nominal values in the MC) or a possible mismodelling of the distribution of the blob energy (under investigation). The ratios between the data-driven and MC-driven efficiencies are used as scaling factors in the simulations of \bb and background events used in the current analysis. The impact of possible residuals in this MC correction has been proven to be negligible by running the statistical analysis described in Sec.~\ref{sec:bb} with a set of biased MC samples.   

\begin{table}[!htb]
\caption{\label{tab:effs}Efficiencies of the cosmogenic, radiogenic, and \bb selections, integrated from 1 MeV to 2.8 MeV. The specific \bb efficiencies for double-electron and single-electron tracks are shown.}
\begin{center}
\begin{tabular}{c|cccc}
\hline
Period  &   Cosmogenic    & Radiogenic & \bb double-electron & \bb single-electron  \\ \hline
Run-V &   54.8$\pm$0.2\%  & 47.7$\pm$0.2\% & 24.7$\pm$0.5\% &  2.24$\pm$0.06\% \\
Run-VI &  54.9$\pm$0.2\%  & 48.3$\pm$0.2\% &  27.5$\pm$0.6\% & 2.34$\pm$0.07\%\\
MC     &  62.6$\pm$0.1\%  & 54.7$\pm$0.1\% &  33.0$\pm$0.4\% & 2.09$\pm$0.05\%\\
\hline
\end{tabular}
\end{center}
\end{table}

\section{Backgrounds}
\label{sec:bg}

The backgrounds in the NEXT-White detector have been measured relying on a model considering both radiogenic and cosmogenic contributions. Although neutron-induced backgrounds from ($\alpha$,$n$) reactions and spontaneous fission are expected to be negligible, small contributions would be naturally embedded in the treatment of the cosmogenic background induced by fast-neutrons. The goal of this measurement is twofold. First, it allows the identification of the main background sources. Second, it offers a handle to assure the time stability of the backgrounds in the two data taking periods considered for the current \bb analyses.

\subsection{Radiogenic background model}
\label{sec:radio}

As presented in \cite{Novella:2019cne}, the expected radiogenic background budget in NEXT-White is derived from a detailed simulation and an extensive radiopurity measurements campaign conducted by the NEXT collaboration \cite{Alvarez:2012as,Alvarez:2014kvs,Cebrian:2017jzb}. The radiogenic background model accounts for four isotopes (\Bi{214}, \Tl{208}, \K{40} and \Co{60}) and 44 detector materials distributed in up to 23 detector volumes in the GEANT4 simulation. Overall, the model consists of 85 background sources (isotope $\times$ GEANT4 volume contributions).

The radiopurity screening of the 44 materials has been mostly conducted by gamma spectroscopy, using the high-purity Germanium detectors of the LSC Radiopurity Service. However, mass spectroscopy techniques (ICP-MS, GDMS) have also been used for some detector materials in order to reach sensitivities below 1~mBq/kg. This is the case of copper, lead, steel, high-density polyethylene and PTFE. For those contributions whose specific activity was measured, the obtained central values are adopted in the background model. In a conservative scenario, the 95\% CL upper limits are considered for those contributions where the specific activity could not be quantified. These specific activity assumptions are multiplied by the material quantities to obtain the total background activity assumptions. These material quantities are obtained from the as-built engineering drawings of NEXT-White and the known material densities. With respect to the background model described in \cite{Novella:2019cne}, new ICP-MS measurements at PNNL for PTFE (light tube) and copper (inner copper shield) have provided the specific activities for \Tl{208} and \Bi{214}, for which only upper limits were available. In addition to the activity assumptions in the various materials, the background model also assumes $3.1$~mBq of \Bi{214} decays uniformly from the cathode plane. This is estimated from the activity of internal radon \Rn{222} measured in NEXT-White using alpha particles, and assuming that all \Rn{222} daughters (charged and chemically active) plate out on the cathode \cite{Novella:2018ewv}.

According to the estimated activities, a full GEANT4-based Monte Carlo simulation has been performed. The materials are associated to 23 GEANT4 volumes describing the components of the NEXT-White detector, as well as the shielding structures, with one or more materials assigned to each volume. In turn, these 23 volumes are grouped at analysis level into the three effective volumes defined in Sec.~\ref{sec:sim} (CATHODE, ANODE and OTHER). With respect to the simulation used in \cite{Novella:2019cne}, a new activity contribution to describe stainless steel frames supporting the anode plate and gate mesh has been added. Overall, $1.2\times 10^7$ events with more than 400~keV of energy deposited in the active volume have been simulated. According to our activity assumptions, this nominal background MC corresponds to an effective exposure of 48.99~yr, well in excess of Run-V and Run-VI exposures. The simulated events have been reconstructed as described in Sec.~\ref{sec:reco}, following the same procedure as for real data. Finally, the background selection presented in Sec.~\ref{sec:sel} has been applied to the reconstructed events in order to derive the radiogenic background expectations.

\begin{figure}[htb]
  \begin{center}
    \includegraphics[width=1\textwidth]{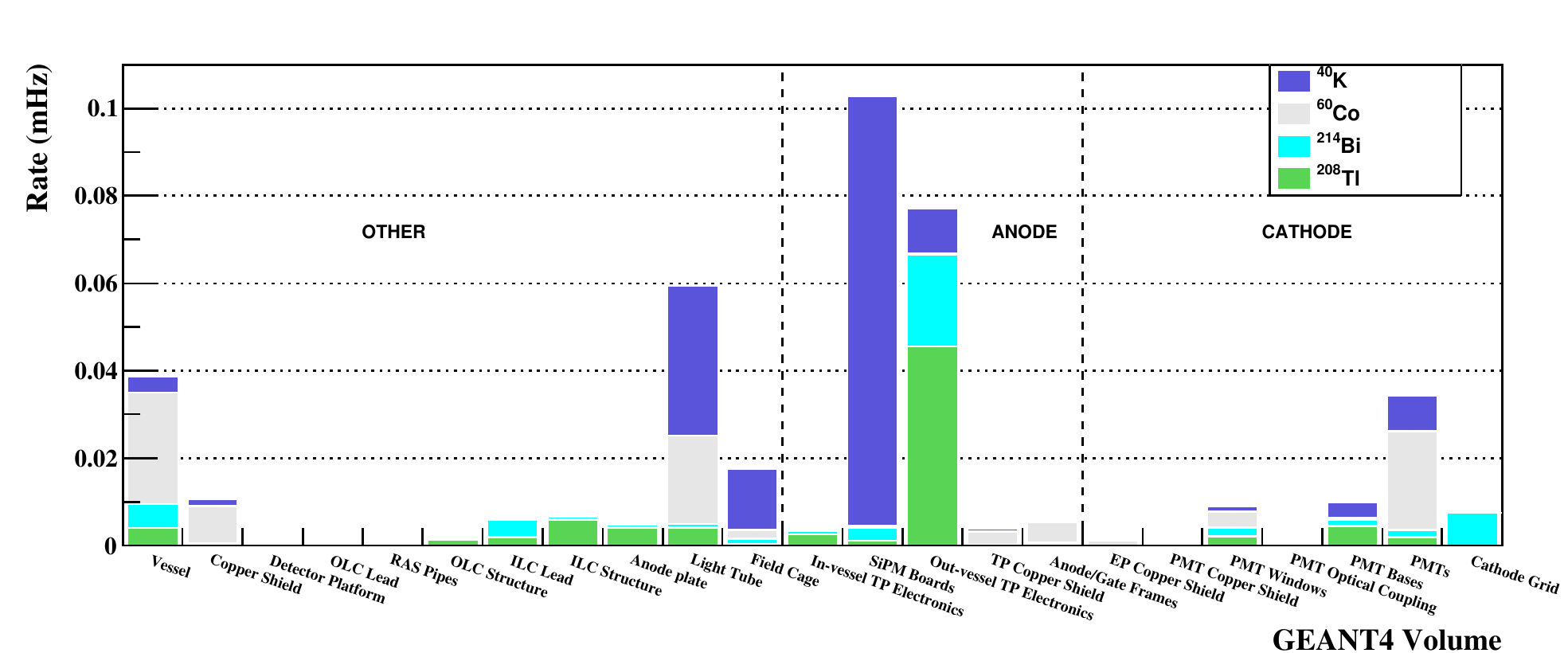}
    \caption{Radiogenic background model according to background selection defined in Sec.~\ref{sec:sel}. The rate of events above 1 MeV from each GEANT4 detector volume is presented, along with the specific \Bi{214}, \Tl{208}, \Co{60} and \K{40} contributions.}
      
    \label{fig:bgconts}
  \end{center}
\end{figure}

The total expected radiogenic background rate above 1 MeV amounts to R$_{BG}$(radio)=0.404 mHz. The specific contributions from \Bi{214}, \Tl{208}, \Co{60} and \K{40} are 0.051, 0.081, 0.094 and 0.178 mHz, respectively. The total rates from the CATHODE, ANODE and OTHER effective volumes are 0.063, 0.198 and 0.143 mHz, respectively. The estimated background rate from each specific volume is presented in Fig.~\ref{fig:bgconts}, where the specific contributions from the four isotopes are also displayed. The dominant backgrounds come from \K{40} in the SiPM dice boards, within the ANODE effective volume, \Co{60} in the vessel materials (OTHER volume) and \Tl{208} from tracking plane readout electronics (ANODE volume). While the light tube might also be a major contributor to the background budget, the activity assumptions for \Co{60} and \K{40} correspond to upper limits. Finally, the PMTs (CATHODE volume) also represent a significant contribution. These dominant background sources do not correspond necessarily to the most active volumes. The probability of a radiogenic interaction in the active volume depends on the spatial location of the volume where the original decay takes place. Thus, the highest probabilities correspond to the innermost volumes (such as the PTFE light tube or the SiPM dice boards), while the outermost simulated volumes (particularly the lead-based shielding) barely contribute to the total background budget.

%----

\begin{figure}[htb]
  \begin{center}
    \includegraphics[width=0.32\textwidth]{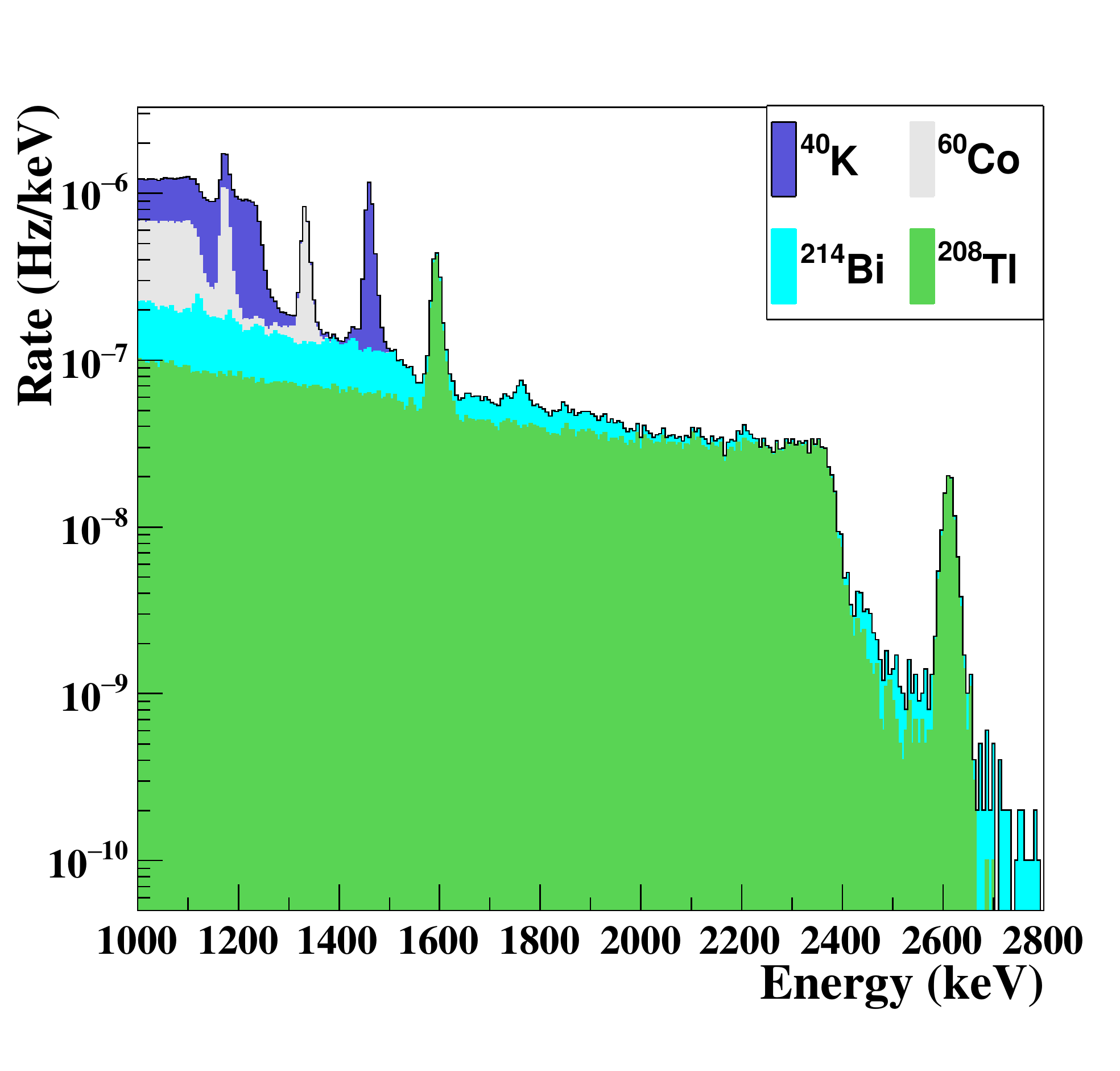} 
    \includegraphics[width=0.32\textwidth]{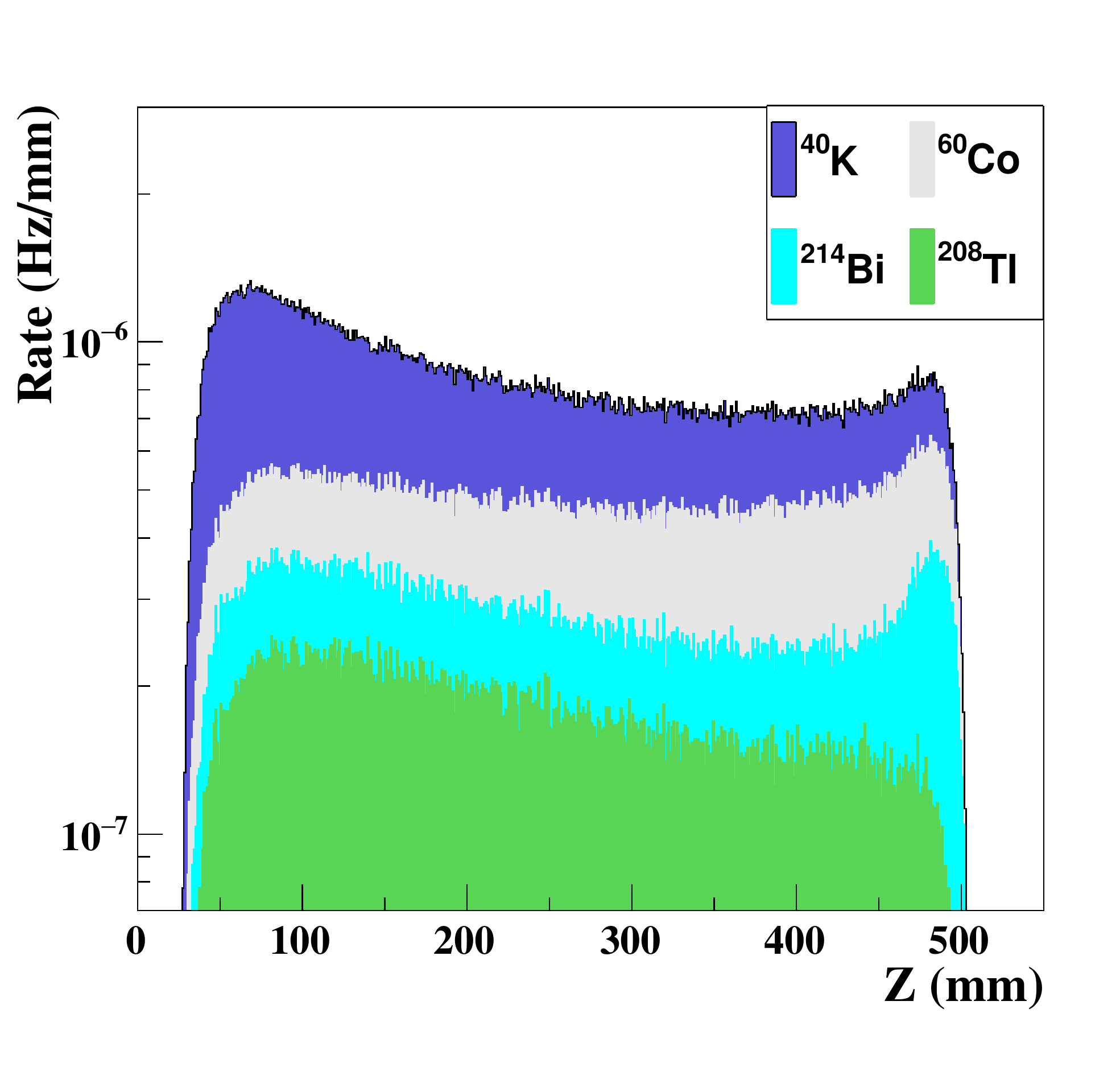}
    \includegraphics[width=0.32\textwidth]{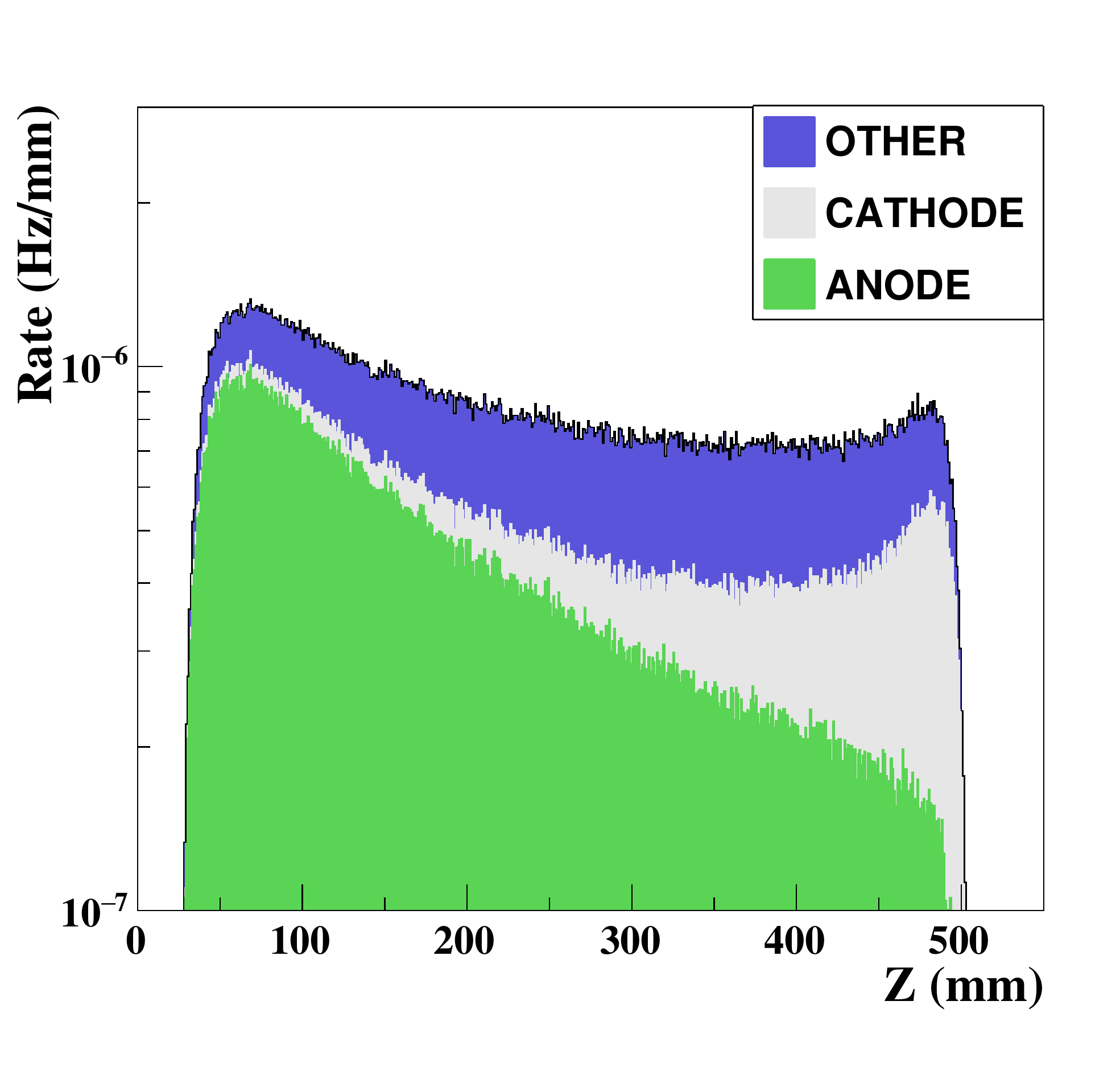}
    \caption{Radiogenic background model according to background selection defined in Sec.~\ref{sec:sel}. The rate of events as a function of the energy (left) and average Z position (middle) are shown for the four isotopes considered in the model. The total rate of events from the three effective volumes defined in Sec.~\ref{sec:sim} (right) is presented as a function of the average Z position. The background contributions are shown stacked to each other.} 
    \label{fig:mc_background}
  \end{center}
\end{figure}

The left panel of Fig.~\ref{fig:mc_background} shows the energy of the four isotopes considered in the background model. The characteristic peaks of \Co{60} (1173 and 1333 keV), \K{40} (1461 keV), and \Tl{208} (1593 and 2615 keV) are clearly visible, offering a handle to measure their specific contributions from real data. According to their energy spectra, while the four isotopes represent a background for the measurement of the \bbtwonu half-life, only \Bi{214} and \Tl{208} contribute to the region of interest for the \bbnonu searches. Defining the Z position of an event as the charge-weighted average over all reconstructed hits, the Z distributions of the four isotopes are shown in the middle panel of Fig.~\ref{fig:mc_background}. The Z-dependence of the expected rate reflects the position of the dominant background sources, peaking in particular around the anode (Z$\rightarrow$0 mm) for \Tl{208} and \K{40}, and around the cathode (Z$\rightarrow$530 mm) for \Bi{214} and \Co{60}. This Z-dependence of the backgrounds provides some sensitivity to the spatial origin of the specific sources. Considering the three effective volumes defined in Sec.~\ref{sec:sim}, as shown in the right panel of Fig.~\ref{fig:mc_background}, this sensitivity is fully exploited in order to measure from real data the specific contributions of these volumes.

\subsection{Cosmogenic background model}
\label{sec:cosmo}

%-- description of cosmogenic backgrounds

The cosmogenic backgrounds are induced by high energy (up to a few TeV) cosmic-ray muons reaching the laboratory. These muons produce fast-neutrons and unstable nuclides when interacting in the rock walls and detector materials. In turn, the fast neutrons travel some distance until they thermalize and get captured. Activation of isotopes upon neutron capture can lead to prompt and delayed signals. Immediately after the capture, prompt-gammas are emitted due to the nuclear de-excitation. The resulting isotope may also be radioactive and decay, according to its specific half-life, inducing a delayed signal. Both types of events can become a background in \bbnonu searches if they deposit an energy around the \Qbb value. In particular, the prompt-$\gamma$s can be suppressed to a negligible level by means of a muon tagger system. These signals are typically emitted within the same trigger window of the muon, so an effective veto can be applied. However, the decay of long-lived radio-isotopes cannot be efficiently correlated to a muon event, and thus becomes an irreducible background. In xenon-based detectors, this is the case of \Xe{137}, undergoing a $\beta$ decay with a half-life of 3.82 minutes and a \Qb value of 4173$\pm$7 keV \cite{Browne:2007nds}.

%---

The muon flux at Hall A of the LSC has been previously measured with a dedicated muon monitor in \cite{Trzaska:2019kuk}. Located at the opposite side of Hall A, a consistent muon flux of $\Phi_{\mu}=4.77\pm0.03(\textrm{stat})\pm0.02(\textrm{sys})\times10^{-7}$ cm$^{-2}$s$^{-1}$ has been measured by NEXT-White. Since NEXT-White is not surrounded by an external muon veto system, the cosmogenic prompt-$\gamma$s are of particular importance. When the muons originating the ($n$,$\gamma$) reactions do not cross the active volume of the detector, they become a significant background. The Monte Carlo simulation of the muon flux and the detector have been used to identify the isotope activations that can lead to backgrounds in the region of interest for \bb decay analyses. As discussed in \cite{NEXT:2020qup}, the simulation relies on the neutron capture cross-section from the {\tt ENDF/B-VII.1} database~\cite{chadwick2011endf}, originally derived from \cite{mughabghab2006atlas} and recently tested experimentally in \cite{Albert:2016vmb} (validated at 20\% for \Xe{136}). In particular, the \Xe{136} cross-section resonances are accounted for. The MC shows that the fast-neutrons responsible for energy deposits in the active volume of NEXT-White are generated mainly by muon interactions in the detector shielding (lead and support structures of the OLC and the ILC). Fast neutrons generated in the rock walls of the LSC account only for $\sim$3\% of the cosmogenic backgrounds. Each muon generates a mean value of $\sim$3.7 fast neutrons, although only $\sim$0.17 induce a background upon a ($n$,$\gamma$) process. The characteristic capture time is different for each isotope, depending on the cross-section and detector volume. However, $>$99\% of the neutrons are captured within a time window of 1 ms following the muon.

\begin{figure}
  \begin{center}
    \includegraphics[width=\columnwidth]{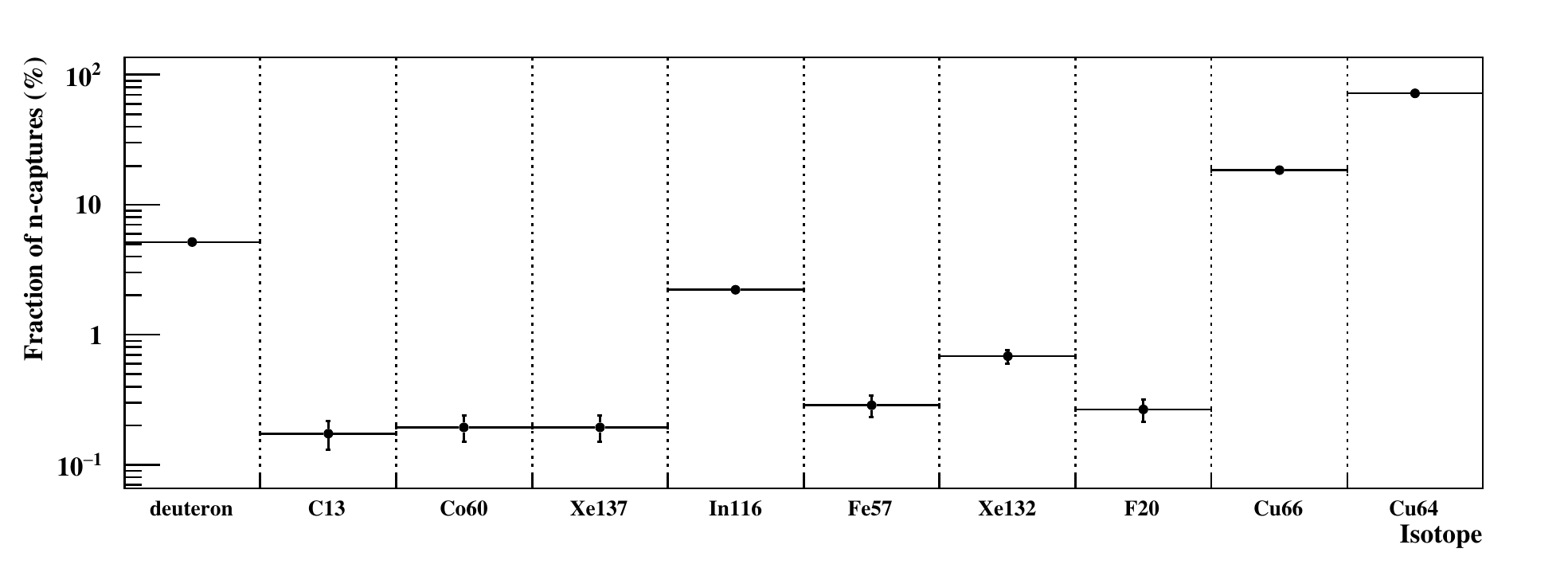}
    \caption{Relative fraction of isotope activations through ($n$,$\gamma$) reactions, for which the de-excitation gamma deposits energy in the active volume of NEXT-White. A fraction of 97\% of the prompt-$\gamma$ backgrounds corresponds to four isotopes: \Cu{64} (72\%), \Cu{66} (18\%),\Deu (5\%) and \In{116} (2\%). Isotopes contributing less than 0.1\% are not shown.}
    \label{fig:ncaptures}
  \end{center}
\end{figure}

As shown in Fig.~\ref{fig:ncaptures}, only four isotopes are responsible for 97\% of the prompt-$\gamma$ background in NEXT-White: \Cu{64} (72\%), \Cu{66} (18\%), \Deu (5\%) and \In{116} (2\%). The contribution of \Xe{137} is negligible in NEXT-White ($\sim$0.2$\%$ in the \Xe{136}-enriched gas) due to the limited xenon mass. The same applies to captures in other Xe isotopes, for both the $^{136}$Xe-enriched and $^{136}$Xe-depleted gasses. The captures on copper occur mostly on the ICS ($\sim$60\%) and the rings of the field cage of the TPC ($\sim$25\%), with smaller contributions from the PMTs support structures. The production of deuteron by neutron capture on H occurs in the high-density polyethylene (HDPE) of the field cage ($>$99\%), although a sub-percent fraction of the captures happen also in the SiPM dice boards. The captures on indium take place on the ITO covering the anode plate. Although a significant fraction of the neutron captures take place in lead (OLC and ILC) and iron (mostly the vessel), the corresponding prompt gammas do not induce a significant background due to the internal copper shielding. The relative capture fractions in each isotope are found to be rather independent of the muon energy and angular distribution of the muon flux.

\begin{figure}
  \begin{center}
    \includegraphics[scale=0.37]{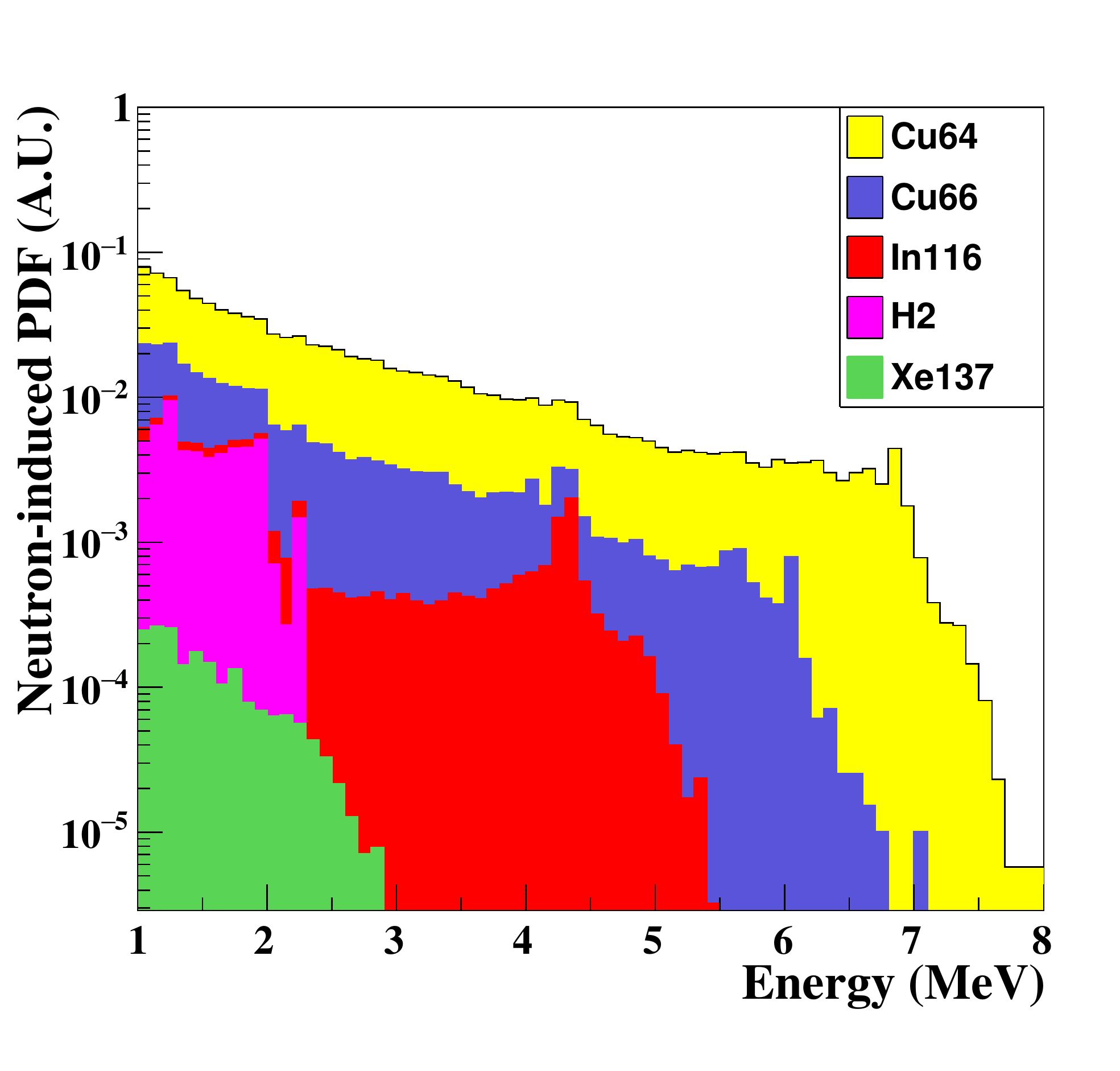}
    \includegraphics[scale=0.37]{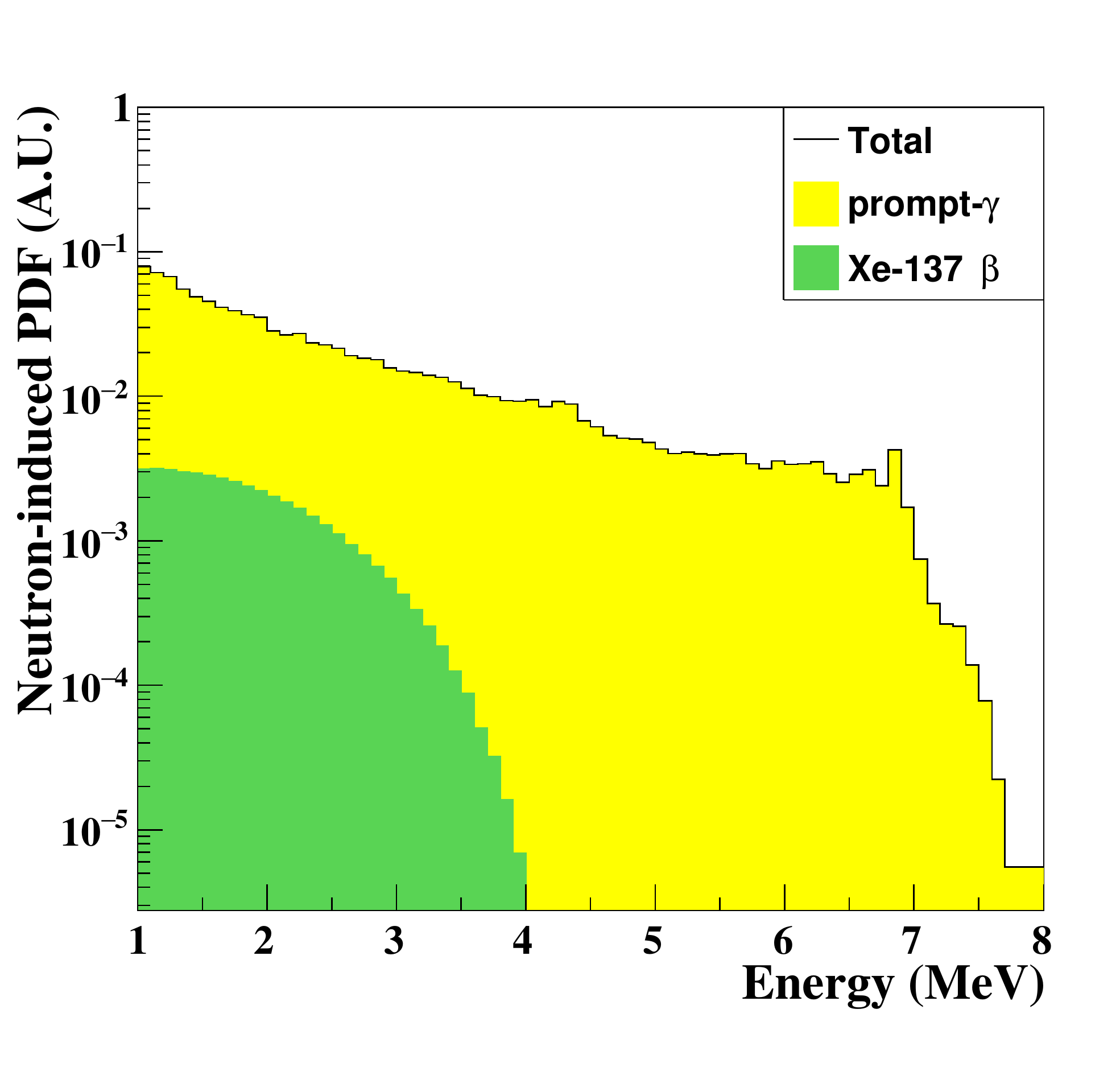}
    \caption{ Cosmogenic background model in \Xe{136}-enriched xenon. Left: energy spectrum of the overall prompt-$\gamma$ background (black line), with the specific contributions of the relevant activated isotopes stacked to each other.
Right: total cosmogenic background (black line) and relative contributions from prompt-$\gamma$ (yellow) and \Xe{137} $\beta$ decay (green) events.}
    \label{fig:cosmopdf}
  \end{center}
\end{figure}

The prompt-$\gamma$s energy spectra of the relevant activated isotopes are also obtained from dedicated simulations of neutron captures in the detector materials. Probability Density Functions (PDFs) are obtained for  \Cu{64}, \Cu{66}, \Deu and \In{116} (dominant backgrounds), as well as for \Xe{137} (illustration purposes), by registering the total energy deposited in the active volume by the prompt-gammas. The PDFs corresponding to the cosmogenic selection (see Sec.~\ref{sec:sel}) are shown in the left panel of Fig.~\ref{fig:cosmopdf}, once normalized according to the relative capture fractions presented in Fig.~\ref{fig:ncaptures}. The overall energy spectrum expands up to $\sim$8 MeV due to the copper activations, which are dominant for all energies. While the fraction of deuteron background is not negligible, the single gamma emitted has an energy of 2.22 MeV, well below the \Qbb of \Xe{136} and thus it does not represent a background for \bbnonu searches. Although the prompt-gammas from \Xe{136} activations expand to energies above the \Qbb, their contribution to the overall background is negligible even in \Xe{136}-enriched gas. The same applies to the activation of other xenon isotopes.  

The simulation of the muon flux at the LSC confirms that the only long-lived activated isotope inducing a delayed signal of high energy in NEXT-White is \Xe{137}. The ratio of the total (all isotope activations considered) prompt-$\gamma$ background rate to the \Xe{137}-$\beta$ background rate is found to be 23.8 (913.6) in \Xe{136}-enriched (\Xe{136}-depleted) xenon. A dedicated simulation of the $\beta$ decays of this isotope in the active volume of the detector has been conducted to obtain the corresponding energy spectrum. The right panel of Fig.~\ref{fig:cosmopdf} shows the contribution of this delayed background to the total cosmogenic background when operating NEXT-White with \Xe{136}-enriched xenon.

\begin{figure}
  \begin{center}
    \includegraphics[scale=0.37]{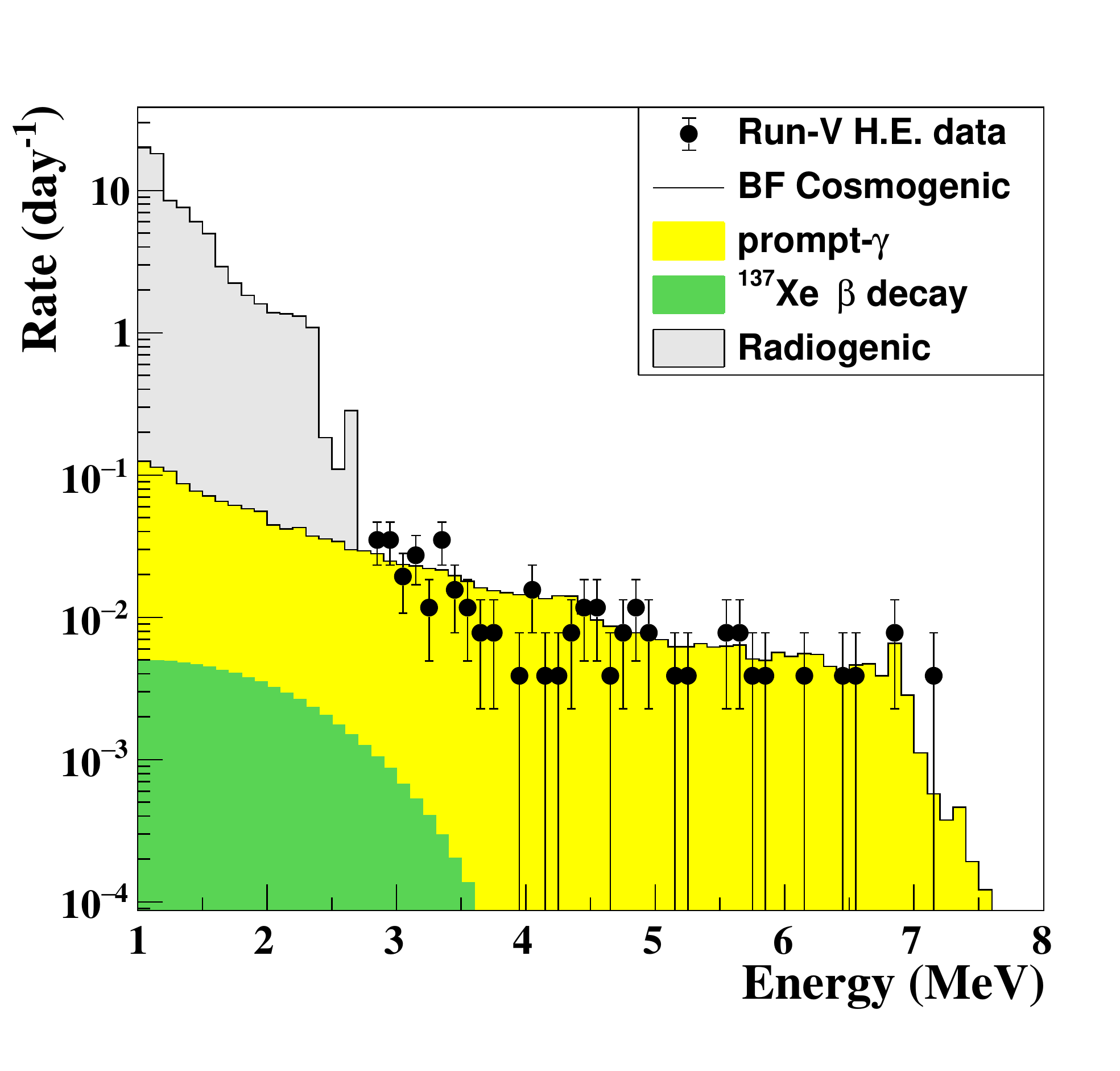}
    \includegraphics[scale=0.37]{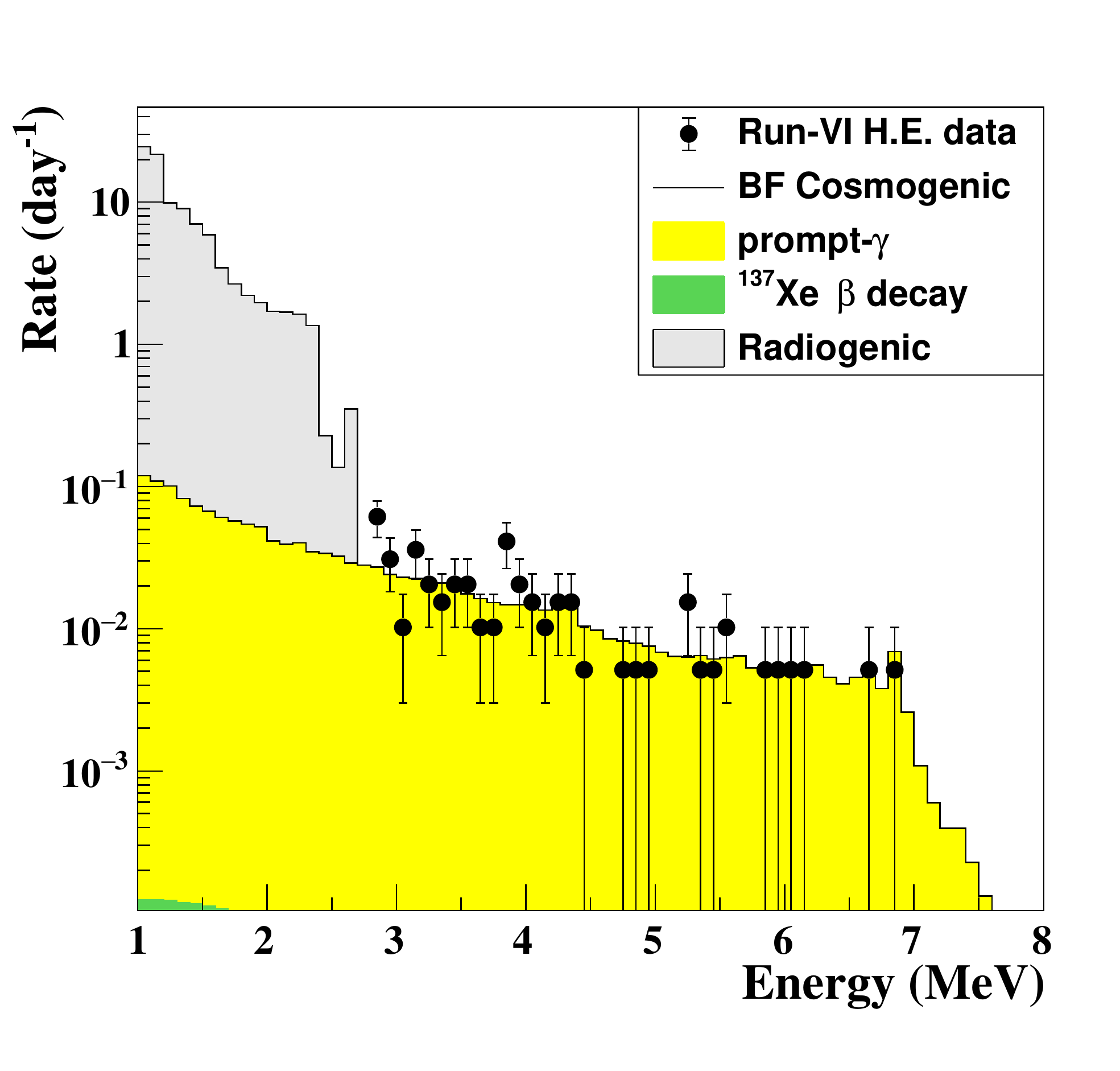}
    \caption{Cosmogenic background fit combining \Xe{136}-enriched xenon (left) and \Xe{136}-depleted xenon (right) data. The rate of events above 2.8 MeV and passing the cosmogenic selection is superimposed to the best-fit normalization of the cosmogenic background model, consisting of prompt-$\gamma$s (yellow) and \Xe{137} $\beta$ decays (green). For illustration purposes, the radiogenic background model for the same selection is also shown (gray).}
    \label{fig:cosmofit}
  \end{center}
\end{figure}

In the absence of a reliable muon simulation accounting for the mountain profile above the LSC, an expectation for the normalization of the cosmogenic background is obtained from an analysis of the high-energy data collected with NEXT-White in Run-V and Run-VI. Since there are no radiogenic backgrounds above 2.7 MeV, samples of cosmogenic events above this energy are obtained by applying the cosmogenic selection. In particular, consistent rates ($1.2\sigma$) of $0.40\pm0.04$ day$^{-1}$ and of $0.48\pm0.05$ day$^{-1}$ are found in Run-V and Run-VI, respectively, for events with $E>$2.8 MeV. This consistency is indeed expected, as the integrated rates of muons crossing the active volume is not significantly different in Run-V (0.921$\pm$0.007 mHz) and Run-VI (0.899$\pm$0.010 mHz). The Run-V and Run-VI data samples are jointly fitted to the cosmogenic background model in order to derive the best-fit normalization of this background. The results of a likelihood fit with only one free parameter (overall normalization of the cosmogenic background considering the prompt-$\gamma$s and the \Xe{137} $\beta$ decay) are shown in Fig.~\ref{fig:cosmofit}. With a $\chi^2/\textrm{dof}=43.4/61$ ($p$-value of 96\%), the best-fit value for the cosmogenic background rate in the 1-8 MeV energy window is 0.018$\pm$0.001 mHz in Run-V and 0.017$\pm$0.001 mHz in Run-VI. According to relative fractions of prompt-$\gamma$ and $\beta$ events derived from MC, this corresponds to a prompt-$\gamma$ rate of 0.017$\pm$0.001 mHz in both runs (negligible contribution of \Xe{137} gammas), and $\beta$-decay rates of 0.77$\pm$0.06 $\mu$Hz in Run-V and 0.019$\pm$0.001 $\mu$Hz in Run-VI. As a validation, a second likelihood fit with independent normalizations for the prompt-$\gamma$s and the $\beta$ decay has been performed, yielding consistent results.

In order to convert this cosmogenic background model into the corresponding one for the radiogenic and \bb selections defined in Sec.~\ref{sec:sel}, the energy-dependent ratios of the different selection efficiencies are considered. According to this MC-driven extrapolation in the energy range between 1 and 2.8 MeV, the Run-V cosmogenic rate in the background selection amounts to R$_{BG}$(cosmo)=9.9$\pm$0.7 $\mu$Hz. In Run-VI, the rate is R$_{BG}$(cosmo)=9.3$\pm$0.7 $\mu$Hz, given the reduction of the \Xe{137} $\beta$ decay contribution. Both rates can be compared with the contribution from the radiogenic background model of Sec.~\ref{sec:radio}, R$_{BG}$(radio)=404 $\mu$Hz.

\subsection{Background measurement and time stability}
\label{sec:bgmeas}

%---- intro

A measurement of the specific radiogenic and cosmogenic backgrounds in Run-V and Run-VI has been performed in order to make a detailed comparison with the expectations and to assure the background time stability between both data taking periods. The measurement relies on the MC background model presented in Sec.~\ref{sec:radio} and Sec.~\ref{sec:cosmo} and follows the statistical approach described in \cite{Novella:2019cne}. The specific expectations for Run-V and Run-VI account for the small differences in the DAQ live times, gas densities, and selection efficiencies.

%---- fit description

An effective fit to the events passing the radiogenic selection (see Sec.~\ref{sec:radio}) in the 1000--2800 keV range has been performed to obtain the normalization of each background source. The fit relies on the minimization of a maximum extended likelihood, considering the energy spectra of both the radiogenic and cosmogenic contributions, as well as the Z distribution of the radiogenic sources. As the dominant cosmogenic background comes from the copper surrounding the active volume, the expected cosmogenic background does not exhibit a particular Z-dependence. The fit considers the contribution of the \Bi{214}, \Tl{208}, \K{40}, and \Co{60} isotopes from the 3 effective volumes, as well as the contribution of the cosmogenic backgrounds (accounting for both prompt-$\gamma$s and the $\beta$ decay of \Xe{137}). This results in a total of 13 fit parameters that provide normalization factors with respect to the nominal model predictions. Since the contribution of \Xe{136} in this sample is small, its normalization has been fixed as corresponding to the \bbtwonu half-life reported in \cite{Albert:2013gpz}. Although negligible with respect to the statistical errors, systematic uncertainties accounting for the energy scale (0.3\%) and total normalization contributions (0.2\%, from DAQ live time and trigger uncertainties) are incorporated into the fit via two nuisance parameters. Provided that the MC samples reproduce the energy resolution measured in data, and that the bin size (25 keV) is larger than the resolution below 2.6 MeV, we assume no uncertainty in the energy resolution model. 

%---- fit results

\begin{figure}[ht]
  \begin{center}
    \includegraphics[scale=0.37]{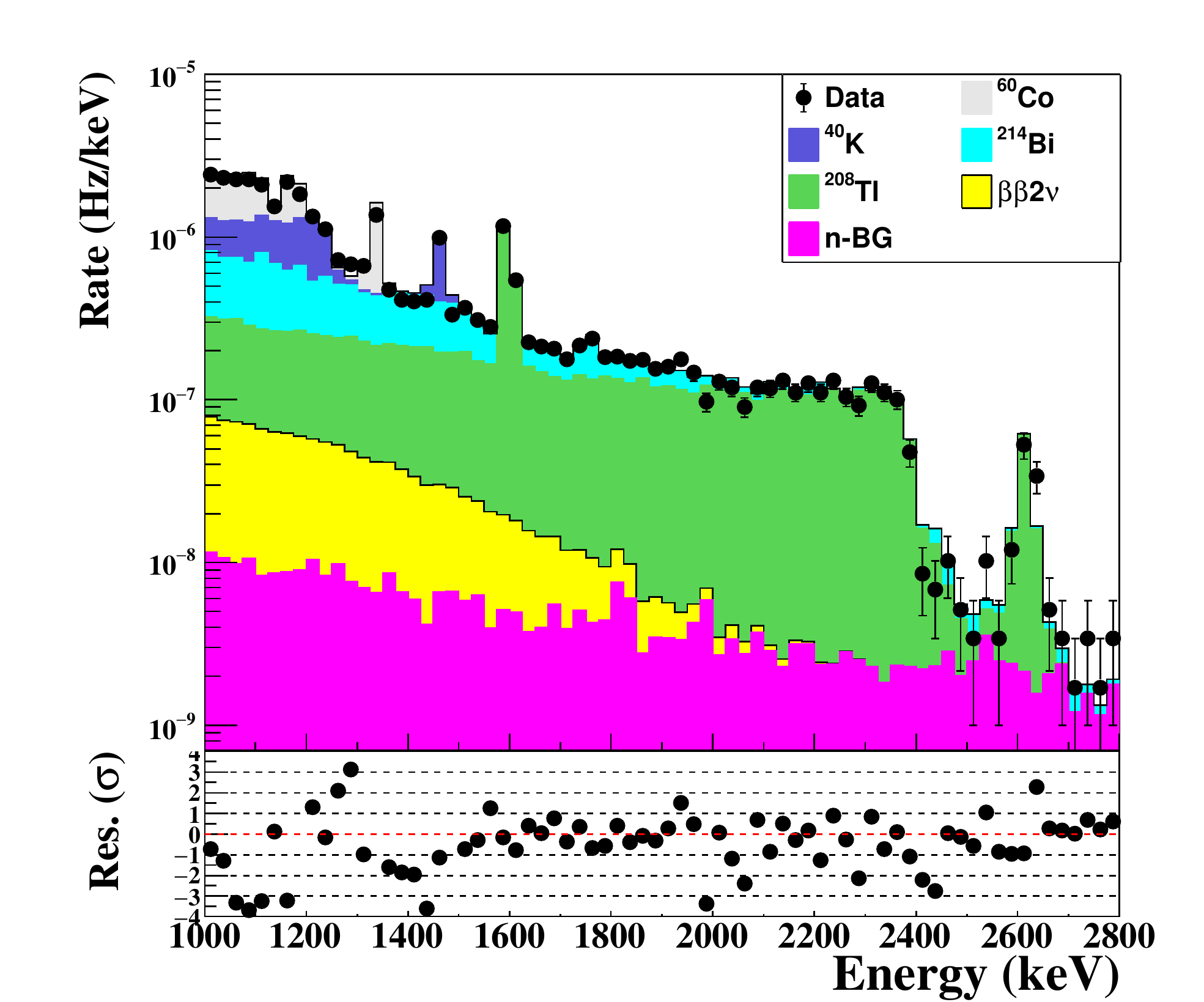}
    \includegraphics[scale=0.37]{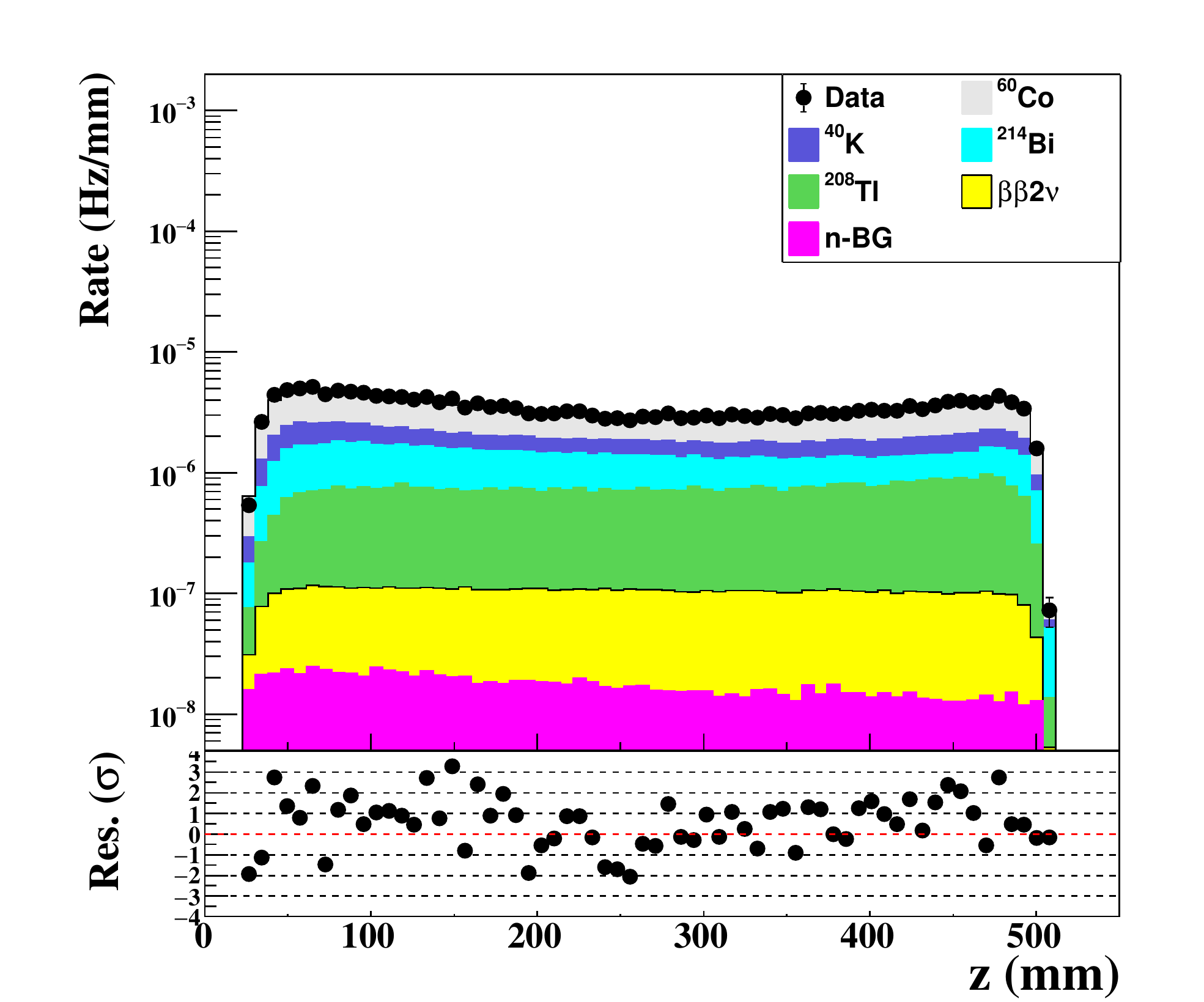}
    \includegraphics[scale=0.37]{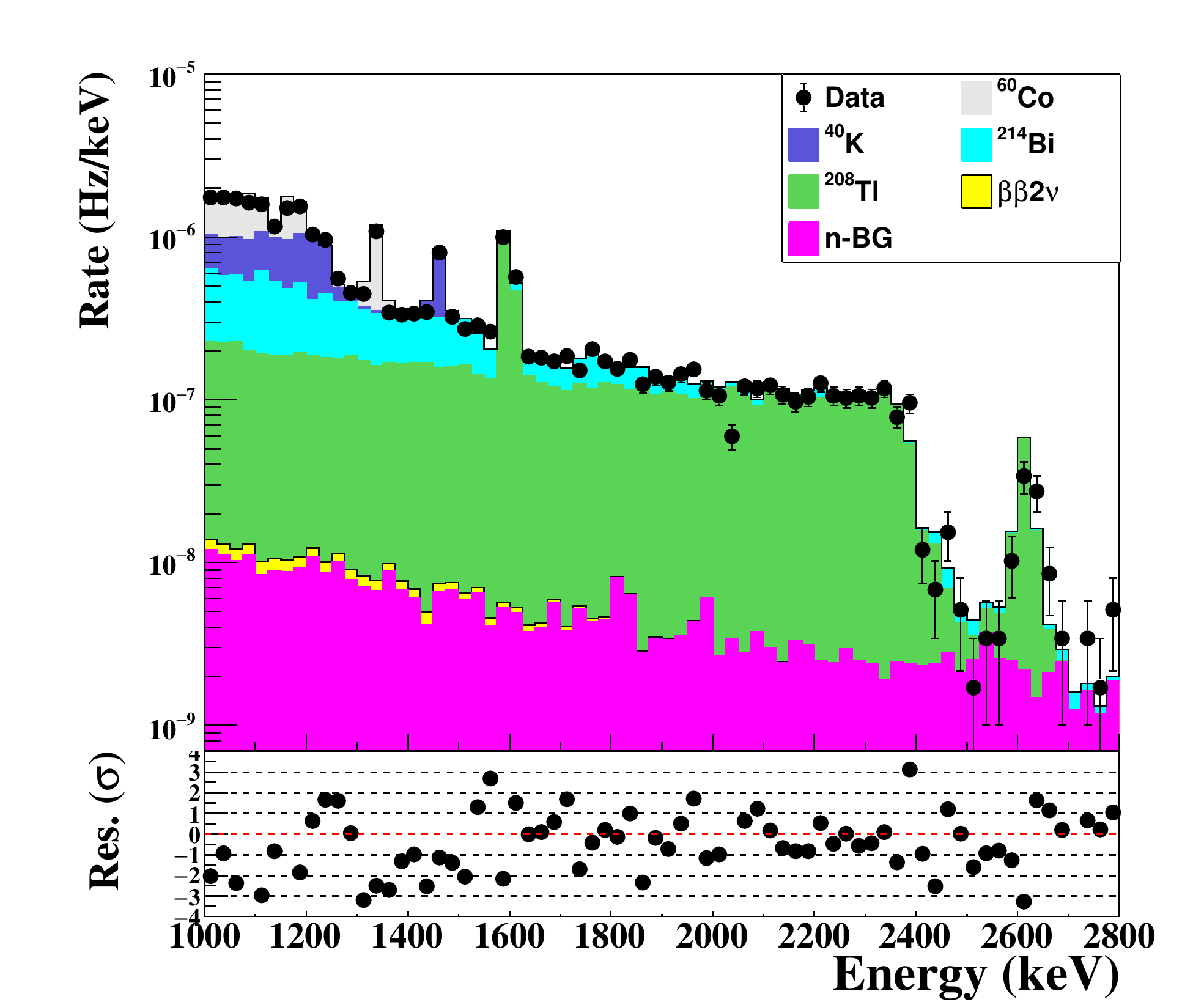}
    \includegraphics[scale=0.37]{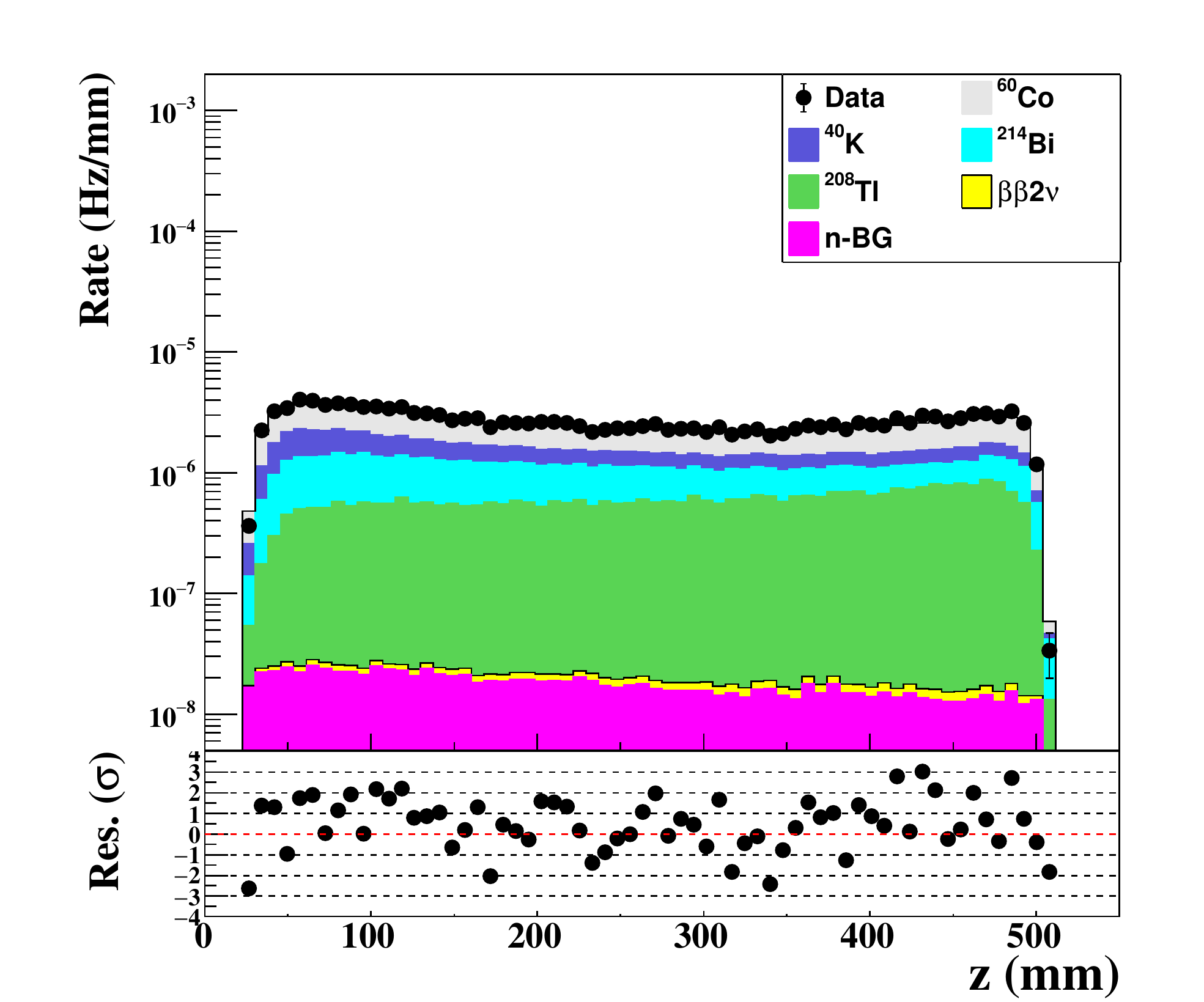}
    \caption{ Run-V (top) and Run-VI (bottom) background fits. Data (black dots) are superimposed to the best-fit background model expectation (solid histograms), for which the different radiogenic (\Bi{214}, \Tl{208}, \K{40}, and \Co{60}) and cosmogenic-neutron-induced (n-BG) contributions are shown stacked to each other. The best-fit residuals are shown below the distributions.}
    \label{fig:bgfit}
  \end{center}
\end{figure}

The results of the Run-V and Run-VI background fits are presented in Fig.~\ref{fig:bgfit}, superimposing the data to the best-fit MC expectations. All the characteristic lines of the considered radiogenic isotopes are well described, as well as the cosmogenic contribution above the Compton edge of \Tl{208} ($E\gtrsim$2.4 MeV). However, the best-fit residuals of the energy and Z distributions exhibit deviations between data and MC (particularly below 1.5 MeV) leading to a poor goodness of fit ($\chi^2$/dof=329.6/131 and $\chi^2$/dof=345.1/131 in Run-V and Run-VI respectively). Although under investigation, these deviations seem to be connected to limitations in the simulation of the detector response and/or of the background model.

\begin{figure}
  \begin{center}
    \includegraphics[scale=0.5]{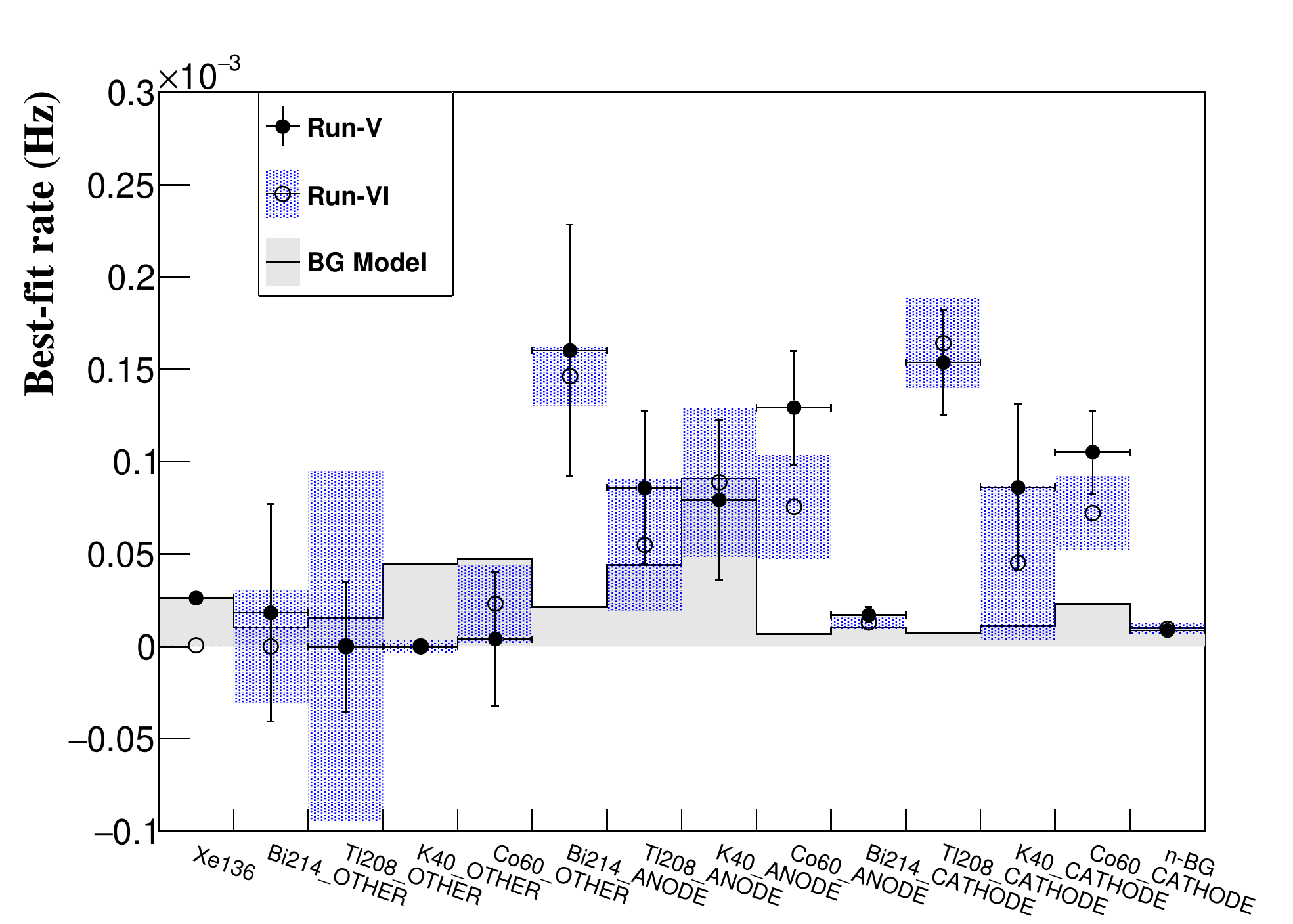}
    \caption{ Best-fit rates of the 13 background contributions considered in the background fit of Run-V (black dots) and Run-VI (empty dots). The expectation from the background model (gray histogram) is superimposed for comparison purposes.}
    \label{fig:bgbfval}
  \end{center}
\end{figure}

The central values and errors for the 13 fit parameters are shown in Fig.~\ref{fig:bgbfval}, superimposed on the corresponding background model expectation. The best-fit rates are found to be fully consistent between Run-V and Run-VI, implying background stability within measurement errors between both data taking periods. The dominant radiogenic background sources are identified to come from the cathode and anode regions, with \Tl{208} (\Bi{214}) being dominant in the former (latter). Significant deviations with respect to the model are observed, in particular in these two regions. Some component installed in the detector may be responsible for a larger activity contribution compared to what inferred from the corresponding screened samples. Possible causes include contamination in the machining and/or installation processes, differences in the cleaning procedures, or non-uniform activities of the bulk materials. While care has been placed into minimizing those differences between screened and installed materials, they cannot be eliminated. Concerning the virtual volume OTHER, a vanishing contribution of \K{40} is measured. This indicates that the upper limits adopted in the model for the \K{40} activity in the PTFE light tube might be significantly larger than the actual contamination. Finally, the measured cosmogenic (neutron-induced) background, labeled as n-BG in Fig.~\ref{fig:bgbfval}, is found to be fully consistent with the expectation. Since the normalization of the cosmogenic model is extracted from data above 2.8 MeV in Sec.~\ref{sec:cosmo}, this validates the reliability of the model at lower energies. Conservatively, the cosmogenic background has not been constrained in the background fit using the $E>2.8$~MeV result in Sec.~\ref{sec:cosmo}, as in absence of calibration data above 2.6 MeV, this normalization has been derived adopting the selection efficiency computed with MC alone.  

%--- TIME VARIATION

Beyond the detailed analysis of the background contributions, the time stability has also been assessed by model-independent means. The observed total rate of events above 1 MeV passing the radiogenic selection is found to be stable on a daily basis across Run-V and Run-VI \cite{NEXT:2021dqj}. The integrated rates for the whole data taking periods are 0.758$\pm$0.006 mHz and 0.742$\pm$0.011 mHz, respectively. The observed difference (0.016$\pm$0.013 mHz) is consistent with the \bbtwonu rate expectation in Run-V ($\sim$0.027 mHz) based on the half-life reported in \cite{Albert:2013gpz}. In the energy window between 2.0 and 2.4 MeV (close to the \bbnonu region of interest and with negligible \bbtwonu contribution) the background rates are also found to be consistent between Run-V (46.7$\pm$1.4 $\mu$Hz) and Run-VI (43.7$\pm$1.4 $\mu$Hz). Finally, specific analyses have been performed to assure the stability of the \Co{60} and \Rn{222}-induced backgrounds, as discussed in the following.

\Co{60} has a cosmogenic origin. Assuming that most of the \Co{59} activations took place while the detector materials were on the surface, the contribution of this background would be decreasing with time according to the half-life of \Co{60} (5.27 year). However, activations keep happening at the LSC overburden, even if at a much lower rate, so the activity evolution may not be an exponential. The evaluation of the integrated variation of the \Co{60} rate between Run-V and Run-VI has been carried out by fitting the 1173 keV gamma line. The intensity of this line in both periods, measured with a $\sim8\%$ precision, is consistent within $<0.5\sigma$, thus indicating secular equilibrium.

The amount of \Rn{222} in the gas evolves in time according to different factors, such as the outgassing rate in the materials or the performance of the hot getter. In order to monitor the time variation of the Rn-induced background, the rate of electron tracks emanating from the cathode surface has been measured. The integrated rates in Run-V and Run-VI are 0.206$\pm$0.003 mHz and 0.214$\pm$0.004 mHz, respectively. The difference of 8$\pm$5 $\mu$Hz is attributed to the variation of the Rn contamination between both periods. According to a MC simulation, this difference translates into a fiducial background difference of 0.12$\pm$0.08 $\mu$Hz between Run-V and Run-VI, a negligible variation with respect to the total observed background. This confirms the results of the background model fit, where the contribution of the \Bi{214} from the cathode region is found to be stable within a 20\% precision (see Fig.~\ref{fig:bgbfval}).

\section{Neutrinoless double beta decay searches}
\label{sec:bb}

Since the \Xe{136}-depleted data sample provides a direct measurement of the backgrounds, the \bbnonu analyses exploit the combination of Run-V and Run-VI data, as done for the \bbtwonu measurement in \cite{NEXT:2021dqj}. Provided that the time stability of the radiogenic and cosmogenic events has been assured, the rate of \bbnonu events can be extracted from the comparison of both samples, together with the rate of \bbtwonu events. In this case, the data and MC events considered are those obtained by applying the $\beta\beta$ selection defined in Sec.~\ref{sec:sel}. This selection boosts the signal over background ratio, thus improving the sensitivity of the \bbnonu search. The combination of the Run-V and Run-VI data is performed in two parallel approaches. In the first one, a background-model-dependent technique is developed along the lines of other \bb decay experiments, where data are confronted to the signal plus background MC expectations. In the second approach, a novel direct background-subtraction technique is developed, suppressing the dependence of the analysis on the background model. As according to the measured energy resolution the expected number of \bbtwonu events above 2.4 MeV is totally negligible ($<$0.006 year$^{-1}$), this decay does not impact the sensitivity of these analyses to the \bbnonu signal. An extensive list of systematic uncertainties has been evaluated. The main normalization uncertainties, presented in previous sections, are the same ones used in our NEXT-White \bbtwonu measurement \cite{NEXT:2021dqj}, and are reported again in Tab.~\ref{tab:sys} for convenience. Although the dominant contribution is that one of the selection efficiencies, all of them have been considered in the \bb analyses. Apart from the rate normalization uncertainty, a 0.3\% error on the energy scale has also been adopted.  

\begin{table}[b]
  \caption{\label{tab:sys} Rate normalization uncertainties in \Xe{136}-enriched and \Xe{136}-depleted data samples. The fourth column indicates whether the uncertainty is correlated between the two periods. No gas density uncertainty is assigned to Run-V as Run-VI density is corrected with respect to this period. The specific uncertainties in the selection efficiencies for single-electron and double-electron events are shown. Sources above the continuous line affect both the signal and background rates, while the sources below have an impact only on the \bb signals (\bbnonu and \bbtwonu).}
\begin{center}
  \begin{tabular}{cccc}
 \hline   
Source  & Run-V (\%) & Run-VI (\%) & Correlated\\
\hline
DAQ live-time & 0.01 & 0.01 & No \\
Gas density & - & 0.6 & No \\
$\beta\beta$ selection for 2e$^{-}$ & 2.1 & 2.1 & No \\
$\beta\beta$ selection for 1e$^{-}$ & 2.8 & 3.0 & No \\ \hline
\Xe{136}-fraction & 0.4 & 0.2 & No \\
Number of Xe atoms & 0.2 & 0.2 & Yes \\
Trigger efficiency & 0.2 & 0.2 & Yes \\\hline
\end{tabular}
\end{center}
\end{table}

 \subsection{Background-model-dependent analysis}
\label{sec:bbmodel}

%--- fit

The search for the \bbnonu process is performed by means of a likelihood fit of the Run-V and Run-VI data samples to the corresponding MC expectations. This joint fit follows the approach presented in Sec.~\ref{sec:bgmeas} for the measurement of the radiogenic and cosmogenic backgrounds. However, the Z distributions of the different sources are not considered, as they are not relevant for the estimation of the \bb event rate. As the time stability of the backgrounds is demonstrated (see Sec.~\ref{sec:bgmeas}), the specific contributions are taken to be the same in both periods. This stability has also been confirmed around the \Qbb value and for the \bb selection, by computing the event rates in a 2.0-2.4 MeV energy window, where the contribution of \bb decays is negligible. Consistent rates of 2.9$\pm$0.4 $\mu$Hz and 2.7$\pm$0.4 $\mu$Hz are found in Run-V and Run-VI, respectively. The best-fit values for the 5 background sources (the four radiogenic isotopes and the total neutron-induced cosmogenic backgrounds) are extracted along with the \bbnonu and \bbtwonu rates, which in turn are translated into best-fit values for the corresponding neutrinoless and two-neutrino mode half-lives. The normalization systematic uncertainties are included in the fit via nuisance parameters constrained by the estimations summarized in Tab.~\ref{tab:sys}. The error on the energy scale ($\sigma_{scl}$=0.3\%) has also been considered with an extra nuisance parameter, although it has been found to have a negligible impact.

%--- results

\begin{figure}
  \begin{center}
    \includegraphics[scale=0.37]{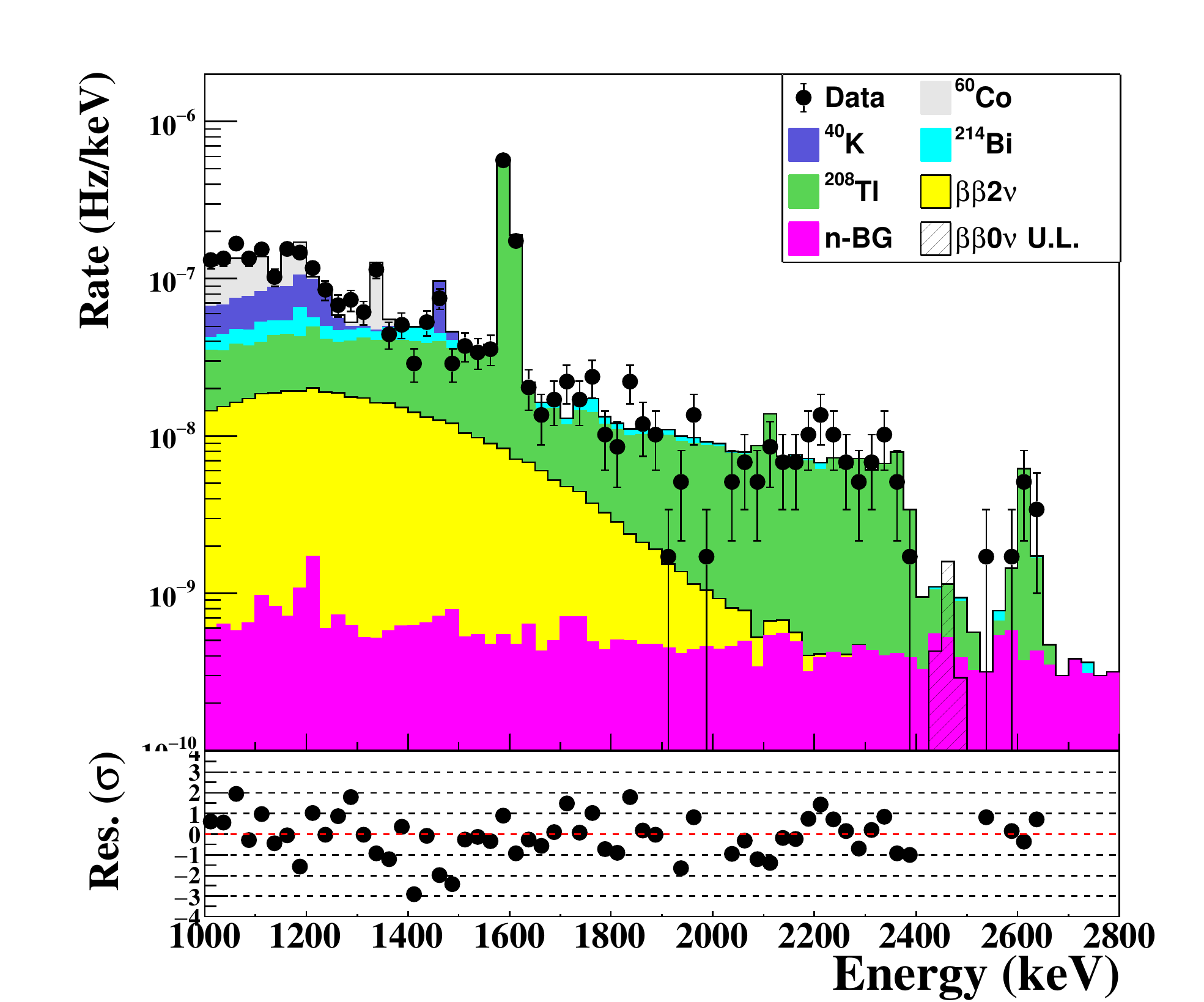}
    \includegraphics[scale=0.37]{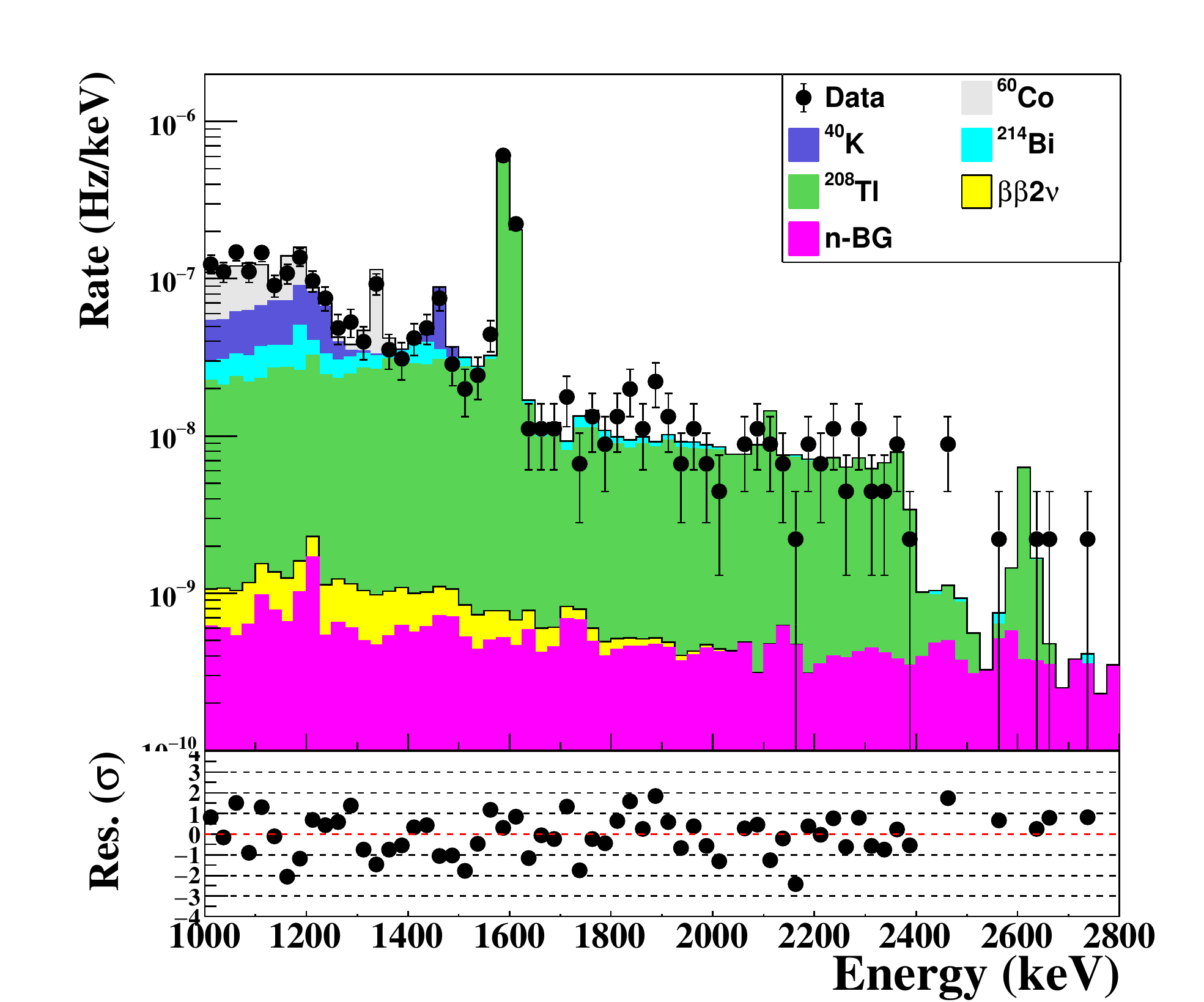}
    \caption{Background-model-dependent \bb fit  combining Run-V (left) and Run-VI (right). Data (black dots) are superimposed to the best-fit background model expectation (solid histograms), for which the different contributions are shown stacked to each other. The \bbnonu signal corresponding to the obtained upper limit rate at 90\% C.L. is also shown in the left panel.}
    \label{fig:bbmodelfit}
  \end{center}
\end{figure}

The result of this background-model-dependent \bb fit is shown in Fig.~\ref{fig:bbmodelfit}, where the Run-V and Run-VI data are superimposed to the post-fit energy distributions. The best-fit values and errors of the different background contributions are presented in Tab.~\ref{tab:bfbg}. The central values of the nuisance parameters in the fit are found to be consistent with their priors within one sigma, reinforcing the reliability of the systematic uncertainties' evaluation.

\begin{table}[!htb]
\caption{\label{tab:bfbg} Background rates extracted from the \bb fit. The middle and right columns show the background contributions in the entire energy range considered in the fit and in a 100 keV window around \Xe{136} \Qbb, respectively.}
\begin{center}
\begin{tabular}{ccc}
\hline
Background & 1000--2800 keV  & 2400--2500 keV \\
           & ($\mu$Hz)       & (yr$^{-1}$) \\ \hline
\Bi{214}   &  5.5 $\pm$ 2.7  &  0.1 $\pm$ 0.1 \\
\Tl{208}   & 39.8 $\pm$ 1.5  &  1.7 $\pm$ 0.3  \\
\Co{60}   &  14.7 $\pm$ 2.2  &  0  \\
\K{40}    &  10.6 $\pm$ 1.8  &  0  \\
Cosmogenic &  1.0 $\pm$ 0.5  &  1.5 $\pm$ 0.9 \\ \hline
\end{tabular}
\end{center}
\end{table}

%---- bb0nu

\begin{figure}
  \begin{center}
    \includegraphics[scale=0.37]{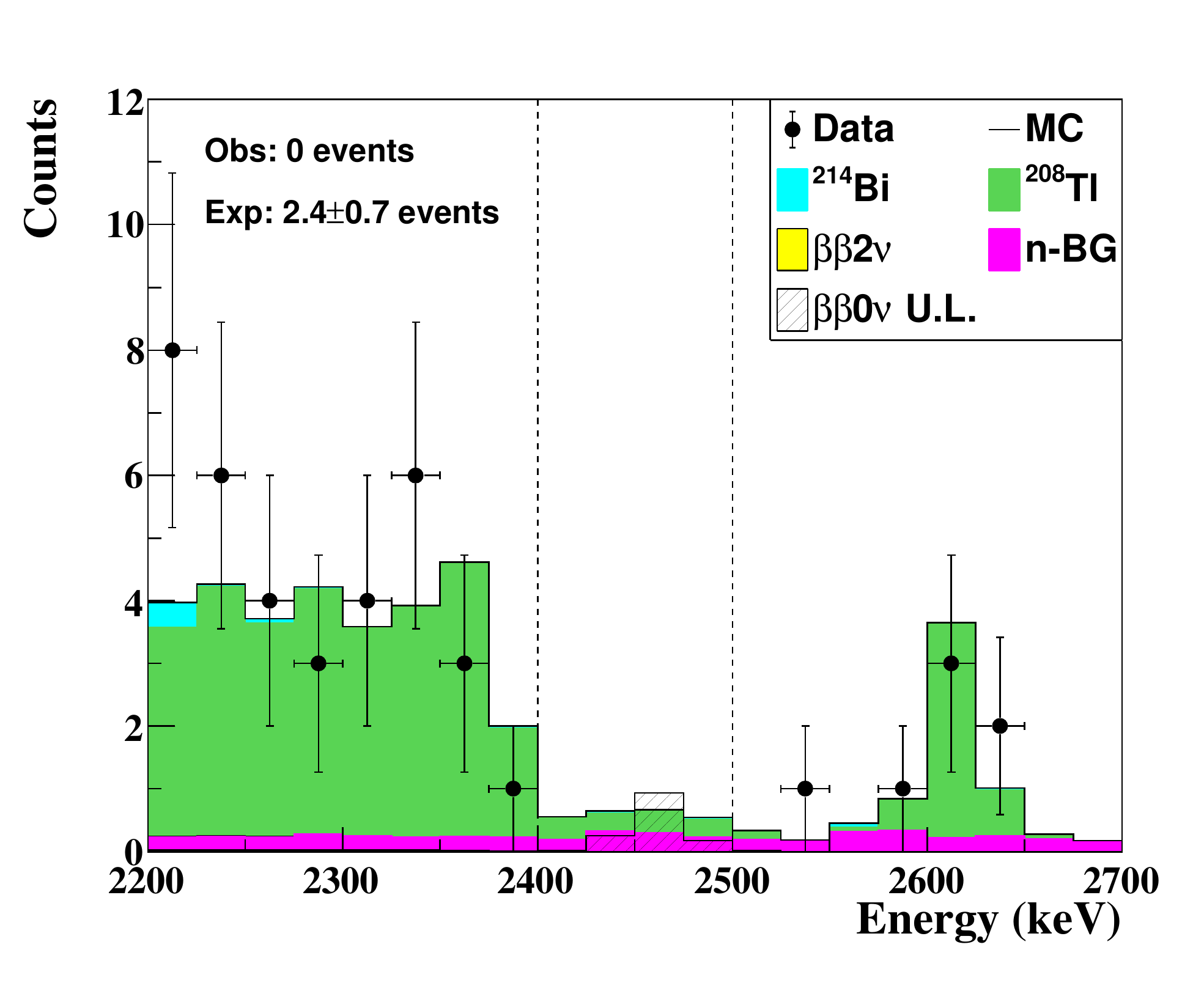}
    \includegraphics[scale=0.37]{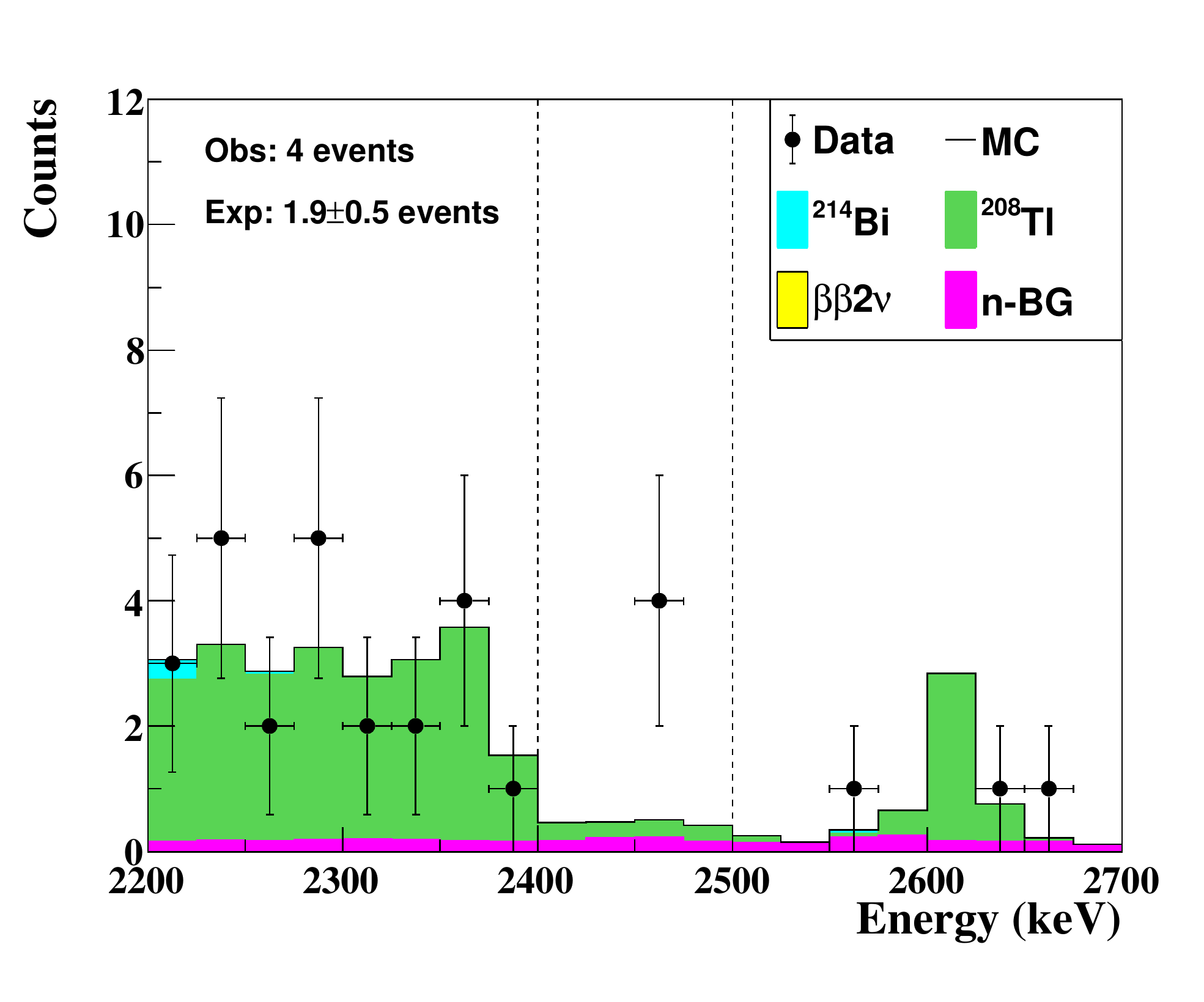}
    \caption{Energy spectra around the \Qbb of \Xe{136} according to the background-model fit presented in the text. Run-V (left) and Run-VI (right) data are superimposed to the best-fit background expectations and the \bbnonu signal corresponding to the obtained upper limit rate at 90\% C.L. The vertical dashed lines define a 100 keV region around \Qbb. The number of observed and expected events within this ROI is also displayed.}
    \label{fig:bbmodelroi}
  \end{center}
\end{figure}

As seen in Fig.~\ref{fig:bbmodelroi}, no events are observed in the 2400-2500 keV energy range in the \Xe{136}-enriched data sample, while 4 events are observed in the \Xe{136}-depleted data sample (displays and further details on these events are presented in Appendix~\ref{sec:displays}). Thus, no evidence of the \bbnonu decay has been found in NEXT-White. According to the fit results, the expected number of Run-V background events in this energy window is 2.4 $\pm$ 0.7, arising from 1.3 $\pm$ 0.2 \Tl{208} events, 1.0 $\pm$ 0.6 cosmogenic events and 0.1 $\pm$ 0.1 \Bi{214} events. For this  best-fit expectation, the probability to observe 0 events is 11.4\%. Owing to the different exposure, the background expectations in Run-VI for \Tl{208}, cosmogenic and \Bi{214} events are 1.0 $\pm$ 0.2, 0.8 $\pm$ 0.5 and 0.1 $\pm$ 0.1, respectively, for a total of 1.9 $\pm$ 0.5 background events. The corresponding probability to observe 4 or more events is 13.8\%. In the absence of a \bbnonu signal, a lower limit on \Tnonu is inferred from the profile likelihood fit, considering the entire energy spectrum. The inferred lower limit on the \bbnonu half-life is \Tnonu$>5.5\times10^{23}$ yr at 90\% C.L., while the expected median sensitivity is \Tnonu$>2.9\times10^{23}$ yr. Relying on the phase space factor of \cite{Kotila:2012zza} and on the nuclear matrix elements summarized in \cite{Agostini:2022zub} (referring to shell model \cite{Menendez:2017fdf,Horoi:2015tkc,Coraggio:2020hwx,Coraggio:2022vgy}, QRPA \cite{Mustonen:2013zu,Hyvarinen:2015bda,Simkovic:2018hiq,Fang:2018tui,Terasaki:2020ndc}, EDF theory \cite{Rodriguez:2010mn,LopezVaquero:2013yji,Song:2017ktj} and IBM \cite{Barea:2015kwa,Deppisch:2020ztt} calculations), this result corresponds to an upper limit on the Majorana neutrino mass of $\mbb\equiv\lvert \sum_i U_{ei}^2 m_i\rvert =$ 0.74--3.19 eV, being $U_{ei}$ and $m_i$ the elements of the neutrino mixing matrix and the neutrino mass eigenvalues, respectively.

%---- bb2nu

This analysis also provides a cross-check of the \bbtwonu measurement reported in \cite{NEXT:2021dqj}. The main difference between the two analyses is to explicitly consider the cosmogenic background contribution in the one described here. As the cosmogenic background contribution is of little importance compared to the radiogenic background contribution in the \bbtwonu energy region of interest, similar results are expected. Indeed, the fit value for the \bbtwonu rate is R(\Xe{136}) = 335 $\pm$ 75(stat) $\pm$ 52(sys) yr$^{-1}$. The measured rate excludes the null hypothesis at 4.1 $\sigma$, reproducing the median sensitivity found in MC studies. The corresponding \bbtwonu half-life is $\Ttwonu= 2.16^{+0.62}_{-0.40}\textrm{(stat)}^{+0.40}_{-0.29}\textrm{(sys)} \times10^{21}~\textrm{yr}$, fully consistent with our measurement \cite{NEXT:2021dqj}.

%--- chi2 discussion

The reduced-$\chi^2$ of the fit, $\chi^2$/dof=172.3/137 ($p$-value = 2.2\%), points at a statistically significant deviation between the data and the MC expectations, as was the case for the radiogenic background fit in Sec.~\ref{sec:bgmeas}. However, this poor goodness of fit is not expected to impact the \bb results, as they are effectively derived from the comparison of the \Xe{136}-enriched and \Xe{136}-depleted data. Indeed, these results are confirmed by the background-subtraction approach described in Sec.~\ref{sec:bbbgsub}, where the impact of a possible mismodelling in the MC would be suppressed.

 \subsection{Background-subtraction \bb analysis}
\label{sec:bbbgsub}

The data sample taken with \Xe{136}-depleted gas allows for a direct background subtraction in the \Xe{136}-enriched data. Provided that the backgrounds are constant in time, as demonstrated in Sec.~\ref{sec:bgmeas}, this subtraction removes all radiogenic backgrounds as well as the cosmogenic ones, except for the \Xe{137}-induced (prompt-$\gamma$s and delayed $\beta$ decay). However, as shown in Sec.~\ref{sec:cosmo} and Sec.~\ref{sec:bgmeas} and due to the limited xenon mass, the contribution of \Xe{136} activations in NEXT-White is totally negligible with respect to all other backgrounds. As a consequence, this background-subtraction allows for a \bb analysis, covering both the measurement of \Ttwonu and the search of a \bbnonu signal, with little dependence on the background model. In particular, this method avoids the possible impact of the imperfections or limitations in the background model assumptions and simulation, providing results that are independent of the particular number of background sources and their spatial origin. The best-fit \bb event rates are extracted from the comparison of the background-subtracted energy spectrum and the expected signal PDF.

%----

\begin{figure}
  \begin{center}
    \includegraphics[scale=0.37]{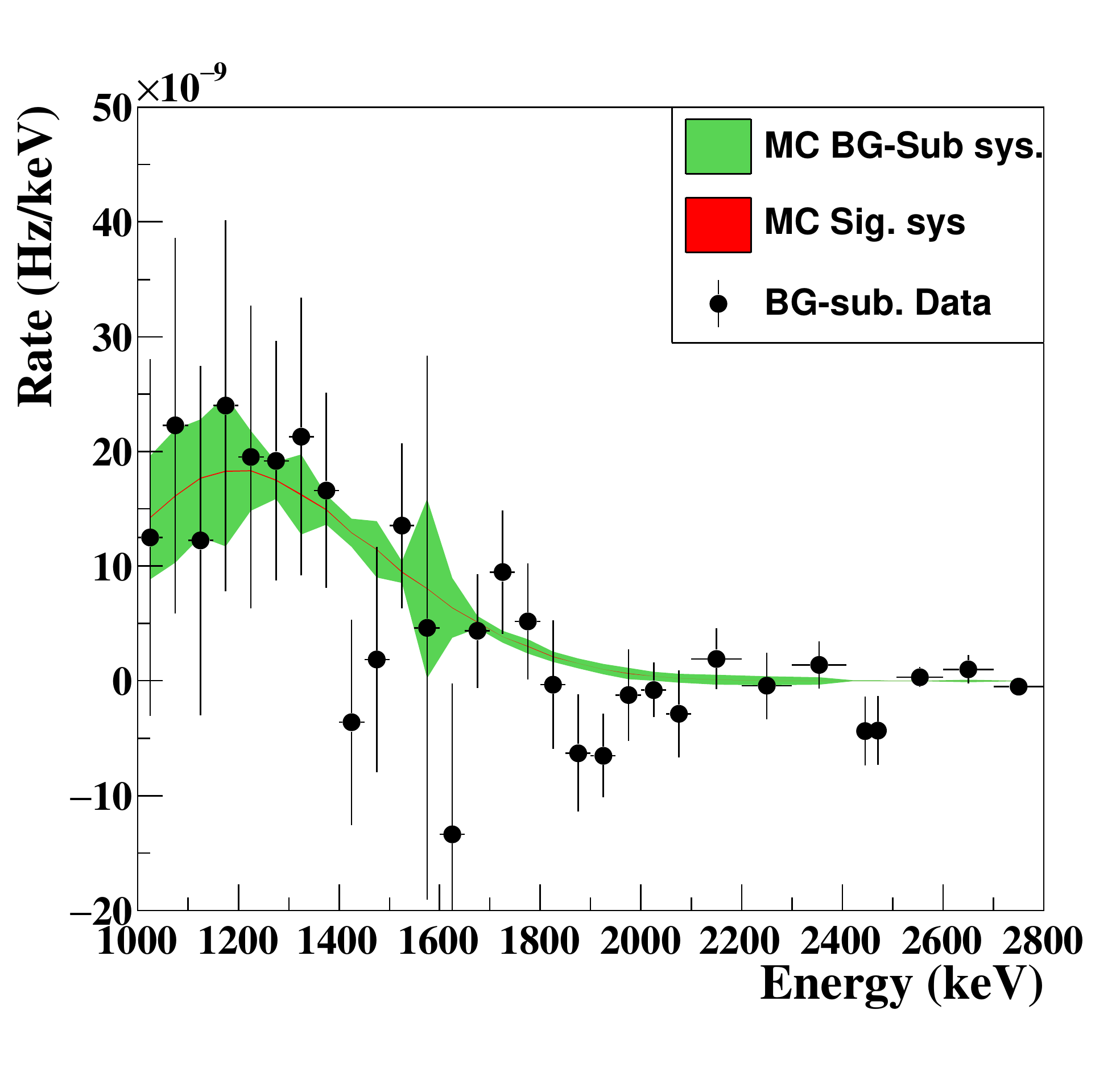}
    \includegraphics[scale=0.37]{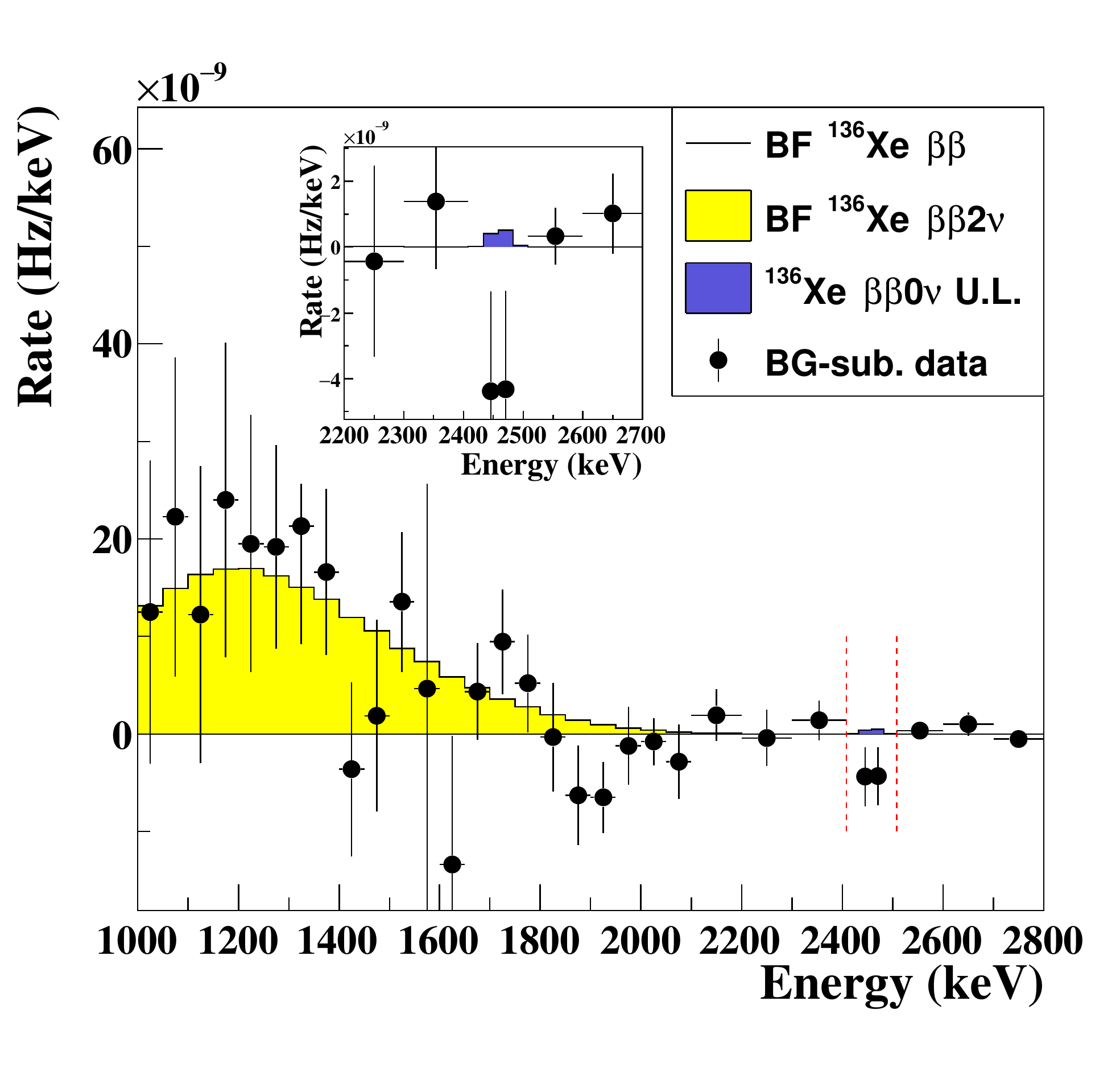}
    \caption{Background-subtracted \bb fit. Left: background-subtracted data superimposed to the \bb expectation according to the \Ttwonu reported in \cite{Albert:2013gpz}. The normalization and background-subtraction systematics are shown with red and green bands, respectively. Right: background-subtracted data superimposed to the post-fit \bbtwonu distribution and the \bbnonu signal corresponding to the obtained upper limit rate at 90\% C.L. The vertical dashed lines define a 100 keV region around \Qbb ($\sim\pm$2$\times$FWHM).}
    \label{fig:bbbgsubfit}
  \end{center}
\end{figure}

Once Run-VI data are corrected by the small differences in the DAQ live time and the selection efficiency with respect to Run-V, the subtraction of both data samples yields a positive value which is attributed to the \Xe{136} \bbtwonu events: R(\Xe{136}) = 244 $\pm$ 83(stat) $\pm$ 29(sys) yr$^{-1}$. As shown in the left panel of Fig.~\ref{fig:bbbgsubfit}, no \bbnonu signal is observed in the background-subtracted data, as a negative rate of -6.9 $\pm$ 3.3(stat) $\pm$ 0.3(sys) yr$^{-1}$ is obtained in the 2400--2500 keV region. The quoted systematic uncertainties account for both the subtraction error (considering the selection efficiencies of single and double-electron events in both periods, and the density correction applied to Run-VI), as well as the signal normalization error (considering the isotopic composition of the gas, the number of xenon atoms and the trigger efficiency). The energy scale systematic uncertainty is found to be negligible.

%---

In order to derive a lower limit to the \Xe{136} \bbnonu half-life from the background subtracted energy spectrum, a fit is performed to the corresponding MC expectation. The normalization of the \bbtwonu and \bbnonu rates stand as the only two free parameters of the fit. Due to the limited statistics, an asymmetric binning is adopted. A region of 100 keV around the \Qbb ($\sim\pm$2$\times$FWHM) is sampled in 25 keV--bins, while bins between 50 and 100 keV are used for the rest of the energy spectrum. The fit relies on the minimization of a Pearson's $\chi^2$ statistic, thus considering the statistical uncertainties in the model. The systematic uncertainty assigned to the expectation accounts for the background subtraction and signal normalization errors. The subtraction systematic uncertainty, derived from the selection efficiencies and gas density correction, is introduced in the fit as a covariance matrix. On the other hand, the signal normalization uncertainty is decomposed into the uncorrelated (isotopic composition) and correlated (number of xenon atoms and trigger efficiency) contributions between Run-V and Run-VI. These contributions are introduced in the fit as three nuisance parameters, with a null prior constrained by the corresponding uncertainties quoted in Tab.~\ref{tab:sys}. 

The pre-fit distributions (left) and the fit outcome (right) are shown in Fig.~\ref{fig:bbbgsubfit}. The fit yields a $\chi^{2}$/dof of 30.2/30 ($p$-value=46\%). The best-fit value for the \bbnonu rate is pushed to the physical limit of zero. Marginalizing with respect to all other parameters in the fit, a lower limit for the corresponding half-life of \Tnonu$>1.3\times10^{24}$ yr at 90\% C.L. is obtained. The expected median sensitivity at the same C.L. is \Tnonu$>2.9\times10^{23}$ yr, significantly below the inferred one. This discrepancy arises from the negative rate observed in the energy region around \Qbb, which deviates from zero (null signal) in $\sim$2.1$\sigma$. According to MC studies, the probability of obtaining a result that is at least as incompatible with a null signal as the obtained one is 0.5\%. Using the same phase space factor and nuclear matrix element assumptions as in Sec.~\ref{sec:bbmodel}, the obtained half-life lower value translates into an upper limit for the Majorana neutrino mass of \mbb$<$0.48--2.07 eV. The fit also yields a best-fit value for the rate of \bbtwonu events which corresponds, as expected, to a \bbtwonu half-life fully consistent with the one reported in \cite{NEXT:2021dqj} using the same background-subtraction analysis methodology.

As the obtained \bbnonu result is driven by the negative rate observed in the ROI, the reliability of the background-subtraction technique has been cross-checked by analyzing in detail the energy region above 2.0 MeV, where the $2\nu\bb$ events are negligible. Regardless of the chosen energy ranges and binning, the background-subtracted rates in energy windows below and above the ROI are found to be consistent with zero. In particular, rates of 5$\pm$12 yr$^{-1}$ and 2.6$\pm$5.0 yr$^{-1}$ are obtained for the $2.1<E<2.4$ MeV and $2.5<E<2.8$ MeV energy ranges, respectively. According to these results, the hypothesis of an unaccounted background source in Run-VI (leading to the observed 4 events in the ROI) seems strongly disfavored, as such a source would only have a discrete gamma line around 2.45 MeV.

\section{Summary and conclusions}
\label{sec:summary}

Although originally beyond its scientific goals, the NEXT-White detector has been fully exploited in order to perform a demonstration of the \bbnonu search capabilities of the NEXT technology. While a competitive result cannot be achieved due to the limited fiducial xenon mass (3.50$\pm$0.01 kg), the excellent performance of NEXT-White has provided a proof-of-concept for \bbnonu searches in future NEXT detectors. The analysis considers 271.6 days of \Xe{136}-enriched data (Run-V) and 208.9 days of \Xe{136}-depleted data (Run-VI). As a preliminary step, a detailed background measurement in both data samples has been conducted to ensure the time stability of the different contributions. The radiogenic-induced backgrounds are modeled upon the screening of the detector materials. Up to 12 radiogenic contributions are accounted for, considering \Bi{214}, \Tl{208}, \Co{60} and \K{40} from three effective volumes. The model for cosmogenic-induced backgrounds is derived from MC simulations and the NEXT-White data collected above 2.7 MeV, the maximum energy of the radiogenic background. Activations of \Cu{63} and \Cu{65} (in the natural copper used as an internal shielding) are found to be the dominant cosmogenic contributions, while the activation of \Xe{136} is negligible. All the background contributions are measured independently in Run-V and Run-VI, yielding consistent results between the two periods. Two different \bb analyses are conducted: 1) a background-model-dependent approach, fitting the Run-V and Run-VI data to the background plus \bb models; and 2) a direct background-subtraction approach, fitting the difference between both data samples to the \bb expectation. No \bbnonu signal is observed in the NEXT-White data. From the background-model-dependent and the background-subtraction fits, 90\% C.L. limits of \Tnonu$>5.5\times10^{23}$ yr and \Tnonu$>1.3\times10^{24}$ yr, respectively, are derived. The background-subtraction result is more sensitive, as the background found in the 2400-2500 keV region is below the model expectation. 

These results demonstrate the unique capabilities of the NEXT technology, in particular exploiting two novel techniques in the field of \bbnonu searches: an efficient background suppression based on the high-definition topology of the electron tracks, and a direct subtraction of the remaining backgrounds by combining \Xe{136}-enriched and \Xe{136}-depleted data samples. A similar approach may be used in NEXT-100 and future detectors to conduct low-background, and virtually background-model-independent, searches for neutrinoless double beta decay.

\acknowledgments
The NEXT Collaboration acknowledges support from the following agencies and institutions: the European Research Council (ERC) under Grant Agreement No. 951281-BOLD; the European Union’s Framework Programme for Research and Innovation Horizon 2020 (2014–2020) under Grant Agreement No. 957202-HIDDEN; the MCIN/AEI of Spain and ERDF A way of making Europe under grants PID2021-125475NB and the Severo Ochoa Program grant CEX2018-000867-S; the Generalitat Valenciana of Spain under grants PROMETEO/2021/087 and CIDEGENT/2019/049; the Department of Education of the Basque Government of Spain under the predoctoral training program non-doctoral research personnel; the Spanish la Caixa Foundation (ID 100010434) under fellowship code LCF/BQ/PI22/11910019; the Portuguese FCT under project UID/FIS/04559/2020 to fund the activities of LIBPhys-UC; the Israel Science Foundation (ISF) under grant 1223/21; the Pazy Foundation (Israel) under grants 310/22, 315/19 and 465; the US Department of Energy under contracts number DE-AC02-06CH11357 (Argonne National Laboratory), DE-AC02-07CH11359 (Fermi National Accelerator Laboratory), DE-FG02-13ER42020 (Texas A\&M), DE-SC0019054 (Texas Arlington) and DE-SC0019223 (Texas Arlington); the US National Science Foundation under award number NSF CHE 2004111; the Robert A Welch Foundation under award number Y-2031-20200401. Finally, we are grateful to the Laboratorio Subterr\'aneo de Canfranc for hosting and supporting the NEXT experiment.

\appendix
\section{\bbnonu candidates in the \Xe{136}-depleted data sample}
\label{sec:displays}

\begin{figure}
  \begin{center}
    \includegraphics[width=\textwidth]{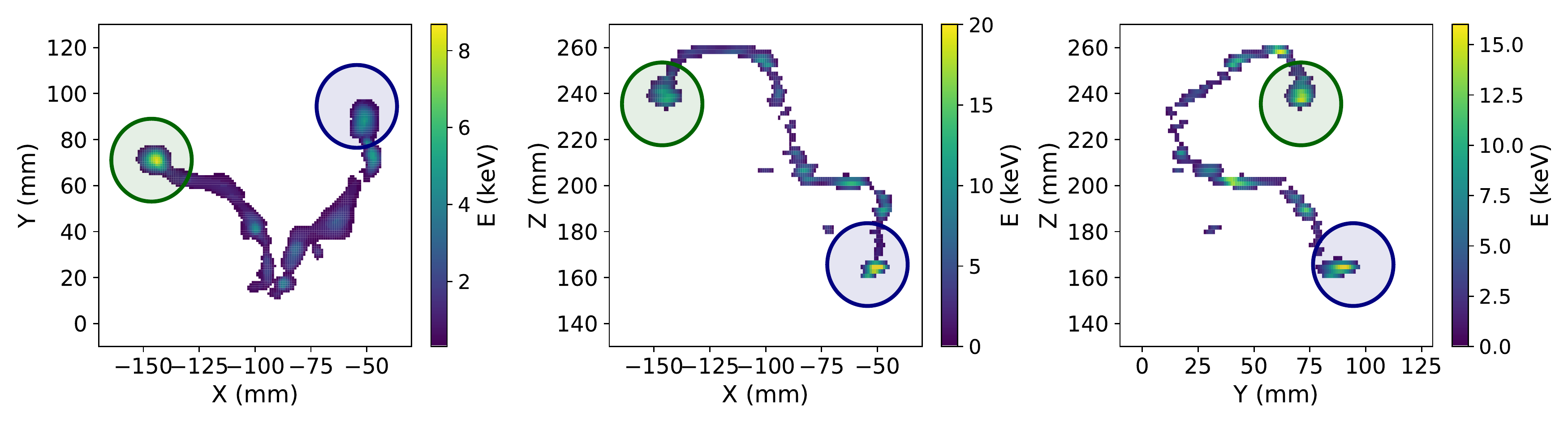}
    \caption{XY (left), XZ (middle), and YZ (right) projections of one of the \bbnonu candidates. The energy of this event is 2.451 MeV, and its blob energies  0.538 MeV (green circle) and 0.449 MeV (blue circle). The event was detected the 14$^{th}$ of November 2020, at 21:35:53 (CET).}
    \label{fig:event1}
  \end{center}
\end{figure}

\begin{figure}
  \begin{center}
    \includegraphics[width=\textwidth]{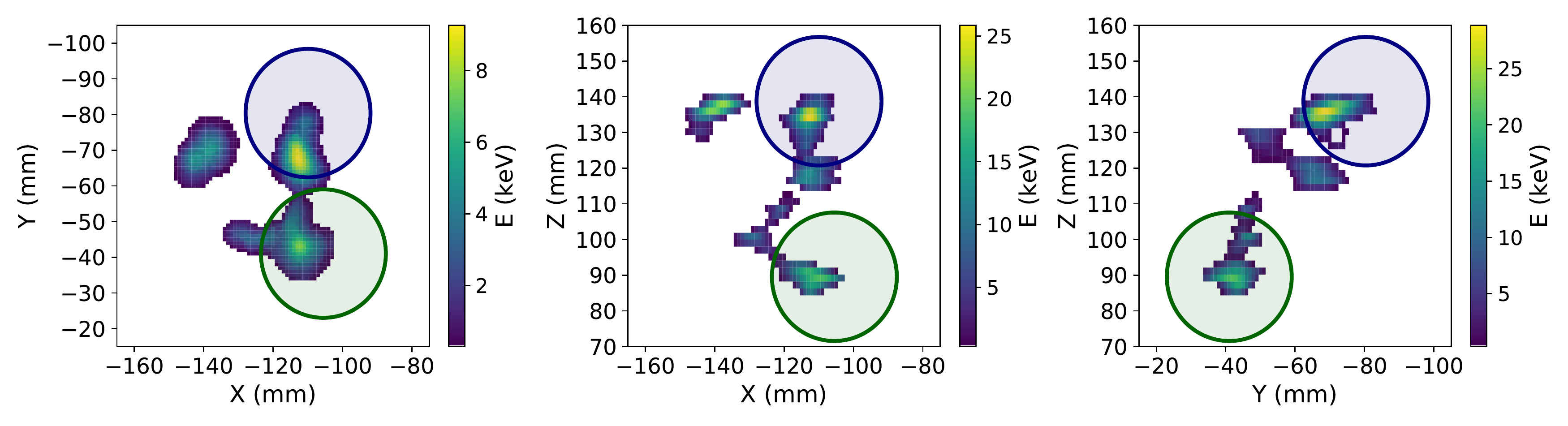}
    \caption{XY (left), XZ (middle), and YZ (right) projections of one of the \bbnonu candidates. The energy of this event is 2.461 MeV, and its blob energies 0.618 MeV (green circle) and 0.563 MeV (blue circle). The event was detected the 28$^{th}$ of December 2020, at 10:41:40 (CET).}
    \label{fig:event2}
  \end{center}
\end{figure}

\begin{figure}
  \begin{center}
    \includegraphics[width=\textwidth]{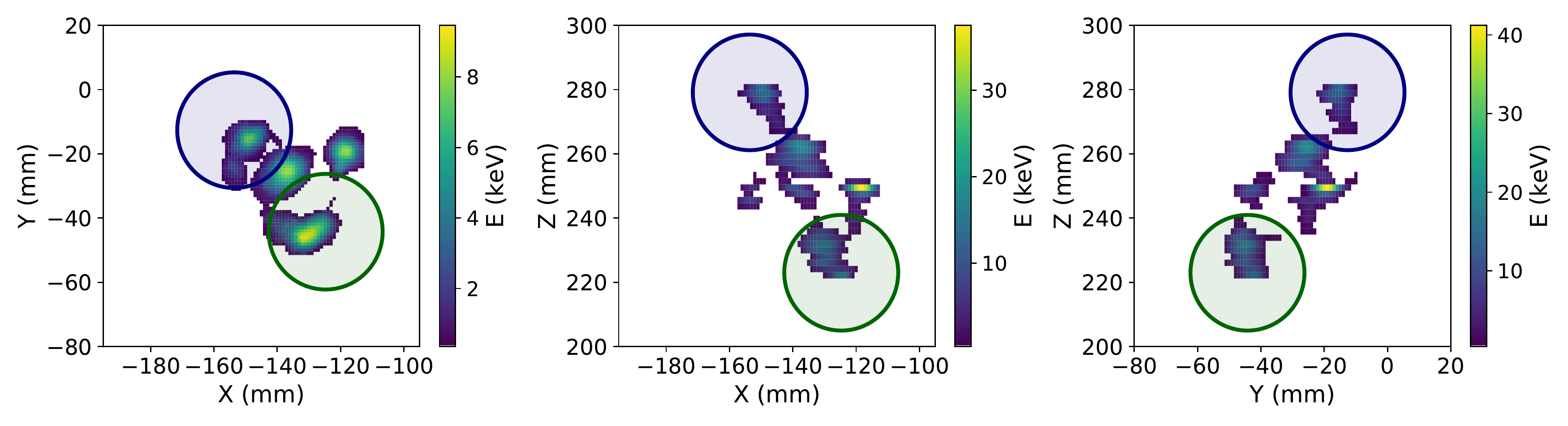}
    \caption{XY (left), XZ (middle), and YZ (right) projections of one of the \bbnonu candidates. The energy of this event is 2.451 MeV, and its blob energies 0.665 MeV (green circle) and 0.398 MeV (blue circle). The event was detected the 10$^{th}$ of March 2021, at 5:14:52 (CET).}
    \label{fig:event3}
  \end{center}
\end{figure}

\begin{figure}
  \begin{center}
    \includegraphics[width=\textwidth]{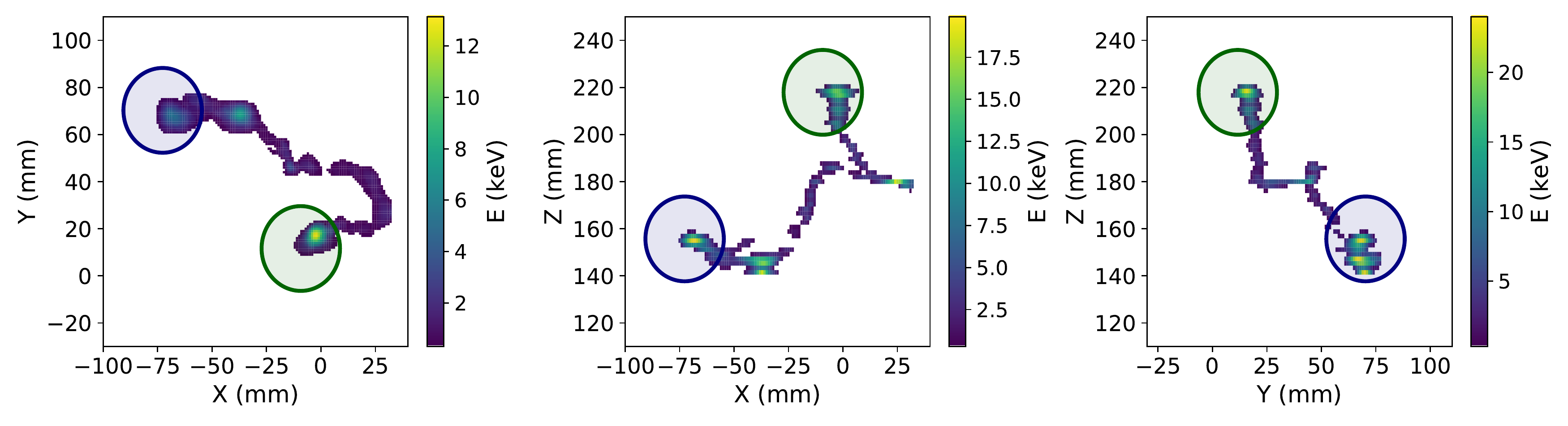}
    \caption{XY (left), XZ (middle), and YZ (right) projections of one of the \bbnonu candidates. The energy of this event is 2.467 MeV, and its blob energies  0.675 MeV (green circle) and 0.551 MeV (blue circle). The event was detected the 27$^{th}$ of May 2021, at 8:12:16 (CET).}
    \label{fig:event4}
  \end{center}
\end{figure}

This appendix provides further information about the 4 \bb candidates observed in the ROI during the \Xe{136}-depleted data taking period. Visual displays are shown in Figures~\ref{fig:event1}-\ref{fig:event4}. The energy of each event, as well as specific energies of the track blobs, are quoted in the corresponding caption. According to the energy of these events, the $E_{\mathrm{b,min}}$ threshold applied to their blob energies is 0.379 MeV. The time stamps of the events, also shown in the caption of the displays, do not reveal any time correlation between them. While the events in Fig.~\ref{fig:event1} and Fig.~\ref{fig:event4} resemble typical double-electron tracks from pair-creation interactions, the ones in Fig.~\ref{fig:event2} and Fig.~\ref{fig:event3} exhibit a more complex topology. From a visual inspection, the hypothesis of multiple isolated energy depositions being wrongly reconstructed as a single track seems plausible.

\bibliographystyle{JHEP}
\bibliography{biblio}

\providecommand{\href}[2]{#2}\begingroup\raggedright\begin{thebibliography}{10}

\bibitem{Barabash:2020nck}
A.~Barabash, \emph{{Precise Half-Life Values for Two-Neutrino
  Double-\ensuremath{\beta} Decay: 2020 Review}},
  \href{https://doi.org/10.3390/universe6100159}{\emph{Universe} {\bfseries 6}
  (2020) 159} [\href{https://arxiv.org/abs/2009.14451}{{\ttfamily
  2009.14451}}].

\bibitem{KamLAND-Zen:2022tow}
{\scshape KamLAND-Zen} collaboration, \emph{{First Search for the Majorana
  Nature of Neutrinos in the Inverted Mass Ordering Region with KamLAND-Zen}},
  \href{https://arxiv.org/abs/2203.02139}{{\ttfamily 2203.02139}}.

\bibitem{GERDA:2020emj}
{\scshape GERDA} collaboration, \emph{{First Search for Bosonic Superweakly
  Interacting Massive Particles with Masses up to 1 MeV/$c^2$ with GERDA}},
  \href{https://doi.org/10.1103/PhysRevLett.125.011801}{\emph{Phys. Rev. Lett.}
  {\bfseries 125} (2020) 011801}
  [\href{https://arxiv.org/abs/2005.14184}{{\ttfamily 2005.14184}}].

\bibitem{NEXT:2018rgj}
{\scshape NEXT} collaboration, \emph{{The Next White (NEW) Detector}},
  \href{https://doi.org/10.1088/1748-0221/13/12/P12010}{\emph{JINST} {\bfseries
  13} (2018) P12010} [\href{https://arxiv.org/abs/1804.02409}{{\ttfamily
  1804.02409}}].

\bibitem{Renner:2019pfe}
{\scshape NEXT} collaboration, \emph{{Energy calibration of the NEXT-White
  detector with 1\% resolution near Q$_{\beta \beta}$ of $^{136}$Xe}},
  \href{https://doi.org/10.1007/JHEP10(2019)230}{\emph{JHEP} {\bfseries 10}
  (2019) 230} [\href{https://arxiv.org/abs/1905.13110}{{\ttfamily
  1905.13110}}].

\bibitem{Redshaw:2007un}
M.~Redshaw, E.~Wingfield, J.~McDaniel and E.G.~Myers, \emph{{Mass and
  double-beta-decay Q value of Xe-136}},
  \href{https://doi.org/10.1103/PhysRevLett.98.053003}{\emph{Phys. Rev. Lett.}
  {\bfseries 98} (2007) 053003}.

\bibitem{NEXT:2019gtz}
{\scshape NEXT} collaboration, \emph{{Demonstration of the event identification
  capabilities of the NEXT-White detector}},
  \href{https://doi.org/10.1007/JHEP10(2019)052}{\emph{JHEP} {\bfseries 10}
  (2019) 052} [\href{https://arxiv.org/abs/1905.13141}{{\ttfamily
  1905.13141}}].

\bibitem{NEXT:2020try}
{\scshape NEXT} collaboration, \emph{{Boosting background suppression in the
  NEXT experiment through Richardson-Lucy deconvolution}},
  \href{https://doi.org/10.1007/JHEP07(2021)146}{\emph{JHEP} {\bfseries 21}
  (2020) 146} [\href{https://arxiv.org/abs/2102.11931}{{\ttfamily
  2102.11931}}].

\bibitem{NEXT:2020jmz}
{\scshape NEXT} collaboration, \emph{{Demonstration of background rejection
  using deep convolutional neural networks in the NEXT experiment}},
  \href{https://doi.org/10.1007/JHEP01(2021)189}{\emph{JHEP} {\bfseries 01}
  (2021) 189} [\href{https://arxiv.org/abs/2009.10783}{{\ttfamily
  2009.10783}}].

\bibitem{Novella:2019cne}
{\scshape NEXT} collaboration, \emph{{Radiogenic Backgrounds in the NEXT Double
  Beta Decay Experiment}},
  \href{https://doi.org/10.1007/JHEP10(2019)051}{\emph{JHEP} {\bfseries 10}
  (2019) 051} [\href{https://arxiv.org/abs/1905.13625}{{\ttfamily
  1905.13625}}].

\bibitem{NEXT:2020amj}
{\scshape NEXT} collaboration, \emph{{Sensitivity of a tonne-scale NEXT
  detector for neutrinoless double beta decay searches}},
  \href{https://doi.org/https://doi.org/10.1007/JHEP08(2021)164}{\emph{JHEP}
  {\bfseries 164} (2021) } [\href{https://arxiv.org/abs/2005.06467}{{\ttfamily
  2005.06467}}].

\bibitem{Jones:2016qiq}
B.J.P.~Jones, A.D.~McDonald and D.R.~Nygren, \emph{{Single Molecule
  Fluorescence Imaging as a Technique for Barium Tagging in Neutrinoless Double
  Beta Decay}},
  \href{https://doi.org/10.1088/1748-0221/11/12/P12011}{\emph{JINST} {\bfseries
  11} (2016) P12011} [\href{https://arxiv.org/abs/1609.04019}{{\ttfamily
  1609.04019}}].

\bibitem{McDonald:2017izm}
A.D.~McDonald et~al., \emph{{Demonstration of Single Barium Ion Sensitivity for
  Neutrinoless Double Beta Decay using Single Molecule Fluorescence Imaging}},
  \href{https://doi.org/10.1103/PhysRevLett.120.132504}{\emph{Phys. Rev. Lett.}
  {\bfseries 120} (2018) 132504}
  [\href{https://arxiv.org/abs/1711.04782}{{\ttfamily 1711.04782}}].

\bibitem{Thapa:2019zjk}
P.~Thapa, I.~Arnquist, N.~Byrnes, A.A.~Denisenko, F.W.~Foss, B.J.P.~Jones
  et~al., \emph{{Barium Chemosensors with Dry-Phase Fluorescence for
  Neutrinoless Double Beta Decay}},
  \href{https://doi.org/10.1038/s41598-019-49283-x}{\emph{Sci. Rep.} {\bfseries
  9} (2019) 15097} [\href{https://arxiv.org/abs/1904.05901}{{\ttfamily
  1904.05901}}].

\bibitem{Rivilla:2020cvm}
I.~Rivilla et~al., \emph{{Fluorescent bicolour sensor for low-background
  neutrinoless double $\beta$ decay experiments}},
  \href{https://doi.org/10.1038/s41586-020-2431-5}{\emph{Nature} {\bfseries
  583} (2020) 48}.

\bibitem{acssensors.0c02104}
P.~Thapa, N.K.~Byrnes, A.A.~Denisenko, J.X.~Mao, A.D.~McDonald, C.A.~Newhouse
  et~al., \emph{Demonstration of selective single-barium ion detection with dry
  diazacrown ether naphthalimide turn-on chemosensors},
  \href{https://doi.org/10.1021/acssensors.0c02104}{\emph{ACS Sensors}
  {\bfseries 6} (2021) 192}
  [\href{https://arxiv.org/abs/https://doi.org/10.1021/acssensors.0c02104}{{\ttfamily
  https://doi.org/10.1021/acssensors.0c02104}}].

\bibitem{NEXT:2022ita}
{\scshape NEXT} collaboration, \emph{{Ba$^{+2}$ ion trapping using organic
  submonolayer for ultra-low background neutrinoless double beta detector}},
  \href{https://doi.org/10.1038/s41467-022-35153-0}{\emph{Nature Commun.}
  {\bfseries 13} (2022) 7741}
  [\href{https://arxiv.org/abs/2201.09099}{{\ttfamily 2201.09099}}].

\bibitem{NEXT:2015wlq}
{\scshape NEXT} collaboration, \emph{{Sensitivity of NEXT-100 to Neutrinoless
  Double Beta Decay}},
  \href{https://doi.org/10.1007/JHEP05(2016)159}{\emph{JHEP} {\bfseries 05}
  (2016) 159} [\href{https://arxiv.org/abs/1511.09246}{{\ttfamily
  1511.09246}}].

\bibitem{NEXT:2021dqj}
{\scshape NEXT} collaboration, \emph{{Measurement of the Xe136 two-neutrino
  double-\ensuremath{\beta}-decay half-life via direct background subtraction
  in NEXT}}, \href{https://doi.org/10.1103/PhysRevC.105.055501}{\emph{Phys.
  Rev. C} {\bfseries 105} (2022) 055501}
  [\href{https://arxiv.org/abs/2111.11091}{{\ttfamily 2111.11091}}].

\bibitem{NEXT:2012lrw}
{\scshape NEXT} collaboration, \emph{{Near-Intrinsic Energy Resolution for 30
  to 662 keV Gamma Rays in a High Pressure Xenon Electroluminescent TPC}},
  \href{https://doi.org/10.1016/j.nima.2012.12.123}{\emph{Nucl. Instrum. Meth.
  A} {\bfseries 708} (2013) 101}
  [\href{https://arxiv.org/abs/1211.4474}{{\ttfamily 1211.4474}}].

\bibitem{NEXT:2015rel}
{\scshape NEXT} collaboration, \emph{{First proof of topological signature in
  the high pressure xenon gas TPC with electroluminescence amplification for
  the NEXT experiment}},
  \href{https://doi.org/10.1007/JHEP01(2016)104}{\emph{JHEP} {\bfseries 01}
  (2016) 104} [\href{https://arxiv.org/abs/1507.05902}{{\ttfamily
  1507.05902}}].

\bibitem{Novella:2018ewv}
{\scshape NEXT} collaboration, \emph{{Measurement of radon-induced backgrounds
  in the NEXT double beta decay experiment}},
  \href{https://doi.org/10.1007/JHEP10(2018)112}{\emph{JHEP} {\bfseries 10}
  (2018) 112} [\href{https://arxiv.org/abs/1804.00471}{{\ttfamily
  1804.00471}}].

\bibitem{Martinez-Lema:2018ibw}
{\scshape NEXT} collaboration, \emph{{Calibration of the NEXT-White detector
  using $^{83m}\mathrm{Kr}$ decays}},
  \href{https://doi.org/10.1088/1748-0221/13/10/P10014}{\emph{JINST} {\bfseries
  13} (2018) P10014} [\href{https://arxiv.org/abs/1804.01780}{{\ttfamily
  1804.01780}}].

\bibitem{G4Phys}
``Geant4 physics reference manual.''
  \url{https://geant4-userdoc.web.cern.ch/UsersGuides/PhysicsReferenceManual/html/index.html}.

\bibitem{NEXT:2018kzp}
{\scshape NEXT} collaboration, \emph{{Electron drift properties in high
  pressure gaseous xenon}},
  \href{https://doi.org/10.1088/1748-0221/13/07/P07013}{\emph{JINST} {\bfseries
  13} (2018) P07013} [\href{https://arxiv.org/abs/1804.01680}{{\ttfamily
  1804.01680}}].

\bibitem{Monteiro:2007vz}
C.M.B.~Monteiro, L.M.P.~Fernandes, J.A.M.~Lopes, L.C.C.~Coelho,
  J.F.C.A.~Veloso, J.M.F.d.~Santos et~al., \emph{{Secondary Scintillation Yield
  in Pure Xenon}},
  \href{https://doi.org/10.1088/1748-0221/2/05/P05001}{\emph{JINST} {\bfseries
  2} (2007) P05001} [\href{https://arxiv.org/abs/physics/0702142}{{\ttfamily
  physics/0702142}}].

\bibitem{Ponkratenko:2000um}
O.A.~Ponkratenko, V.I.~Tretyak and Y.G.~Zdesenko, \emph{{The Event generator
  DECAY4 for simulation of double beta processes and decay of radioactive
  nuclei}}, \href{https://doi.org/10.1134/1.855784}{\emph{Phys. Atom. Nucl.}
  {\bfseries 63} (2000) 1282}
  [\href{https://arxiv.org/abs/nucl-ex/0104018}{{\ttfamily nucl-ex/0104018}}].

\bibitem{Cormen2001_intro_algorithms}
T.~Cormen, C.~Stein, R.~Rivest and C.~Leiserson, \emph{{Introduction to
  algorithms, 2nd ed}}, McGraw-Hill Higher Education, U.S.A. (2001).

\bibitem{Alvarez:2012as}
V.~Alvarez et~al., \emph{{Radiopurity control in the NEXT-100 double beta decay
  experiment: procedures and initial measurements}},
  \href{https://doi.org/10.1088/1748-0221/8/01/T01002}{\emph{JINST} {\bfseries
  8} (2013) T01002} [\href{https://arxiv.org/abs/1211.3961}{{\ttfamily
  1211.3961}}].

\bibitem{Alvarez:2014kvs}
{\scshape NEXT} collaboration, \emph{{Radiopurity assessment of the tracking
  readout for the NEXT double beta decay experiment}},
  \href{https://doi.org/10.1088/1748-0221/10/05/P05006}{\emph{JINST} {\bfseries
  10} (2015) P05006} [\href{https://arxiv.org/abs/1411.1433}{{\ttfamily
  1411.1433}}].

\bibitem{Cebrian:2017jzb}
{\scshape NEXT} collaboration, \emph{{Radiopurity assessment of the energy
  readout for the NEXT double beta decay experiment}},
  \href{https://doi.org/10.1088/1748-0221/12/08/T08003}{\emph{JINST} {\bfseries
  12} (2017) T08003} [\href{https://arxiv.org/abs/1706.06012}{{\ttfamily
  1706.06012}}].

\bibitem{Browne:2007nds}
E.~Browne and J.K.~Tuli, \emph{{Nuclear data sheets for $A = 137$}},
  {\emph{Nuclear data sheets} {\bfseries 108} (2007) 2173}.

\bibitem{Trzaska:2019kuk}
W.H.~Trzaska et~al., \emph{{Cosmic-ray muon flux at Canfranc Underground
  Laboratory}},
  \href{https://doi.org/10.1140/epjc/s10052-019-7239-9}{\emph{Eur. Phys. J. C}
  {\bfseries 79} (2019) 721}
  [\href{https://arxiv.org/abs/1902.00868}{{\ttfamily 1902.00868}}].

\bibitem{NEXT:2020qup}
{\scshape NEXT} collaboration, \emph{{Mitigation of backgrounds from cosmogenic
  $^{137}$Xe in xenon gas experiments using $^{3}$He neutron capture}},
  \href{https://doi.org/10.1088/1361-6471/ab8915}{\emph{J. Phys. G} {\bfseries
  47} (2020) 075001} [\href{https://arxiv.org/abs/2001.11147}{{\ttfamily
  2001.11147}}].

\bibitem{chadwick2011endf}
M.~Chadwick, M.~Herman, P.~Oblo{\v{z}}insk{\`y}, M.E.~Dunn, Y.~Danon, A.~Kahler
  et~al., \emph{Endf/b-vii. 1 nuclear data for science and technology: cross
  sections, covariances, fission product yields and decay data}, {\emph{Nuclear
  data sheets} {\bfseries 112} (2011) 2887}.

\bibitem{mughabghab2006atlas}
S.F.~Mughabghab, \emph{Atlas of Neutron Resonances: Resonance Parameters and
  Thermal Cross Sections. Z= 1-100}, Elsevier (2006).

\bibitem{Albert:2016vmb}
J.B.~Albert, S.J.~Daugherty, T.N.~Johnson, T.~O'Conner, L.~Kaufman, A.~Couture
  et~al., \emph{{Measurement of neutron capture on $^{136}$Xe}},
  \href{https://doi.org/10.1103/PhysRevC.94.034617}{\emph{Phys. Rev.}
  {\bfseries C94} (2016) 034617}
  [\href{https://arxiv.org/abs/1605.05794}{{\ttfamily 1605.05794}}].

\bibitem{Albert:2013gpz}
{\scshape EXO-200} collaboration, \emph{{Improved measurement of the
  $2\nu\beta\beta$ half-life of $^{136}$Xe with the EXO-200 detector}},
  \href{https://doi.org/10.1103/PhysRevC.89.015502}{\emph{Phys. Rev. C}
  {\bfseries 89} (2014) 015502}
  [\href{https://arxiv.org/abs/1306.6106}{{\ttfamily 1306.6106}}].

\bibitem{Kotila:2012zza}
J.~Kotila and F.~Iachello, \emph{{Phase space factors for double-$\beta$
  decay}}, \href{https://doi.org/10.1103/PhysRevC.85.034316}{\emph{Phys. Rev.
  C} {\bfseries 85} (2012) 034316}
  [\href{https://arxiv.org/abs/1209.5722}{{\ttfamily 1209.5722}}].

\bibitem{Agostini:2022zub}
M.~Agostini, G.~Benato, J.A.~Detwiler, J.~Men\'endez and F.~Vissani,
  \emph{{Toward the discovery of matter creation with neutrinoless double-beta
  decay}},  \href{https://arxiv.org/abs/2202.01787}{{\ttfamily 2202.01787}}.

\bibitem{Menendez:2017fdf}
J.~Men\'endez, \emph{{Neutrinoless $\beta\beta$ decay mediated by the exchange
  of light and heavy neutrinos: The role of nuclear structure correlations}},
  \href{https://doi.org/10.1088/1361-6471/aa9bd4}{\emph{J. Phys. G} {\bfseries
  45} (2018) 014003} [\href{https://arxiv.org/abs/1804.02105}{{\ttfamily
  1804.02105}}].

\bibitem{Horoi:2015tkc}
M.~Horoi and A.~Neacsu, \emph{{Shell model predictions for $^{124}$Sn
  double-$\beta$ decay}},
  \href{https://doi.org/10.1103/PhysRevC.93.024308}{\emph{Phys. Rev. C}
  {\bfseries 93} (2016) 024308}
  [\href{https://arxiv.org/abs/1511.03711}{{\ttfamily 1511.03711}}].

\bibitem{Coraggio:2020hwx}
L.~Coraggio, A.~Gargano, N.~Itaco, R.~Mancino and F.~Nowacki,
  \emph{{Calculation of the neutrinoless double-$\beta$ decay matrix element
  within the realistic shell model}},
  \href{https://doi.org/10.1103/PhysRevC.101.044315}{\emph{Phys. Rev. C}
  {\bfseries 101} (2020) 044315}
  [\href{https://arxiv.org/abs/2001.00890}{{\ttfamily 2001.00890}}].

\bibitem{Coraggio:2022vgy}
L.~Coraggio, N.~Itaco, G.~De~Gregorio, A.~Gargano, R.~Mancino and F.~Nowacki,
  \emph{{Shell-model calculation of $^{100}$Mo double-$\beta$ decay}},
  \href{https://doi.org/10.1103/PhysRevC.105.034312}{\emph{Phys. Rev. C}
  {\bfseries 105} (2022) 034312}
  [\href{https://arxiv.org/abs/2203.01013}{{\ttfamily 2203.01013}}].

\bibitem{Mustonen:2013zu}
M.T.~Mustonen and J.~Engel, \emph{{Large-scale calculations of the
  double-\ensuremath{\beta} decay of $^{76}Ge,^{130}Te,^{136}Xe$, and
  $^{150}Nd$ in the deformed self-consistent Skyrme quasiparticle random-phase
  approximation}},
  \href{https://doi.org/10.1103/PhysRevC.87.064302}{\emph{Phys. Rev. C}
  {\bfseries 87} (2013) 064302}
  [\href{https://arxiv.org/abs/1301.6997}{{\ttfamily 1301.6997}}].

\bibitem{Hyvarinen:2015bda}
J.~Hyv\"arinen and J.~Suhonen, \emph{{Nuclear matrix elements for
  $0\nu\beta\beta$ decays with light or heavy Majorana-neutrino exchange}},
  \href{https://doi.org/10.1103/PhysRevC.91.024613}{\emph{Phys. Rev. C}
  {\bfseries 91} (2015) 024613}.

\bibitem{Simkovic:2018hiq}
F.~\v{S}imkovic, A.~Smetana and P.~Vogel, \emph{{$0\nu\beta\beta$ nuclear
  matrix elements, neutrino potentials and $\mathrm{SU}(4)$ symmetry}},
  \href{https://doi.org/10.1103/PhysRevC.98.064325}{\emph{Phys. Rev. C}
  {\bfseries 98} (2018) 064325}
  [\href{https://arxiv.org/abs/1808.05016}{{\ttfamily 1808.05016}}].

\bibitem{Fang:2018tui}
D.-L.~Fang, A.~Faessler and F.~Simkovic, \emph{{0\ensuremath{\nu}$\beta\beta$
  -decay nuclear matrix element for light and heavy neutrino mass mechanisms
  from deformed quasiparticle random-phase approximation calculations for
  $^{76}$Ge, $^{82}$Se, $^{130}$Te, $^{136}$Xe , and $^{150}$Nd with isospin
  restoration}}, \href{https://doi.org/10.1103/PhysRevC.97.045503}{\emph{Phys.
  Rev. C} {\bfseries 97} (2018) 045503}
  [\href{https://arxiv.org/abs/1803.09195}{{\ttfamily 1803.09195}}].

\bibitem{Terasaki:2020ndc}
J.~Terasaki, \emph{{Strength of the isoscalar pairing interaction determined by
  a relation between double-charge change and double-pair transfer for double-
  $\beta$ decay}},
  \href{https://doi.org/10.1103/PhysRevC.102.044303}{\emph{Phys. Rev. C}
  {\bfseries 102} (2020) 044303}
  [\href{https://arxiv.org/abs/2003.03542}{{\ttfamily 2003.03542}}].

\bibitem{Rodriguez:2010mn}
T.R.~Rodriguez and G.~Martinez-Pinedo, \emph{{Energy density functional study
  of nuclear matrix elements for neutrinoless $\beta\beta$ decay}},
  \href{https://doi.org/10.1103/PhysRevLett.105.252503}{\emph{Phys. Rev. Lett.}
  {\bfseries 105} (2010) 252503}
  [\href{https://arxiv.org/abs/1008.5260}{{\ttfamily 1008.5260}}].

\bibitem{LopezVaquero:2013yji}
N.~L\'opez~Vaquero, T.R.~Rodr\'\i{}guez and J.L.~Egido, \emph{{Shape and
  pairing fluctuations effects on neutrinoless double beta decay nuclear matrix
  elements}}, \href{https://doi.org/10.1103/PhysRevLett.111.142501}{\emph{Phys.
  Rev. Lett.} {\bfseries 111} (2013) 142501}
  [\href{https://arxiv.org/abs/1401.0650}{{\ttfamily 1401.0650}}].

\bibitem{Song:2017ktj}
L.S.~Song, J.M.~Yao, P.~Ring and J.~Meng, \emph{{Nuclear matrix element of
  neutrinoless double-$\beta$ decay: Relativity and short-range correlations}},
  \href{https://doi.org/10.1103/PhysRevC.95.024305}{\emph{Phys. Rev. C}
  {\bfseries 95} (2017) 024305}
  [\href{https://arxiv.org/abs/1702.02448}{{\ttfamily 1702.02448}}].

\bibitem{Barea:2015kwa}
J.~Barea, J.~Kotila and F.~Iachello, \emph{{$0\nu\beta\beta$ and
  $2\nu\beta\beta$ nuclear matrix elements in the interacting boson model with
  isospin restoration}},
  \href{https://doi.org/10.1103/PhysRevC.91.034304}{\emph{Phys. Rev. C}
  {\bfseries 91} (2015) 034304}
  [\href{https://arxiv.org/abs/1506.08530}{{\ttfamily 1506.08530}}].

\bibitem{Deppisch:2020ztt}
F.F.~Deppisch, L.~Graf, F.~Iachello and J.~Kotila, \emph{{Analysis of light
  neutrino exchange and short-range mechanisms in $0\nu\beta\beta$ decay}},
  \href{https://doi.org/10.1103/PhysRevD.102.095016}{\emph{Phys. Rev. D}
  {\bfseries 102} (2020) 095016}
  [\href{https://arxiv.org/abs/2009.10119}{{\ttfamily 2009.10119}}].

\end{thebibliography}\endgroup

\end{document}